\documentclass[5p,times]{elsarticle}

\usepackage{nidanfloat}
\usepackage{graphicx}
\usepackage{lineno,hyperref}
\usepackage{here}
\modulolinenumbers[1]

\begin{document}

\begin{frontmatter}

\title{Development and operational experience of magnetic horn system for T2K experiment}

\author[KEKaddress]{T. Sekiguchi\corref{mycorrespondingauthor}\fnref{mytel}}
\cortext[mycorrespondingauthor]{Corresponding author}
\fntext[mytel]{Tel: +81-29-864-5200, ext. 4655}
\ead{tetsuro.sekiguchi@kek.jp}


\author[KEKaddress]{K. Bessho}
\author[KEKaddress]{Y. Fujii}
\author[KEKaddress]{M. Hagiwara}
\author[KEKaddress]{T. Hasegawa}
\author[KEKaddress]{K. Hayashi}
\author[KEKaddress]{T. Ishida}
\author[KEKaddress]{T. Ishii}
\author[KEKaddress]{H. Kobayashi}
\author[KEKaddress]{T. Kobayashi}
\author[KEKaddress]{S. Koike\fnref{koike}}
\fntext[koike]{Deceased.}
\author[KEKaddress]{K. Koseki}
\author[KEKaddress]{T. Maruyama}
\author[KEKaddress]{H. Matsumoto}
\author[KEKaddress]{T. Nakadaira}
\author[KEKaddress]{K. Nakamura}
\author[KEKaddress]{K. Nakayoshi}
\author[KEKaddress]{K. Nishikawa}
\author[KEKaddress]{Y. Oyama}
\author[KEKaddress]{K. Sakashita}
\author[KEKaddress]{M. Shibata}
\author[KEKaddress]{Y. Suzuki}
\author[KEKaddress]{M. Tada}
\author[KEKaddress]{K. Takahashi}
\author[KEKaddress]{T. Tsukamoto}
\author[KEKaddress]{Y. Yamada}
\author[KEKaddress]{Y. Yamanoi}
\author[KEKaddress]{H. Yamaoka}

\author[Kyotoaddress]{A.K. Ichikawa}
\author[Kyotoaddress]{H. Kubo}

\author[CUaddress]{Z. Butcher\fnref{cu1}}
\fntext[cu1]{Now at University of Massachusetts, Amherst, Massachusetts, USA} 
\author[CUaddress]{S. Coleman}
\author[CUaddress]{A. Missert}
\author[CUaddress]{J. Spitz\fnref{cu2}}
\fntext[cu2]{Now at Massachusetts Institute of Technology, Cambridge, Massachusetts, USA}
\author[CUaddress]{E.D. Zimmerman}

\author[LSUaddress]{M. Tzanov}

\author[BEaddress]{L. Bartoszek}

\address[KEKaddress]{High Energy Accelerator Research Organization (KEK), Tsukuba, Ibaraki, Japan}
\address[Kyotoaddress]{Kyoto University, Department of Physics, Kyoto, Japan}
\address[CUaddress]{University of Colorado Boulder, Department of Physics, Boulder, Colorado, USA}
\address[LSUaddress]{Louisiana State University, Department of Physics and Astronomy, Baton Rouge, Louisiana, USA}
\address[BEaddress]{Bartoszek Engineering, Aurora, Illinois, USA}

\begin{abstract}
A magnetic horn system to be operated at a pulsed current of
320 kA and to survive high-power proton beam operation at 750 kW was developed for the T2K experiment.
The first set of T2K magnetic horns was operated for over 12 million pulses 
during the four years of operation from 2010 to 2013, under a maximum beam power of 230 kW,
and $6.63\times10^{20}$ protons were exposed to the production target.
No significant damage was observed throughout this period.
This successful operation of the T2K magnetic horns led to the discovery of the $\nu_{\mu}\rightarrow\nu_e$
oscillation phenomenon in 2013 by the T2K experiment.
In this paper, details of the design, construction, and operation experience of the T2K magnetic horns are described.
\end{abstract}

\begin{keyword}
Magnetic horn\sep Neutrino\sep Beamline\sep T2K\sep J-PARC  
\end{keyword}
\end{frontmatter}



%
%
\section{Introduction}
\label{sec:intro}

%
%
T2K is the Tokai-to-Kamioka long-baseline neutrino oscillation experiment \cite{T2K-LOI}.
The physical motivations behind the T2K experiment are the determination of the value of oscillation parameter $\theta_{13}$,
by searching for the $\nu_{\mu}\rightarrow \nu_{e}$ oscillation phenomenon,
and a precise measurement of oscillation parameter $\theta_{23}$ by measuring $\nu_{\mu}\rightarrow\nu_{\mu}$ oscillation.
The T2K experiment utilizes a high-intensity neutrino beam from proton accelerators at the Japan
Proton Accelerator Research Complex (J-PARC) \cite{J-PARC}, which have a design beam power of 750 kW,
and the Super-Kamiokande detector \cite{SK-NIM} as a far detector, which is at a 295-km distance from J-PARC.
An important feature of the T2K neutrino beam is that it is narrow-band and uses an off-axis beam method \cite{off-axis}. 
The beam axis direction is oriented at 2.5$^{\circ}$ with respect
to the direction of the Super-Kamiokande detector. In this case, the neutrino energy distribution at the Super-Kamiokande
detector has a narrow distribution peaking at 0.6 GeV, for which the oscillation probability at the 295-km baseline is at a maximum.
    
\begin{figure}[H]
\centering
\includegraphics[clip,width=0.5\textwidth,bb=70 50 800 540]{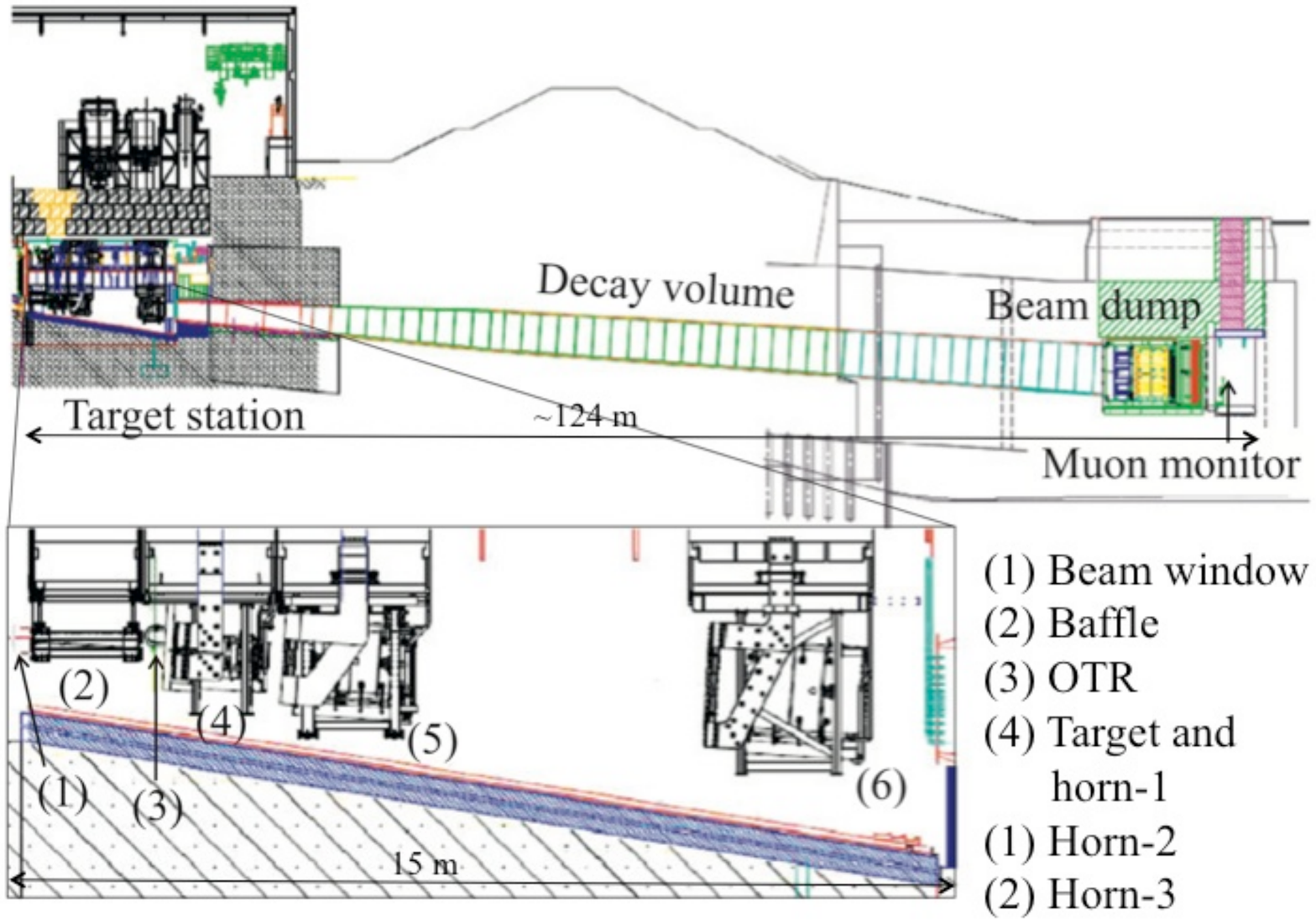}
\caption{Schematic drawing of T2K secondary beamline.}
\label{fig:secondaryline}
\end{figure}

A proton beam with kinetic energy of 30 GeV (design intensity = $3.3 \times 10^{14}$ protons/pulse) 
is extracted from the Main Ring to the T2K neutrino beamline once every 2.1 s, 
and is exposed to a production target composed of a graphite rod (2.6 cm in diameter and 90 cm in length). 
Secondary particles such as pions and kaons are focused in the forward direction by three magnetic horns,
where the target is embedded in the magnetic horn located in the furthest upstream.
The secondary particles decay in flight and produce 
muon neutrinos (purity $\sim$ 99\%) when they travel along a 100-m-long decay volume. 
A beam dump, which is composed of segmented graphite blocks, is positioned at the end of the decay volume
to absorb all the hadrons. The target, magnetic horns, decay volume, and beam dump are enclosed in
a gigantic iron vessel ($\sim$1,500 m$^3$), the inner volume of which is filled with 1 atm of helium gas in order to reduce
tritium ($^3$H) and nitrogen oxide (NOx) production. The helium atmosphere also works to greatly reduce the
oxidation of the graphite material of the beam dump. A schematic drawing of the T2K secondary neutrino beamline
is shown in Fig. \ref{fig:secondaryline} and additional details of the T2K neutrino beamline are given in \cite{T2K-NIM}.

In this paper, details of the T2K magnetic horn system are provided. Introductory information is given
in Section \ref{sec:horns}, electrical properties in Section \ref{sec:electrical}, cooling performance in Section \ref{sec:cooling}, 
mechanical properties in Section \ref{sec:mechanical}, details of remote maintenance
issues in Section \ref{sec:remote} and, finally, operational experience during beam operation for physics 
data-taking is reported in Section \ref{sec:operation}.

\section{T2K magnetic horns}
\label{sec:horns}

The T2K magnetic horn system consists of three horns: horn-1, horn-2, and horn-3, in order from upstream.
Secondary particles are emitted from the target with high angles of a few hundred mrad and low momenta of 1-4 GeV/c.
A field integral of 1.3 T$\cdot$m is required in order to focus those particles efficiently. 
The target is inserted inside the inner conductor of horn-1 so as to provide high collection efficiency, and 
a peak current of 320 kA is adopted for the T2K magnetic horns in order to achieve high magnetic field strength.
The design philosophy and design optimization of the T2K magnetic horns are described in \cite{T2KhornDesign}.

\begin{figure}
\centering
\includegraphics[clip,width=0.5\textwidth,bb=0 0 721 378]{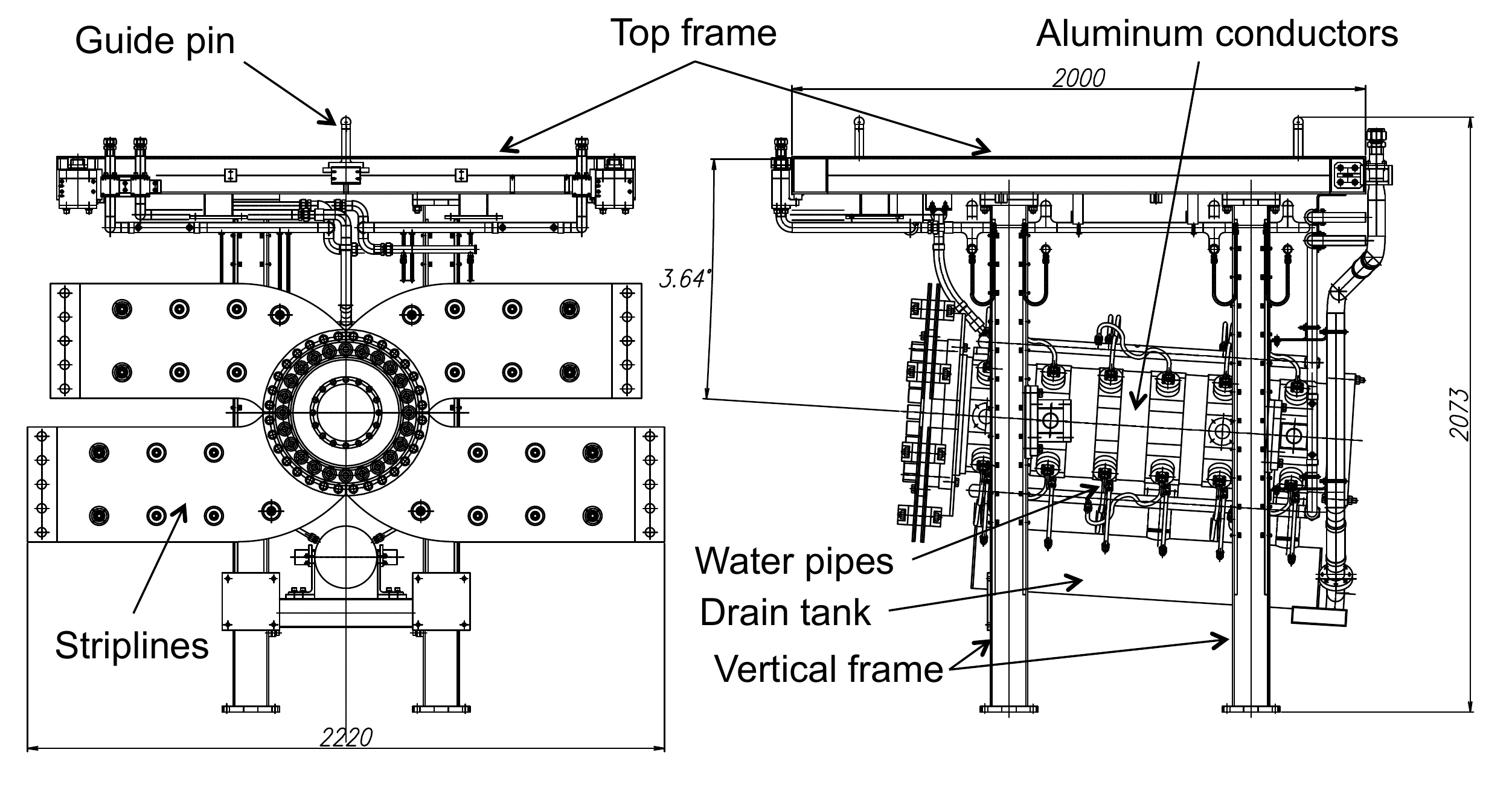}
\includegraphics[clip,width=0.5\textwidth,bb=0 0 721 358]{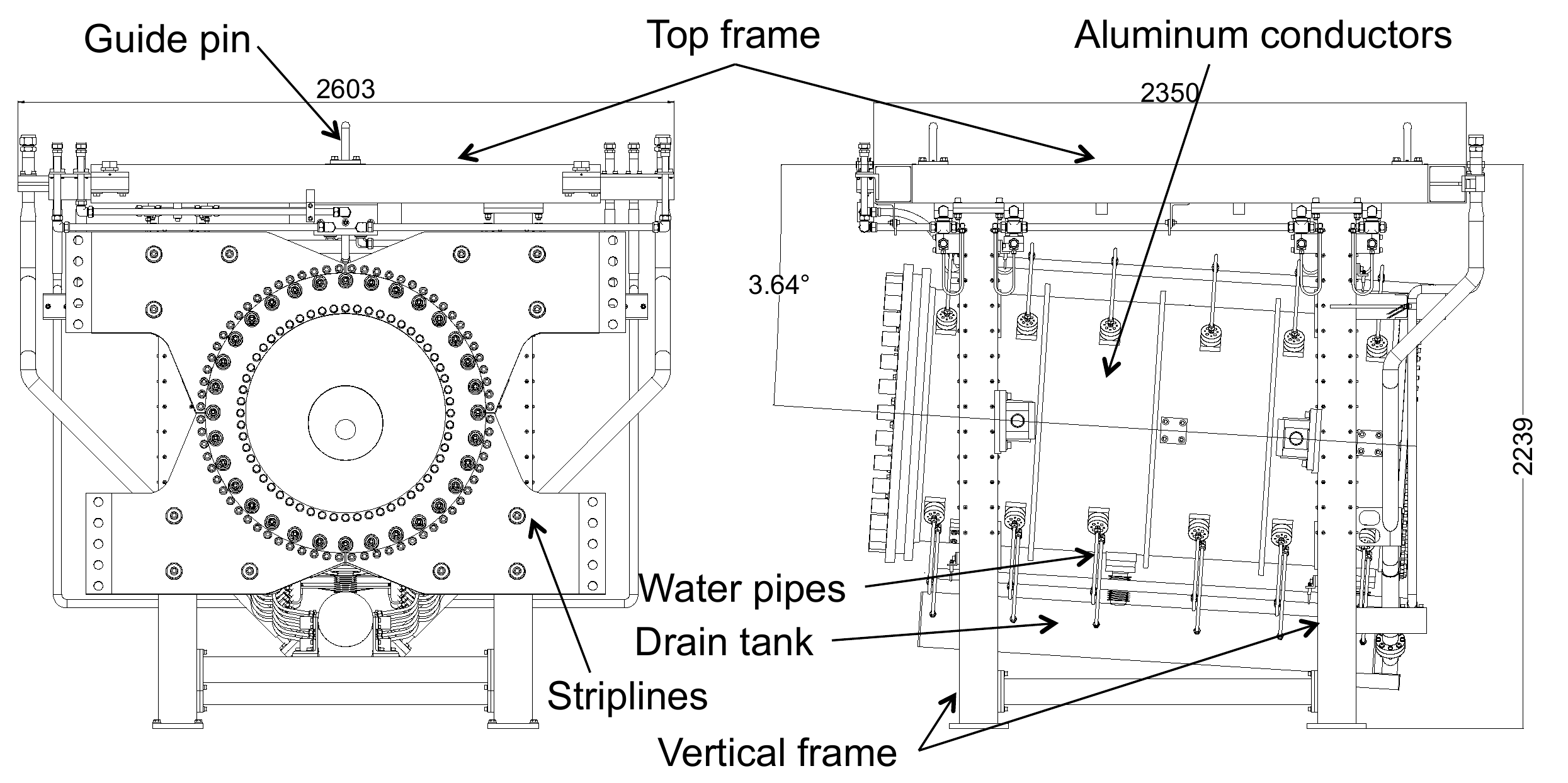}
\includegraphics[clip,width=0.5\textwidth,bb=0 0 720 442]{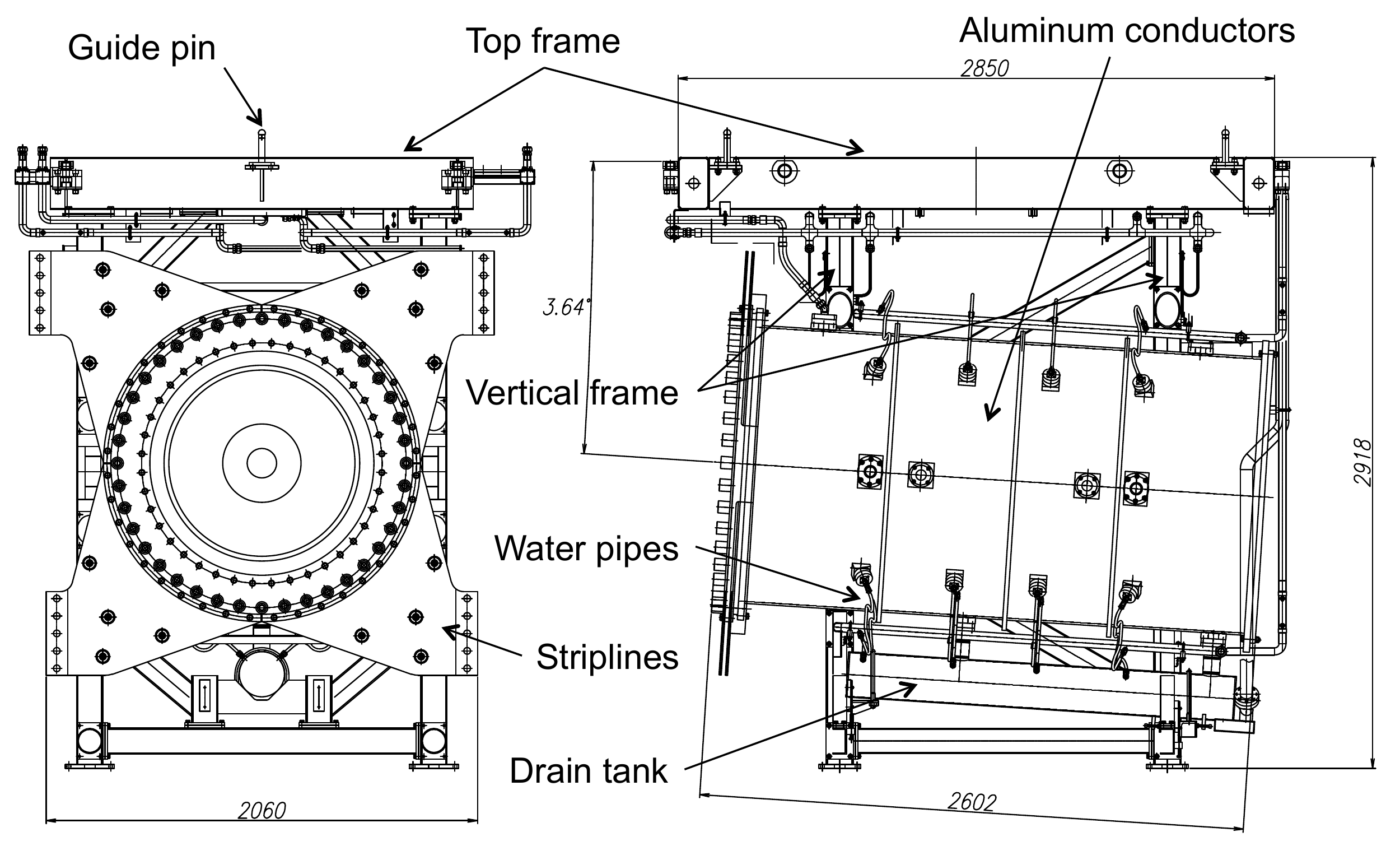}
\caption{Drawings of T2K magnetic horns: horn-1 front view (top left), horn-1 side view (top right),
horn-2 front view (middle left), horn-2 side view (middle right), horn-3 front view (bottom left), and horn-3 side view (bottom right).}
\label{fig:horn_drawing}
\end{figure}

\begin{figure}
\centering
\includegraphics[clip,width=0.5\textwidth,bb=50 20 720 341]{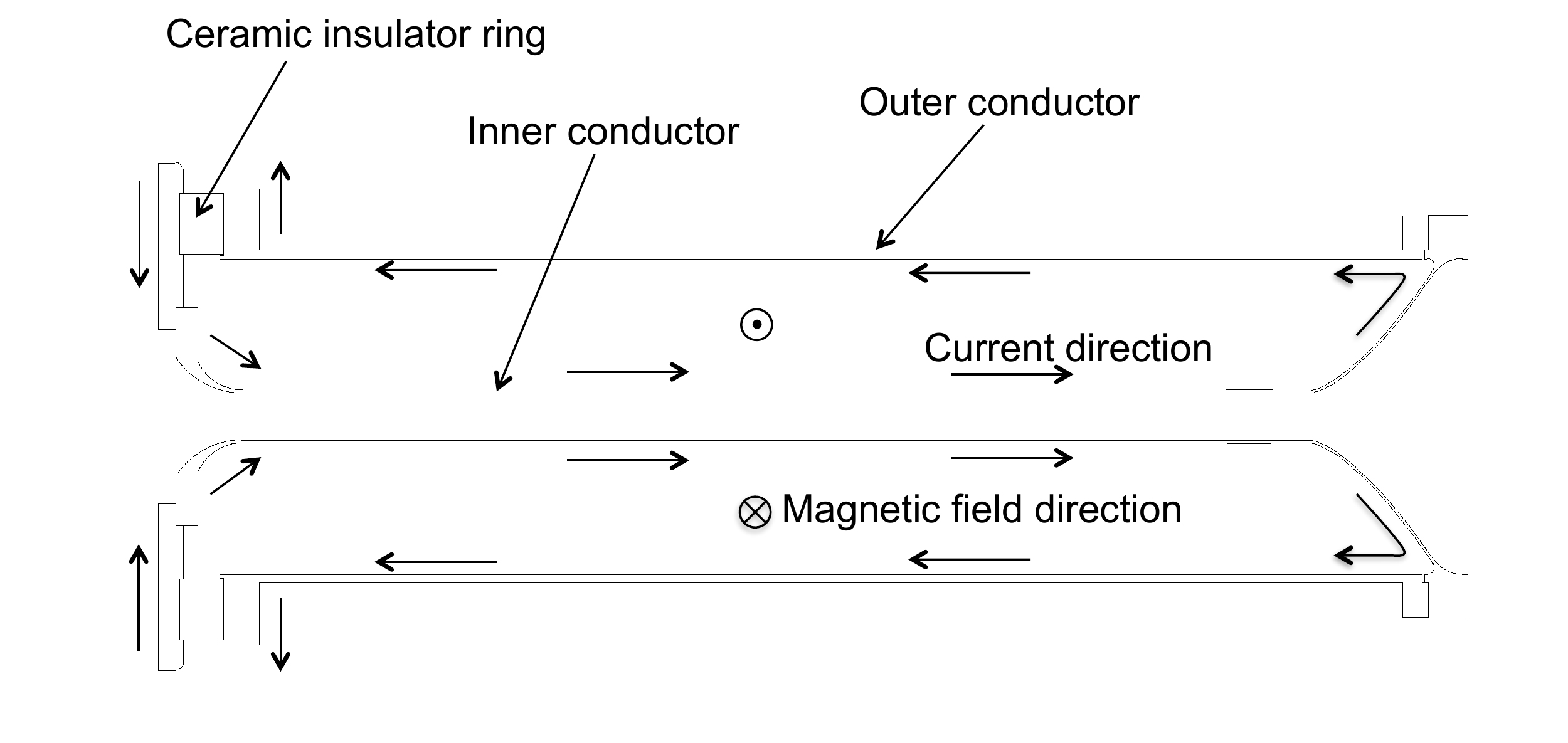}
\caption{Cross section of horn-1 conductors showing current and magnetic field directions.}
\label{fig:horn1cross}
\end{figure}

\begin{figure}
\centering
\includegraphics[clip,width=0.5\textwidth,bb=0 100 595 692]{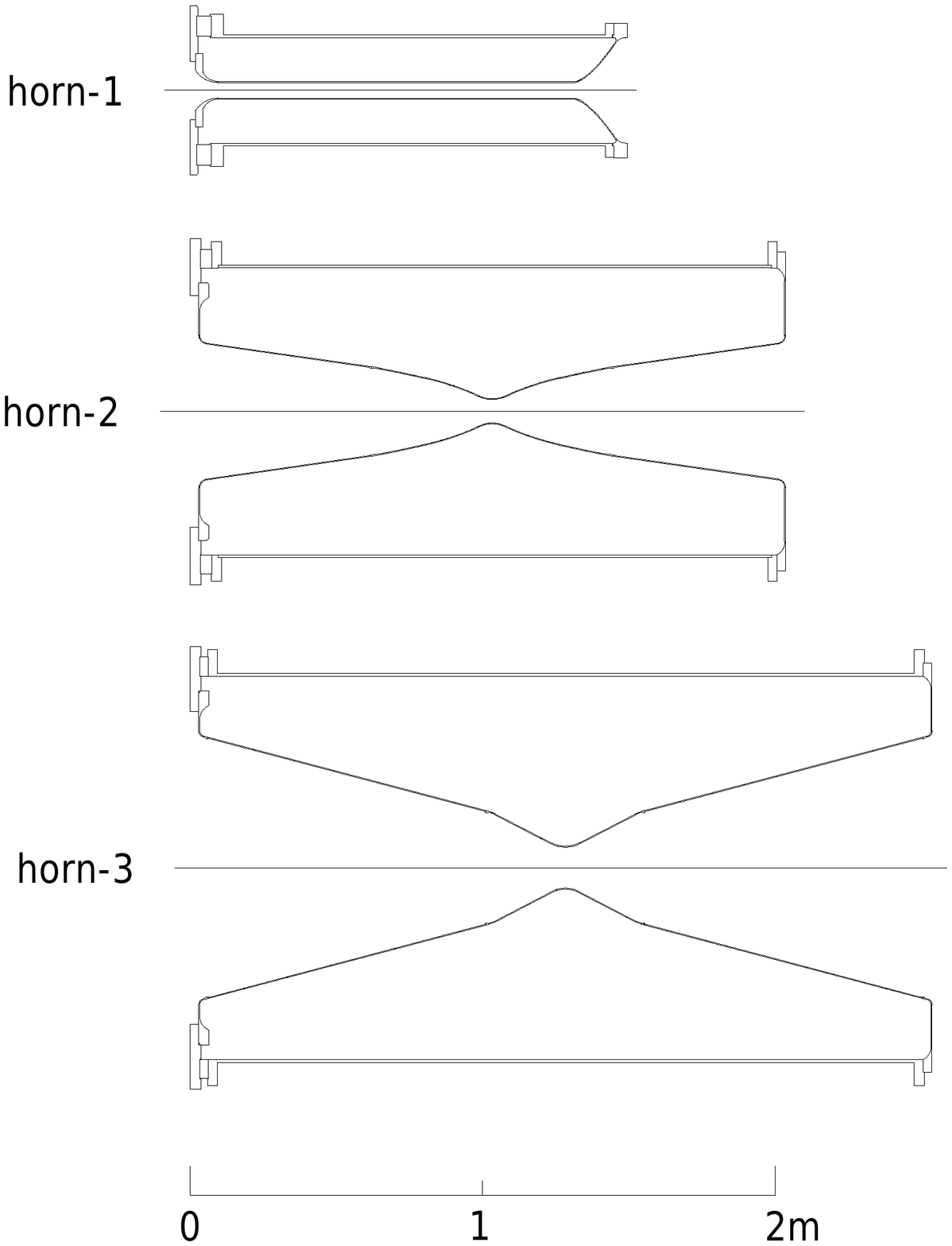}
\caption{Aluminum conductors of horn-1, horn-2, and horn-3.}
\label{fig:horn-shape}
\end{figure}

The T2K horns consist of aluminum conductors, current feeding striplines, water and gas plumbing, and support frames.
Figure \ref{fig:horn_drawing} shows drawings of the T2K magnetic horns.
The aluminum conductors have coaxial conductor structures, with inner and outer conductors
that are connected at the downstream end but
insulated at the upstream end, as shown in Fig. \ref{fig:horn1cross}.
The current first enters the inner conductor
from the upstream end, returns through the outer conductor, and exits from
the upstream end. A toroidal magnetic field is generated in the region between
the inner and outer conductors.
Drawings of the aluminum conductors of horn-1, horn-2, and horn-3 are shown in Fig. \ref{fig:horn-shape},
while typical dimensions are also listed in Table \ref{tab:horndim}.
\begin{table}[htb]
\centering
\caption{Typical dimensions of T2K magnetic horns.}
\begin{tabular}{lccc}
\hline
Parameters & horn-1 & horn-2 & horn-3 \\ \hline
Inner diameter & 54 mm & 80 mm & 140 mm \\
Outer diameter & 400 mm & 1,000 mm & 1,400 mm \\
Length & 1.5 m & 2.0 m & 2.5 m \\ \hline
\end{tabular}
\label{tab:horndim}
\end{table}
The inner conductor thickness is set to 3 mm in order to reduce interactions of the secondary particles, while the thickness
of the outer conductors is 10 mm.

The T2K magnetic horns are designed for a peak current of 320 kA. 
A sinusoidal pulsed current, with a pulse width of 2.4-3.6 ms, is applied to the conductors, and
the beam is exposed for 5 $\mu$s at the peak of the sinusoidal waveform.
The magnetic field inside the conductors is given by
\begin{equation}
B = \frac{\mu I}{2\pi r}, \label{for:bfield} \label{eq:Bfield}
\end{equation}
where $I$ is the current, $\mu$ is the magnetic permeability, and $r$ is the radial distance from the central axis.
The peak current of 320 kA produces a maximum magnetic field of 2.1 T at the outer surface
of the horn-1 inner conductor ($r = 3$ cm). The maximum magnetic fields at horn-2 and horn-3 are
1.5 and 0.9 T, respectively. The electrical properties such as the current generation (power supply, transfer line, etc.)
and the magnetic field measurements are described in Section \ref{sec:electrical}.

The aluminum conductors (especially the inner conductors) are pressurized 
by the Lorentz force created by the current flow and magnetic field. The inner conductors are 3-mm thick, 
however, they must withstand the Lorentz force and also the thermal shock that will be described below.
Magnetic horn conductors should have as low resistivity as possible for high-current operation
and as low material density as possible to reduce interactions of secondary particles,
however, they should also have high resistance to the Lorentz force and the heat shock from the beam exposure and Joule loss.
Thus, the aluminum alloy, A6061-T6, which is often used for magnetic horns in various neutrino experiments, 
is selected as a conductor material.
This alloy has tensile (yield) strength of 310 MPa (275 MPa), which is a significantly higher value
than that of pure aluminum (tensile strength $\sim$70 MPa), 
and has resistivity of 4.0 $\times 10^{-8} \Omega\cdot$m (233\% (161\%) of that of pure copper (pure aluminum)).
A characteristic of the alloy under fatigue shows a reduction in tensile strength to 95 MPa after $5\times10^8$
 cycle repetitive force is applied \cite{Alhandbook}.
The mechanical properties of the T2K magnetic horns are explained in Section \ref{sec:mechanical}.

\begin{figure}
\centering
\includegraphics[clip,width=0.45\textwidth,bb=10 10 724 470]{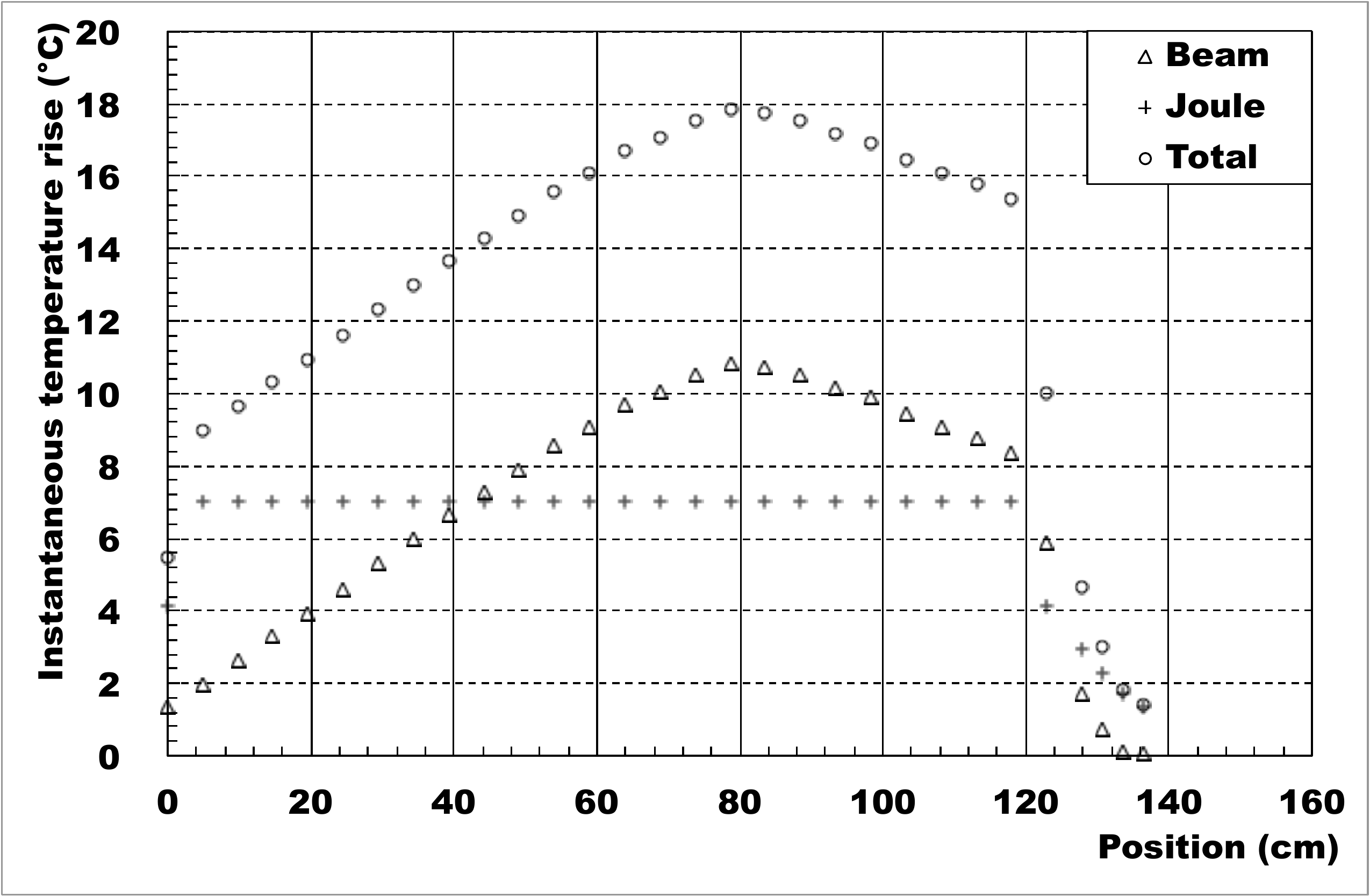}
\includegraphics[clip,width=0.45\textwidth,bb=10 10 724 470]{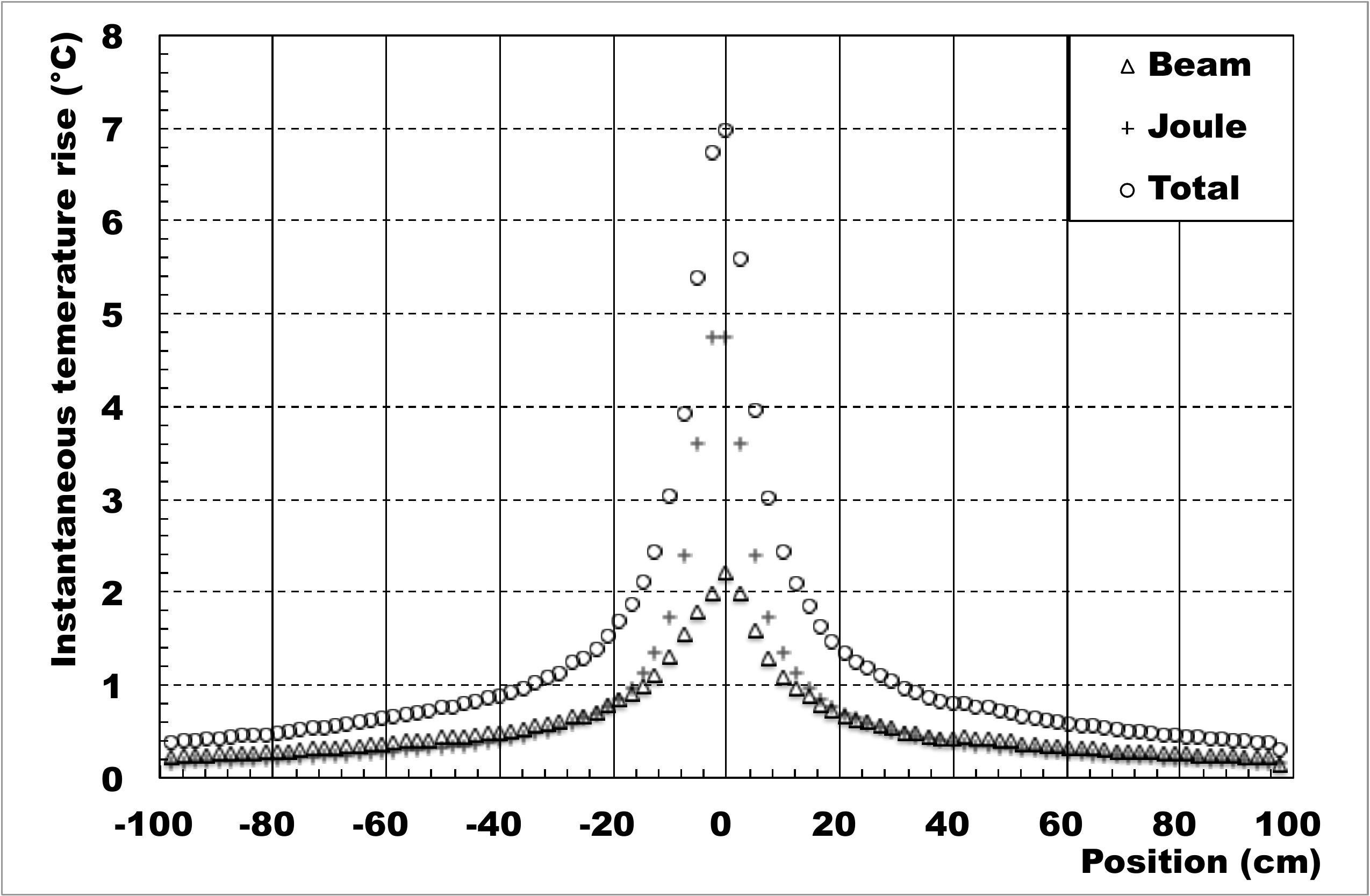} 
\includegraphics[clip,width=0.45\textwidth,bb=10 10 724 470]{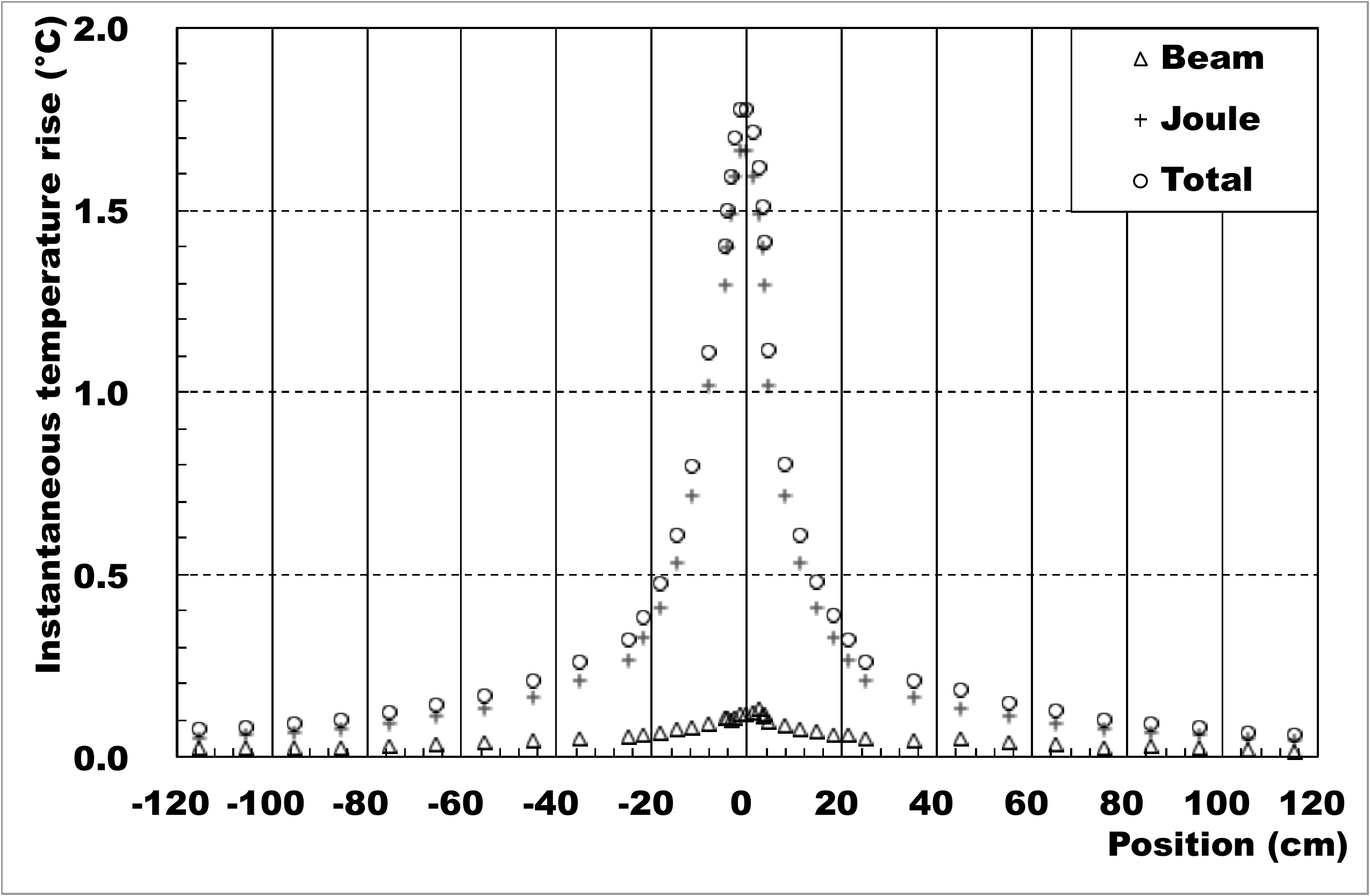}
\caption{Instantaneous temperature increase at inner conductors as a function of longitudinal position for horn-1 (top), horn-2 (middle), 
and horn-3 (bottom). The instantaneous temperature increases due to beam exposure and Joule heating are represented by
triangles and crosses, respectively. The total instantaneous increases in temperature are also indicated by circles.}
\label{fig:temprise}
\end{figure}

When the secondary particles pass through the horn conductors, a large amount of heat deposit arises
at the conductors. In addition, the 320-kA pulsed current also produces a heat deposit due to Joule heating.
The heat deposits in each horn from both beam exposure and Joule heating are summarized in Table \ref{tab:heatdepo}.
\begin{table}[htb]
\centering
 \caption{Summary of heat deposit in each horn. Heat deposit from beam exposure is based on the design intensity
 of $3.3\times 10^{14}$ protons/pulse. Joule heating for each horn is estimated for pulse widths of 2.4 (horn-1) 
 and 3.6 ms (horn-2 and horn-3). The calculation of the total heat deposit in units of kW is based on a 2.1-s cycle.}
\begin{tabular}{l|c|c|c|c|c|c}
\hline
 & \multicolumn{2}{c|}{horn-1} & \multicolumn{2}{c|}{horn-2} & \multicolumn{2}{c}{horn-3} \\ \cline{2-7}
 & inner & outer & inner & outer & inner & outer \\ \hline
 Beam (kJ) & 14.7  & 9.7  & 4.1  & 7.6  & 1.1  & 2.4  \\
 Joule (kJ) & 11.7  & 0.6  & 6.1  & 0.4  & 4.1  & 0.2  \\ \hline
 Total (kJ)         & \multicolumn{2}{c|}{36.7} & \multicolumn{2}{c|}{18.2} & \multicolumn{2}{c}{7.8} \\
 Total (kW)       & \multicolumn{2}{c|}{17.5} & \multicolumn{2}{c|}{8.7} & \multicolumn{2}{c}{3.7} \\ \hline
 \end{tabular}
 \label{tab:heatdepo}
 \end{table}
 For 750-kW beam operation (30 GeV, $3.3\times 10^{14}$ protons/pulse, 2.1-s cycle), the heat deposit in each horn
is 17.5, 8.7, and 3.7 kW for horn-1, horn-2, and horn-3, respectively.
The 5-$\mu$s beam exposure creates an instantaneous increase in temperature, which then causes thermal stress.
Joule heating also creates an instantaneous temperature increase.
The instantaneous rise in temperature at each horn is shown in Fig. \ref{fig:temprise}.
These heat deposits are removed by water cooling, as water
nozzles attached to the outer conductors spray water onto the inner conductors.
Details of the cooling performance are given in Section \ref{sec:cooling}.

\begin{figure}
\centering
\includegraphics[clip,width=0.35\textwidth,bb=0 0 586 394]{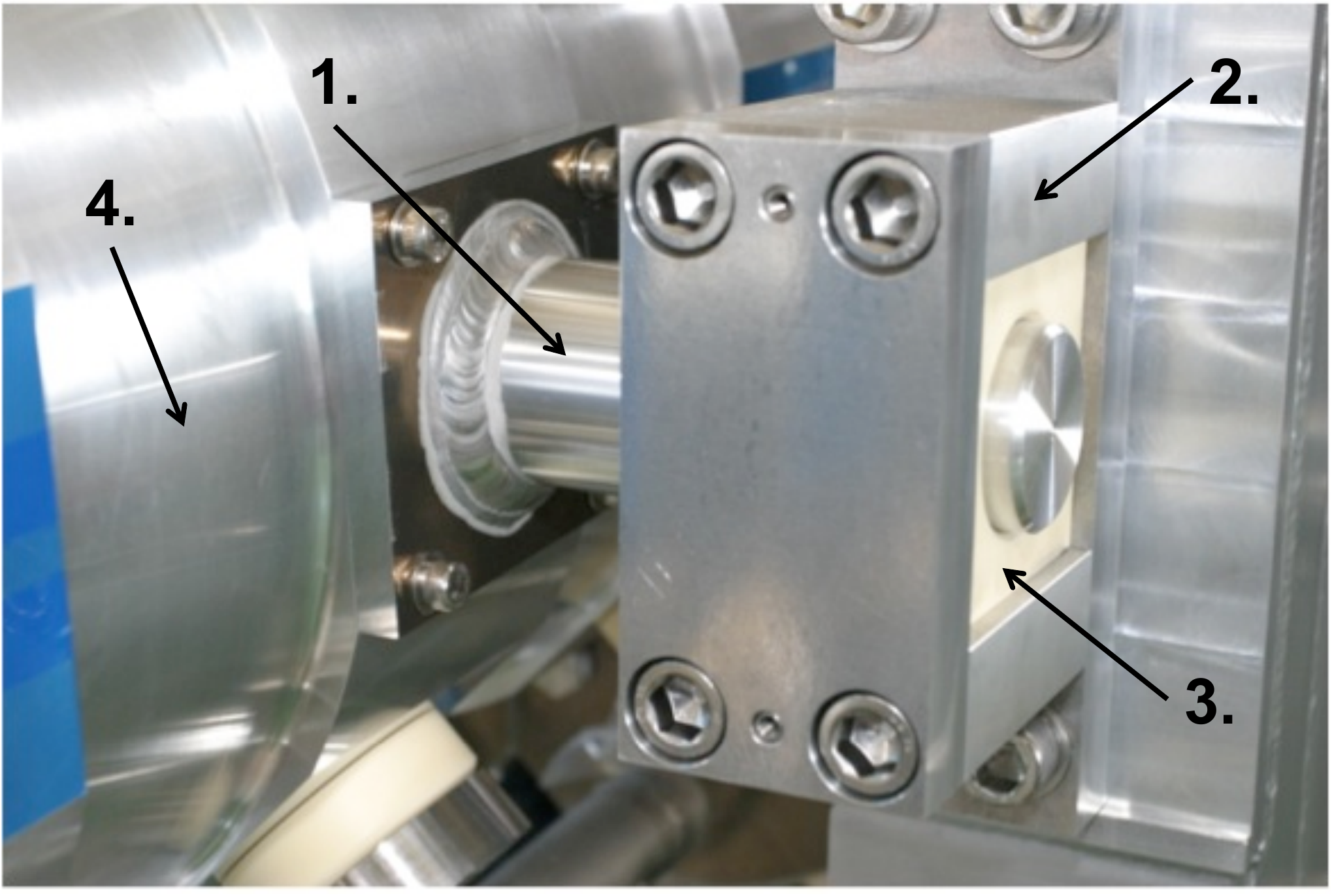}
\caption{Horn support components: 1. Support pin; 2. Support fixture attached to
vertical frame; 3. Ceramic insulation block; and 4. Outer conductor.}
\label{fig:insulation_support}
\end{figure}

The inner and outer conductor assembly is supported by aluminum frames.
The central axis of the conductors is angled downward by 3.637$^\circ$ in order
to obtain an off-axis angle for the Super-Kamiokande detector of 2.5$^\circ$.
The aluminum support structures consist of four vertical columns and a top frame.
There are water cooling channels inside the vertical columns.
The outer conductors are both supported and insulated by four alumina ceramic blocks
embedded in the vertical columns, as shown in Fig. \ref{fig:insulation_support}.
The ceramic blocks in the upstream columns are fixed, while those in the downstream columns are free to
slip along the axis to absorb the thermal expansion of the outer conductor.

\begin{figure}
\centering
\includegraphics[clip,width=0.5\textwidth,bb=0 0 721 541]{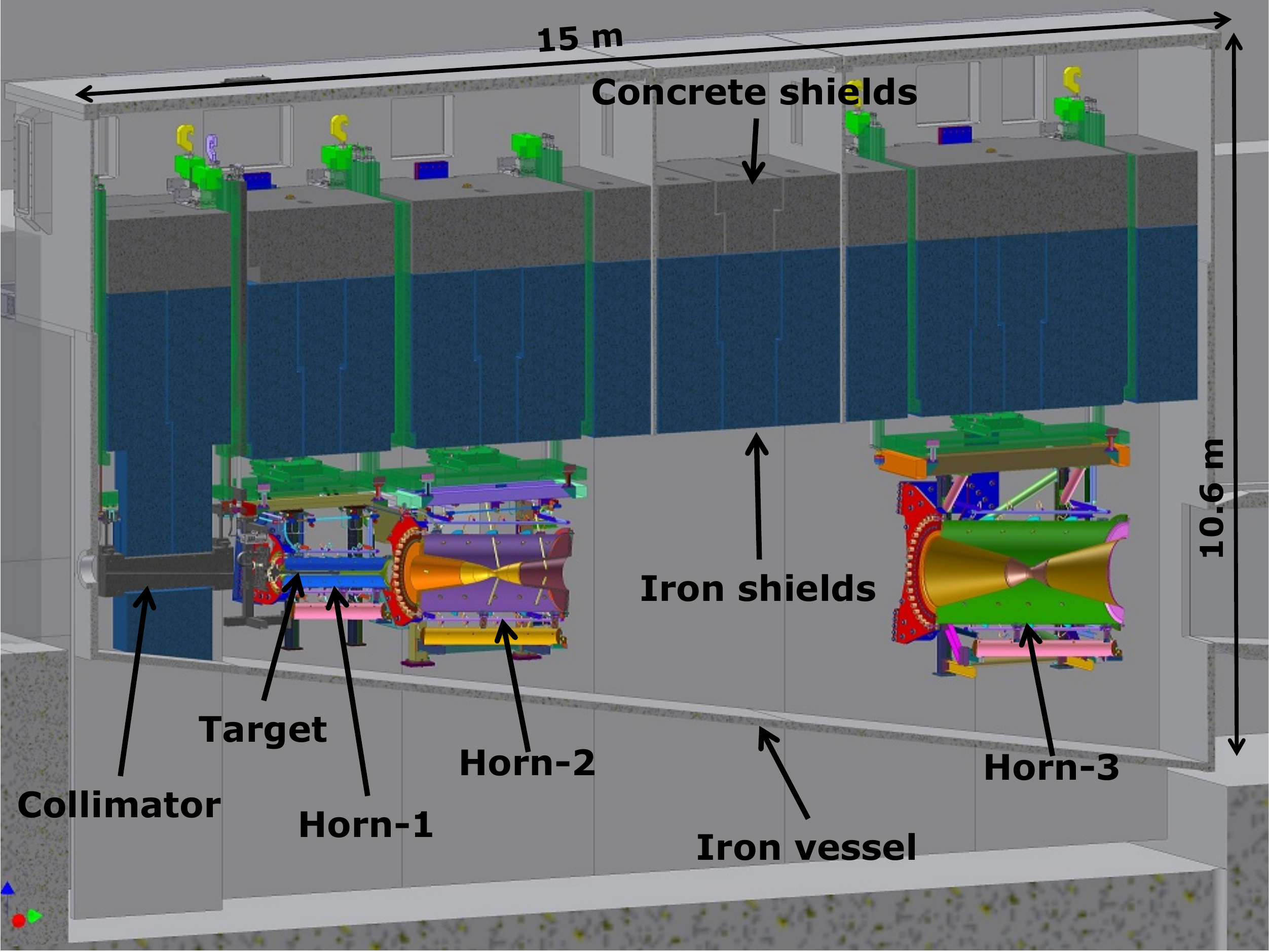}
\caption{Schematic diagram of the HV. The magnetic horns are also shown in this figure.}
\label{fig:Hvessel}
\end{figure}

The T2K magnetic horns are located inside the upstream section of the Helium Vessel (HV), and
a schematic figure of the HV including the magnetic horns is shown in Fig. \ref{fig:Hvessel}.
The magnetic horns are suspended by iron structures called support modules and can be
disconnected from them remotely.
\begin{figure}
\centering
\includegraphics[clip,width=0.35\textwidth,bb=0 0 355 538]{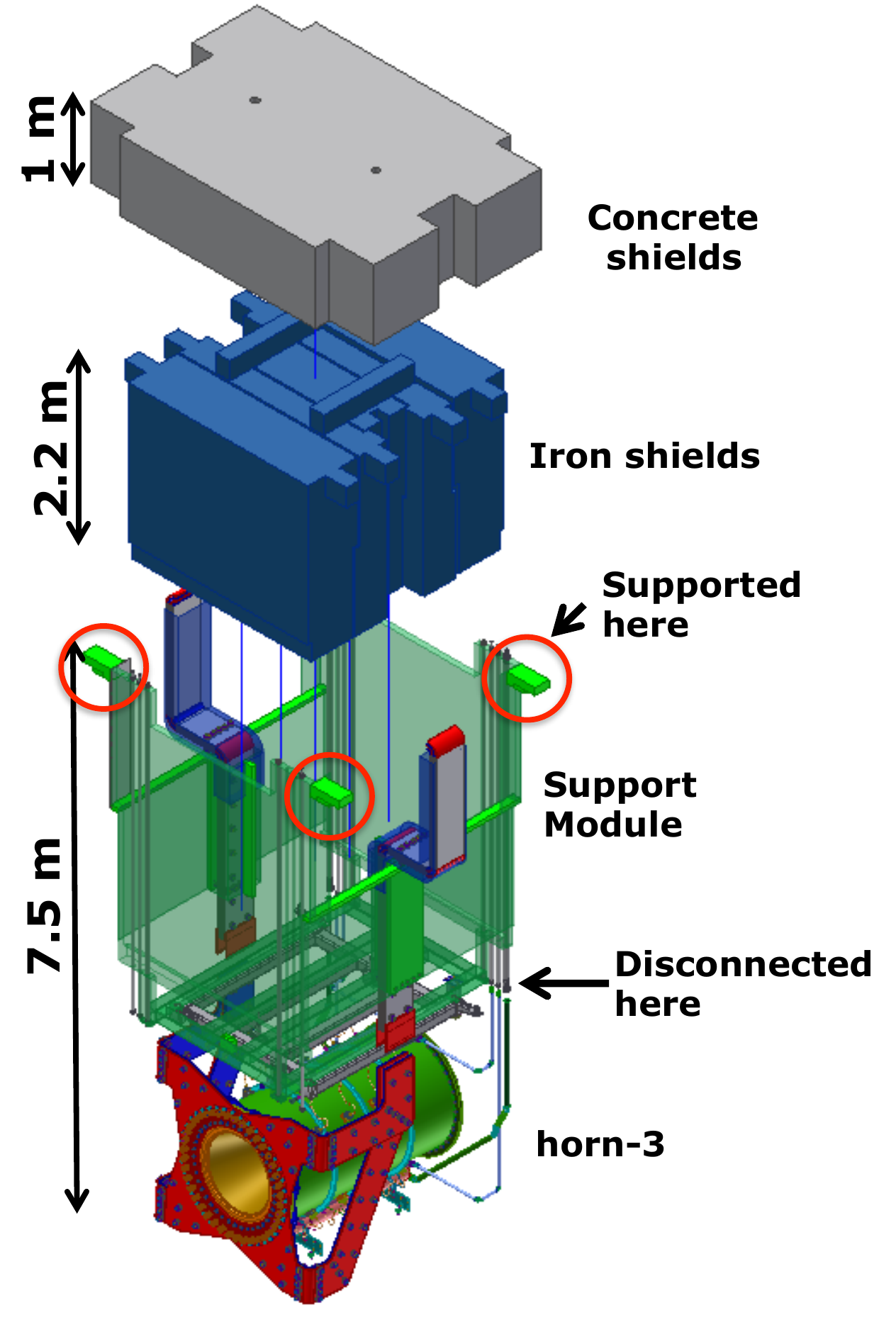}
\caption{Schematic diagram of horn-3 and its support module. Support points are enclosed by circles.}
\label{fig:ModulePic}
\end{figure}
The support modules are placed inside the HV with four support points.   
Iron shielding blocks are inserted inside the box-shape support module, and a concrete shielding
block is placed on top of the iron blocks, as shown in Fig. \ref{fig:ModulePic}. All the shielding
blocks are independently supported by the HV at the upper position.
In the case of one-year beam operation with the design beam power of 750 kW, all the equipment below the shielding
blocks becomes highly irradiated to the order of a few tens of Sv/h. Therefore, one cannot access
the radioactive equipment and perform maintenance or repair work.
If an instrument is broken, the shielding blocks are moved using a remotely operated
overhead crane, the broken equipment is removed from the HV, and new apparatus is installed.
This movement of the radioactive equipment is also performed by remote operation.
More information on the remote maintenance is provided in Section \ref{sec:remote}.


%
%
\section{Electrical properties}
\label{sec:electrical}

\begin{figure}
\centering
\includegraphics[clip,width=0.49\textwidth,bb=50 50 750 400]{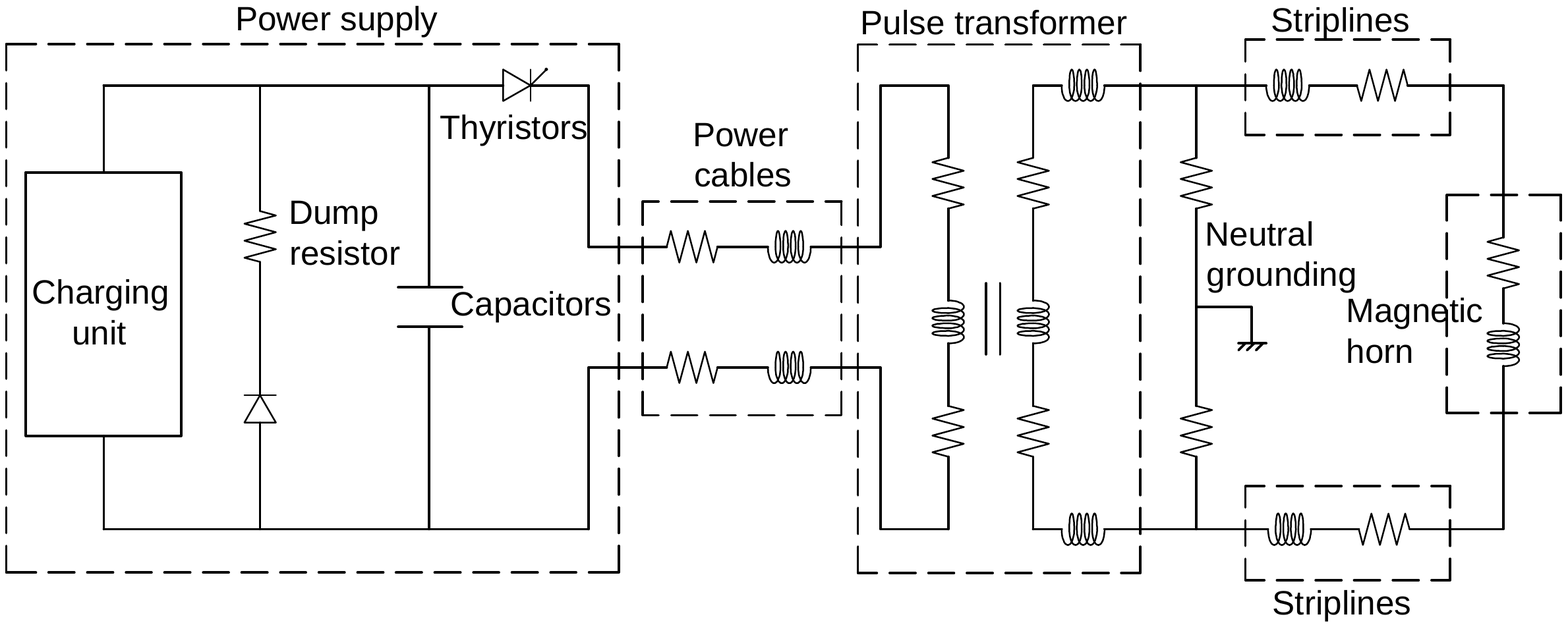}
\caption{Schematic diagram of electric circuit for T2K magnetic horns. }
\label{fig:elec-config}
\end{figure}

The electrical properties of the T2K magnetic horns are described in this section.
The rated peak current of these horns is 320 kA and two power supplies that were used to operate 
the horns in the K2K experiment \cite{K2K} (produced in 1994 and 1998)
were refurbished for use in the T2K experiment.
Since there are three magnetic horns in the T2K experiment, one power supply is used to drive horn-1 only,
while another is used to drive both horn-2 and horn-3 connected in series. 
In the transition from the K2K to the T2K experiments, some improvements were made to the power supplies, e.g.,
modifications to increase the rated current from the 250 kA used in K2K to the 320 kA required for the T2K experiment.
A schematic diagram of the electric circuit used for the T2K magnetic horns is shown in Fig. \ref{fig:elec-config}.

\subsection{Current flow and power supply}
A charging unit in the power supply charges a set of capacitors, at 5 mF (6 mF),
up to 5.8 kV (6.9 kV) for horn-1 (horn-2 and horn-3), and the stored energy of 84 kJ for horn-1 (143 kJ for horn-2 and horn-3) 
is released by turning on a series of three thyristor switches simultaneously. This outputs a pulsed current of 32 kA.
The output current is transferred to a pulse transformer through six pairs of 100-m-long high-voltage cables
(twisted-pair cables with a rated voltage of V$_{DC}$= 11 kV; one cable is used as a supply line and another as a return line).
The pulse transformer amplifies the pulsed current by a factor of ten. Four pairs of aluminum striplines
(eight parallel plates; four for the supply line and four for the return line), made of A6061-T6,
are used as a transfer line between the pulse transformer and the magnetic horn, and they carry a 320-kA pulsed current
to the horns. Horn-2 and horn-3 are connected in series in the secondary circuit of the pulse transformer. 
The pulsed current then returns to the capacitor bank in the power supply.  Because of Joule loss, approximately 50\% of 
the stored energy is returned to the capacitor bank and the capacitors are inversely charged.
In this power supply system, however, the recovered energy is not recycled for the next pulse,
but is totally consumed by a series of dump resistors located inside the power supply. 
The maximum rate of the power supply cycle is 0.5 Hz.

\subsection{Striplines}
\begin{figure}
\centering
\includegraphics[clip,width=0.49\textwidth,bb=0 0 720 500]{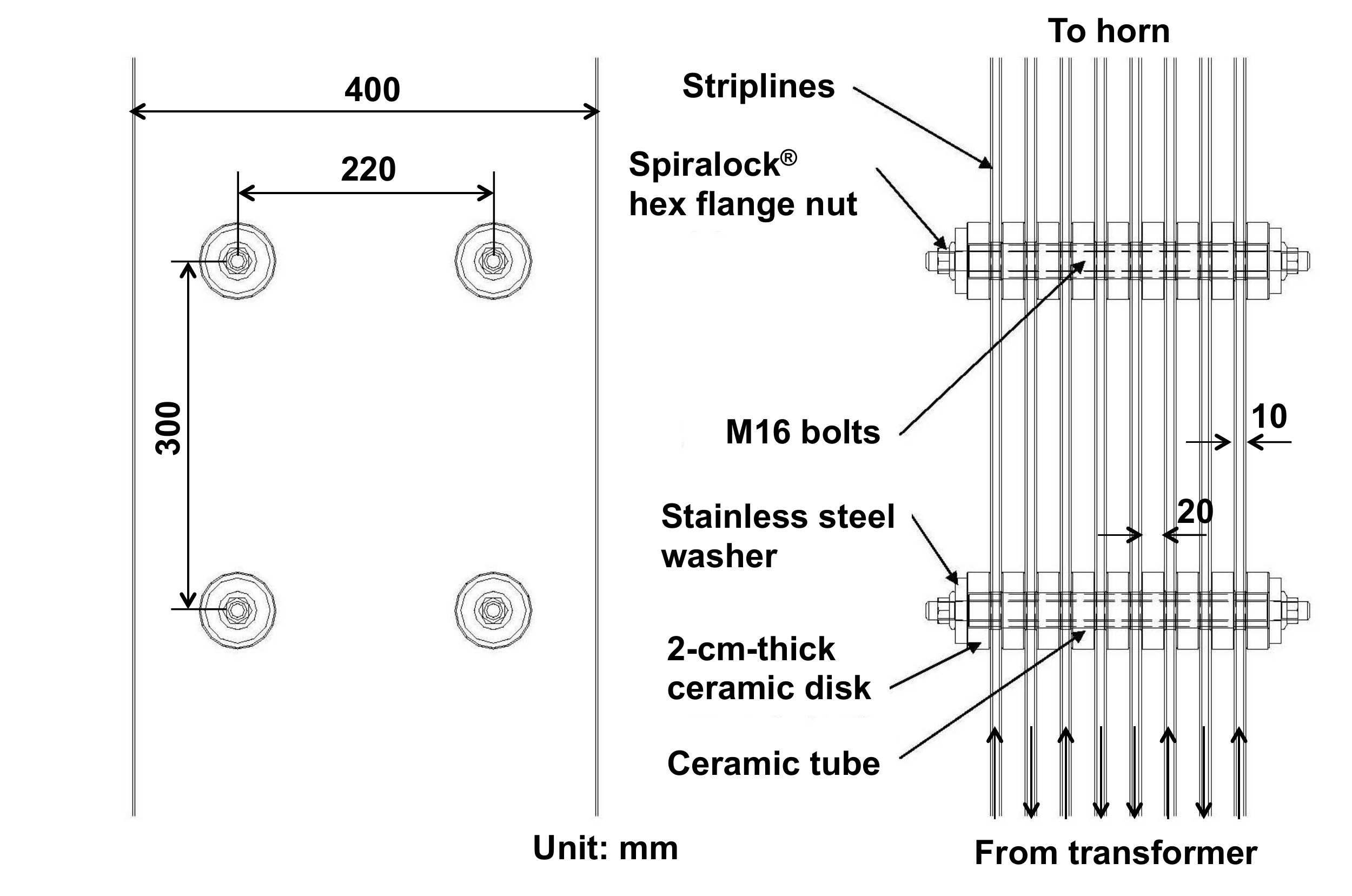}
\includegraphics[clip,width=0.45\textwidth,bb=0 0 722 395]{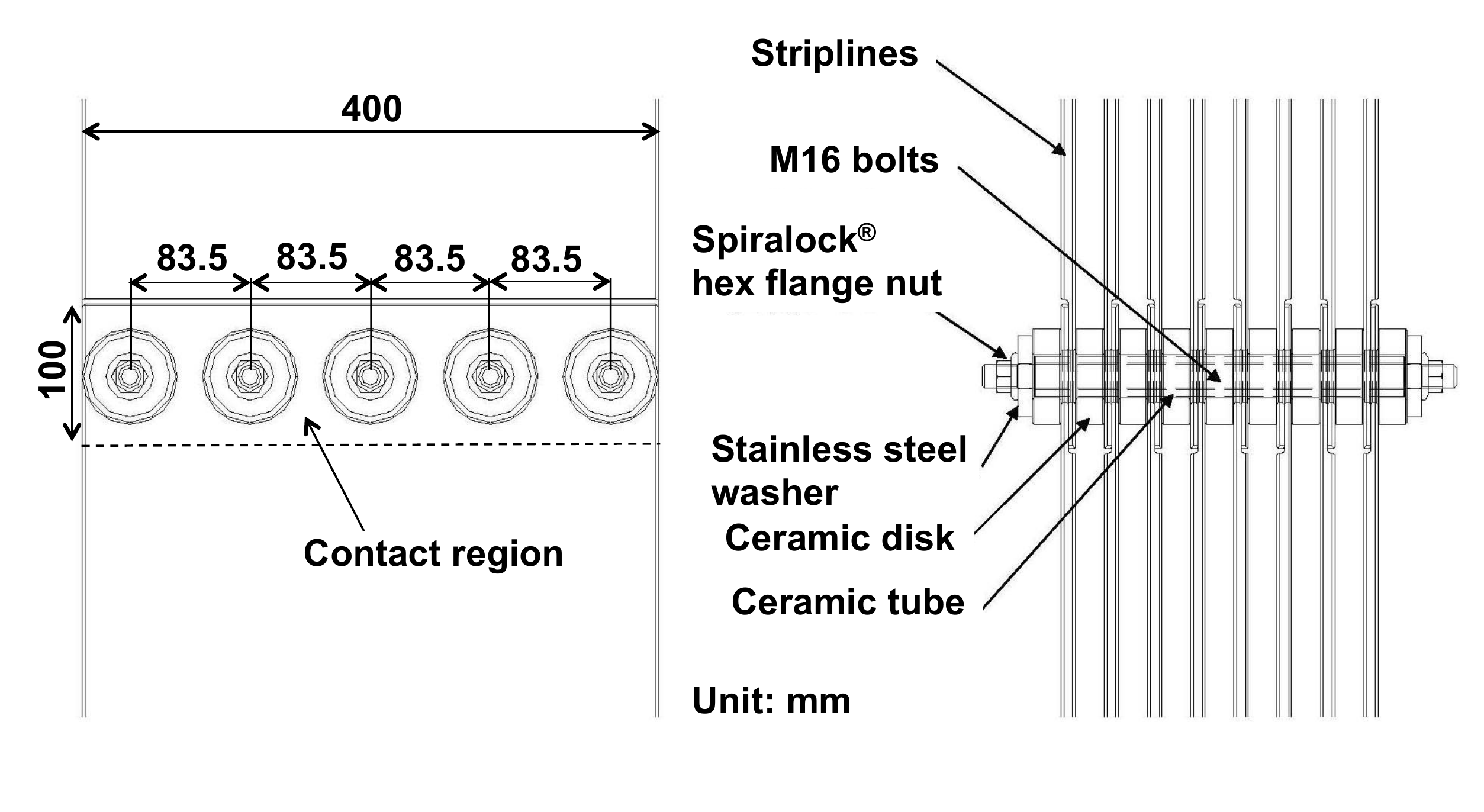}
\caption{Striplines of T2K magnetic horn system. The top (bottom) figure shows the striplines at a normal (joint) section.
The direction of current flow at each stripline plate is shown by the arrows in the upper right figure.}
\label{fig:stripline}
\end{figure}

The striplines consist of eight parallel aluminum conductors with 1-cm thickness, 40-cm width, and 2-cm gap spacing.
Eight stripline plates are bundled by long M16 bolts and Spiralock$^{\textrm{\scriptsize{\textregistered}}}$ hex flange nuts, 
which are insulated with 2-cm-thick disks and tubes composed of alumina ceramic, as shown in Fig. \ref{fig:stripline}.
The striplines are clamped by insulators placed at 30-cm intervals. The spacing was determined, 
considering the required mechanical strength against the Lorentz force, using a finite element stress analysis. 
In this configuration, the inductance and dc resistance per unit length are 0.063 $\mu$F/m and 10 $\mu\Omega$/m, respectively.
The length of the striplines is approximately 18 m (36 m) for horn-1 (horn-2 and horn-3), respectively. 
The serial striplines for horn-2 and horn-3 are designed to have the same lengths in both horns. 
The stripline plate junctions are bolted connections, with the same scheme as that of the stripline clamping,
as shown in Fig. \ref{fig:stripline}. A silver plating is applied to the contact surfaces to reduce joint resistance.

Since the pulse transformers and magnetic horns are located outside and inside the HV, respectively,
the striplines penetrate the wall of the HV via feedthroughs.
Each magnetic horn has its own feedthrough, and the feedthroughs are located in the upper part of the HV,
so the radiation level is not high during beam exposure ($\sim$30 Gy for 1 year of operation).
The stripline plates of the feedthroughs are insulated from their steel flanges by glass-epoxy laminated plate (G10). 
The edges of all the contact surfaces of the G10 insulators and the steel flange
are sealed with an adhesive bond. The definition of vacuum tightness is a leak rate of
$< 10^{-7}$ Pa$\cdot$m$^3$/s at the 10-Pa level.

Since the withstanding voltage is expected to be reduced in a helium atmosphere, 
the withstanding voltage for the striplines with 2-cm spacing and using ceramic insulators
was measured using mockup striplines. The results revealed that the withstanding voltage was 3.5 kV 
in the helium atmosphere.
Because all the aluminum conductors (the magnetic horns and striplines) are operated below 1 kV,   
the withstanding voltage value of 3.5 kV is sufficiently high for insulation between conductors and between
a conductor and the ground.
The supply and return striplines are connected via 7.5-$\Omega$ resistors and the neutral point
is grounded, as shown in Fig. \ref{fig:elec-config}. 
The supply (return) stripline-to-ground voltage is maintained below a maximum of +500 V (-500 V).
In the early stages of the experiment, this neutral grounding did not exist, which caused a voltage breakdown 
of the striplines inside the HV.
This neutral grounding scheme is essential to preventing a voltage breakdown in the helium atmosphere.

\subsection{Magnetic horns}
The four striplines are attached to the most upstream part of each magnetic horn, as shown in Fig. \ref{fig:horn_drawing}.
The four supply striplines are connected to the inner conductor and the four return
striplines to the outer conductor. In order to focus positively charged secondary particles, 
the current enters the inner conductor and exits from the outer conductors as shown in Fig. \ref{fig:horn1cross}. 
The magnetic field is proportional to both $I$ and $1/r$, as expressed
in Equation (\ref{eq:Bfield}). The inner and outer conductors are insulated at the upstream end
with a 30-mm-thick large ceramic ring, with inner (outer) diameters of 370 (530) mm, 980 (1,110) mm and 
1,310 (1,430) mm for horn-1, horn-2, and horn-3, respectively, while the inner and outer conductors are connected
at the downstream end by bolted joints consisting of M16 bolts and 
Spiralock$^{\textrm{\scriptsize{\textregistered}}}$ hex flange nuts, both made of A6061-T6.

The typical circuit parameters of the T2K magnetic horn system are listed in Table \ref{tab:circuit_const}.
\begin{table}[htb]
\centering
\caption{Summary of circuit parameters for T2K magnetic horn system.}
\small
\begin{tabular}{l|r|r|r|r}
\hline
Components & \multicolumn{2}{c|}{horn-1} & \multicolumn{2}{c}{horn-2 + horn-3} \\ \cline{2-5}
  & $L$ & $R$ & $L$ & $R$  \\
  & ($\mu$H) & (m$\Omega$) & ($\mu$H) & (m$\Omega$) \\ \hline
Horn & 0.47 & 0.100 & 0.46 & 0.035 \\
& & & +0.53 & +0.023 \\
Striplines &  0.28 & 0.100 & 0.60 & 0.210 \\ 
Transformer & 0.30 & 0.040 & 0.30 & 0.040 \\ \hline
Total (secondary) & 1.05 & 0.240 & 1.89 & 0.308 \\
Total (seen from primary) & 105 & 24.0 & 208 & 34.0 \\
Cable & 15 & 4.0 & 17 & 4.3 \\
Total (primary) & 120 & 28.0 & 223 & 38.3 \\ \hline
Pulse width & \multicolumn{2}{c|}{2.4 ms.} & \multicolumn{2}{c}{3.6 ms.} \\
Peak current & \multicolumn{2}{c|}{320 kA} & \multicolumn{2}{c}{320 kA} \\
Capacitance of capacitors  & \multicolumn{2}{c|}{5 mF} & \multicolumn{2}{c}{6 mF} \\
Charging voltage at capacitors & \multicolumn{2}{c|}{5.8 kV} & \multicolumn{2}{c}{6.9 kV} \\
Stored energy at capacitors & \multicolumn{2}{c|}{84 kJ} & \multicolumn{2}{c}{143 kJ} \\ \hline
\end{tabular}
\label{tab:circuit_const}
\end{table}

\subsection{Current monitoring}
\begin{figure}[htb]
\centering
\includegraphics[clip,width=0.4\textwidth,bb=10 10 729 470]{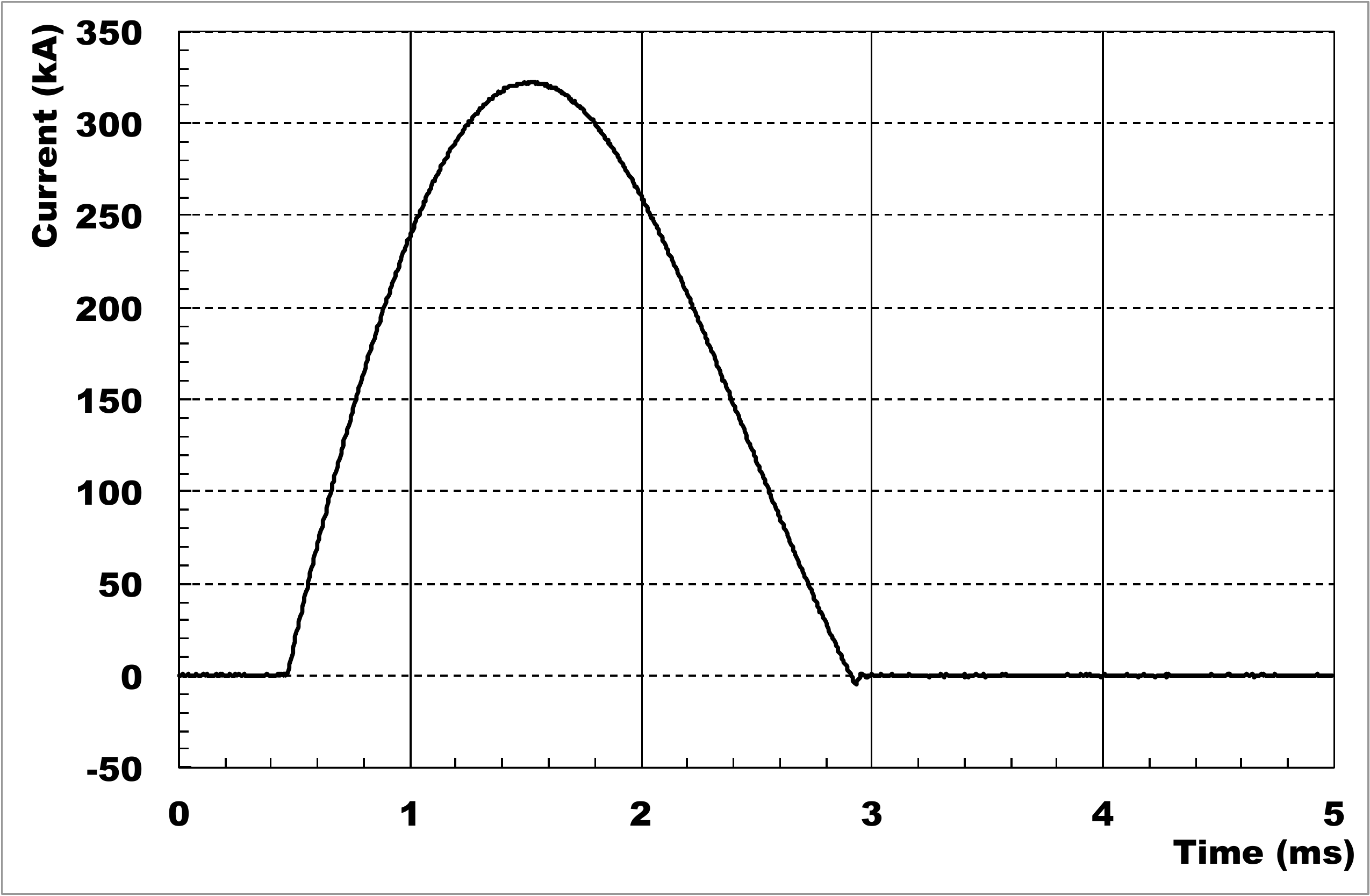}
\includegraphics[clip,width=0.4\textwidth,bb=10 10 729 470]{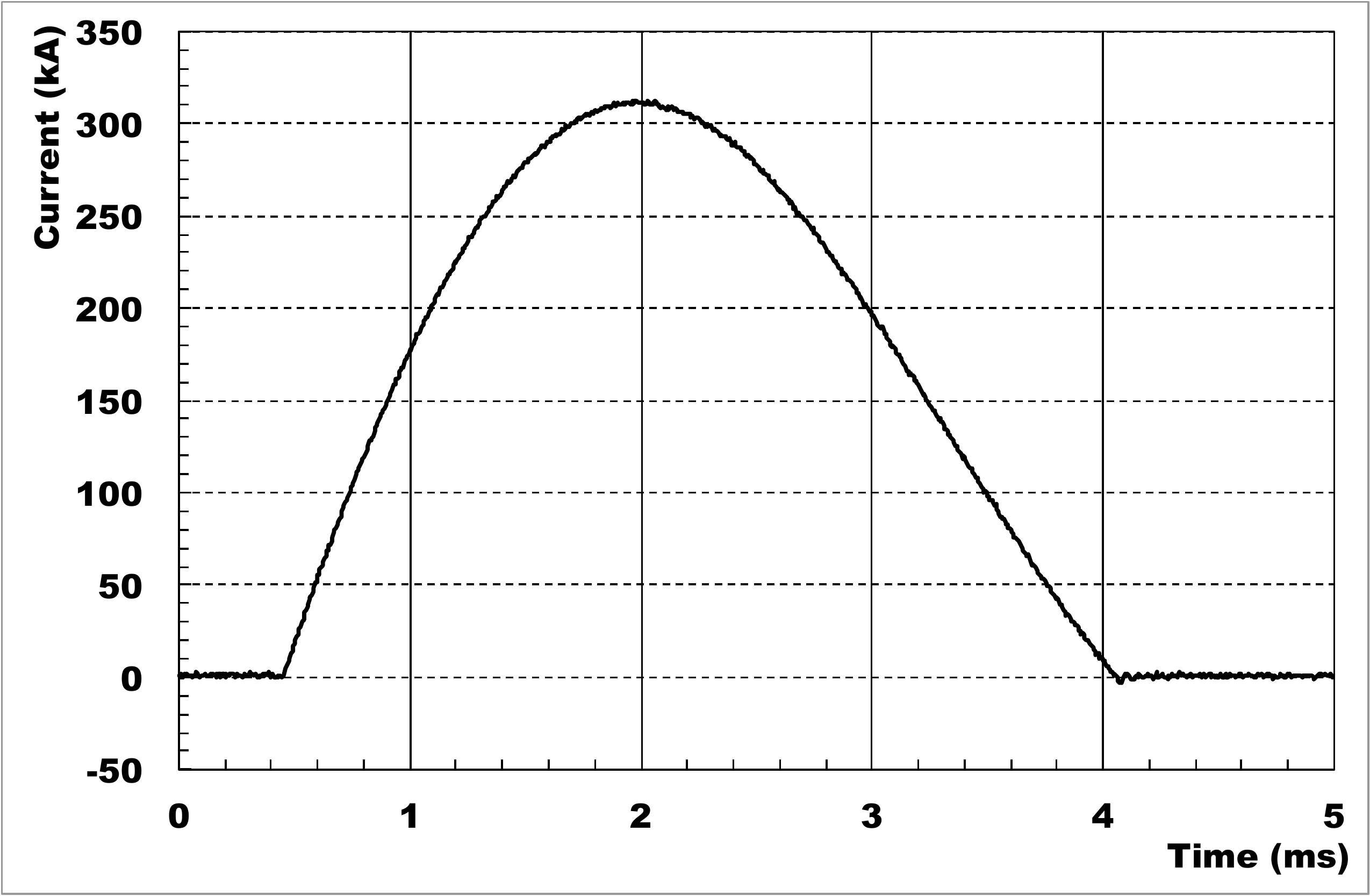}
\caption{Typical current waveform for horn-1 (top) and for horn-2 and horn-3 (bottom).}
\label{fig:waveform}
\end{figure}

The pulsed current sent to each horn is monitored with custom Rogowski coils\footnote{Manufactured by
Power Electronic Measurements (PEM) Ltd.}
whose signal is digitized by 200 kHz FADCs.
One Rogowski coil is wound around each stripline plate and four independently measured current values
are summed to obtain a total current for each horn. If any current deviation ($\pm$3\% from the nominal current) is observed
by one Rogowski coil or more, then the power supply operation is stopped by an interlock system.
Typical current waveforms for horn-1 and for horn-2 and horn-3 are shown in Fig. \ref{fig:waveform}.

The Rogowski coils are calibrated by the production company with $\pm$1\% precision.
The shape of the "loop" of the Rogowski coil may cause a 1\% change in gain.
FADCs and related electronics are calibrated with better than 1\% precision.
During long-term beam operation, a peak current variation of 2\% was observed, mainly due to temperature
variation, since the power cables partly run along building exteriors ($\sim$50 m of the 100-m total). 
Uncertainties on the absolute horn-current measurement is summarized in Table \ref{tab:cur_err} \cite{nuflux}.
\begin{table}[H]
\centering
\caption{Uncertainties on the absolute horn-current measurement. In the total error calculation, full width (FW)
errors are scaled by 1/$\sqrt{12}$ to estimate 1$\sigma$ uncertainty.}
\begin{tabular}{ll}
\hline
 & Uncertainty \\ \hline
 Coil calibration & $\pm$1\% (FW) \\
 Coil setting & $\pm$1\% (FW) \\
 Electronics calibration & $<$ 1\% \\
 Monitor stability & 2\% (FW) \\ \hline
 Total & 1.3\% \\ \hline
\end{tabular}
\label{tab:cur_err}
\end{table}
The uncertainty on the absolute horn-current measurement is estimated to be $\pm$1.3\%, however, this uncertainty contributes
the uncertainty of less than 1\% to the neutrino flux prediction, which is less than the uncertainties from other sources \cite{nuflux}.

\subsection{Magnetic field measurement}
The outer conductors have holes (4 for horn-1 and horn-3 and 9 for horn-2)
for some service uses, which are covered by aluminum plates for beam operation. A magnetic field probe can be inserted
into the inner volume of the horns through the service holes to measure the magnetic fields inside the horns.
The magnetic field probe used is a 3-axis hall probe\footnote{Manufactured by SENIS AG (The Three-Axis
Field Transducer x-H3x-xxE3A-25kHz-0.1-2T).}. Its specifications are a linear range of $\pm$2 T, an accuracy of 0.5\%,
and a bandwidth of 25 kHz.
A custom-made jig for magnetic field measurement was produced to insert the field probe into the interior of the device.
The field probe is enclosed inside housing attached to the end of a G10 rod.
The G10 rod can be moved in a radial direction so that field measurements can be conducted in various positions. 
A fixture for the G10 rod is attached to the service ports at the outer conductors.

\begin{figure}
\centering
\includegraphics[clip,width=0.4\textwidth,bb=15 5 730 475]{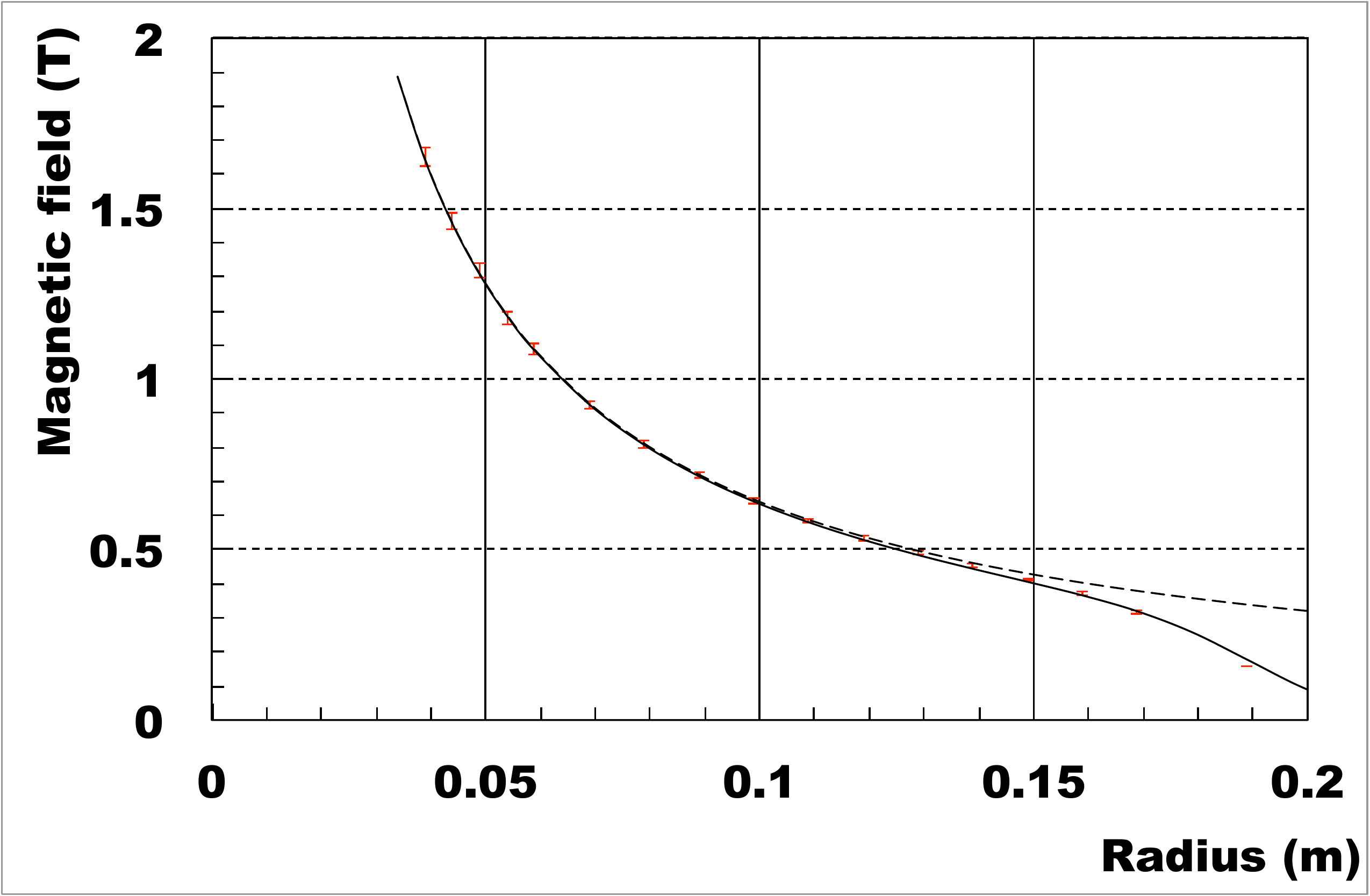}
\includegraphics[clip,width=0.4\textwidth,bb=15 5 730 475]{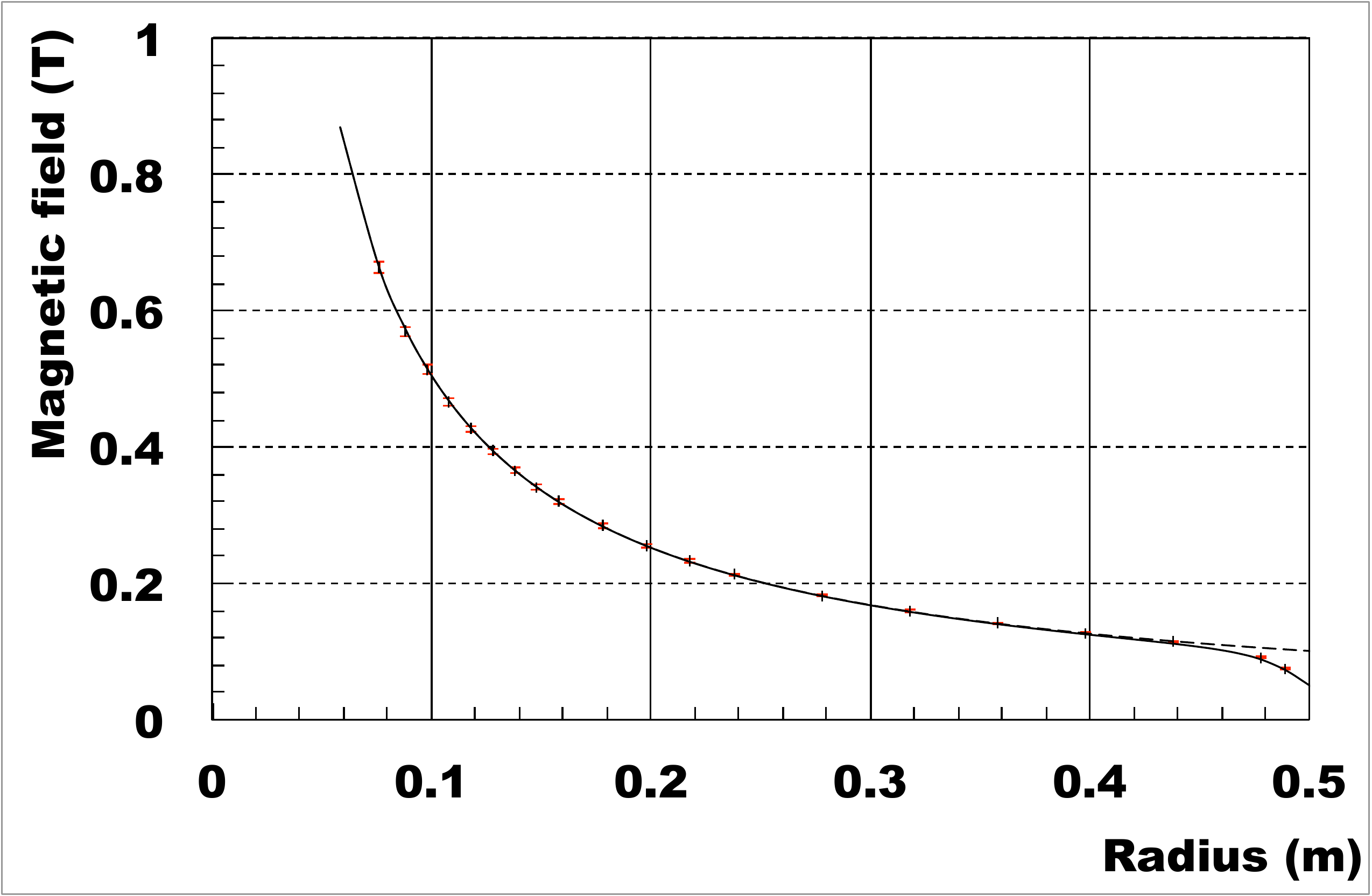}
\includegraphics[clip,width=0.4\textwidth,bb=15 5 730 475]{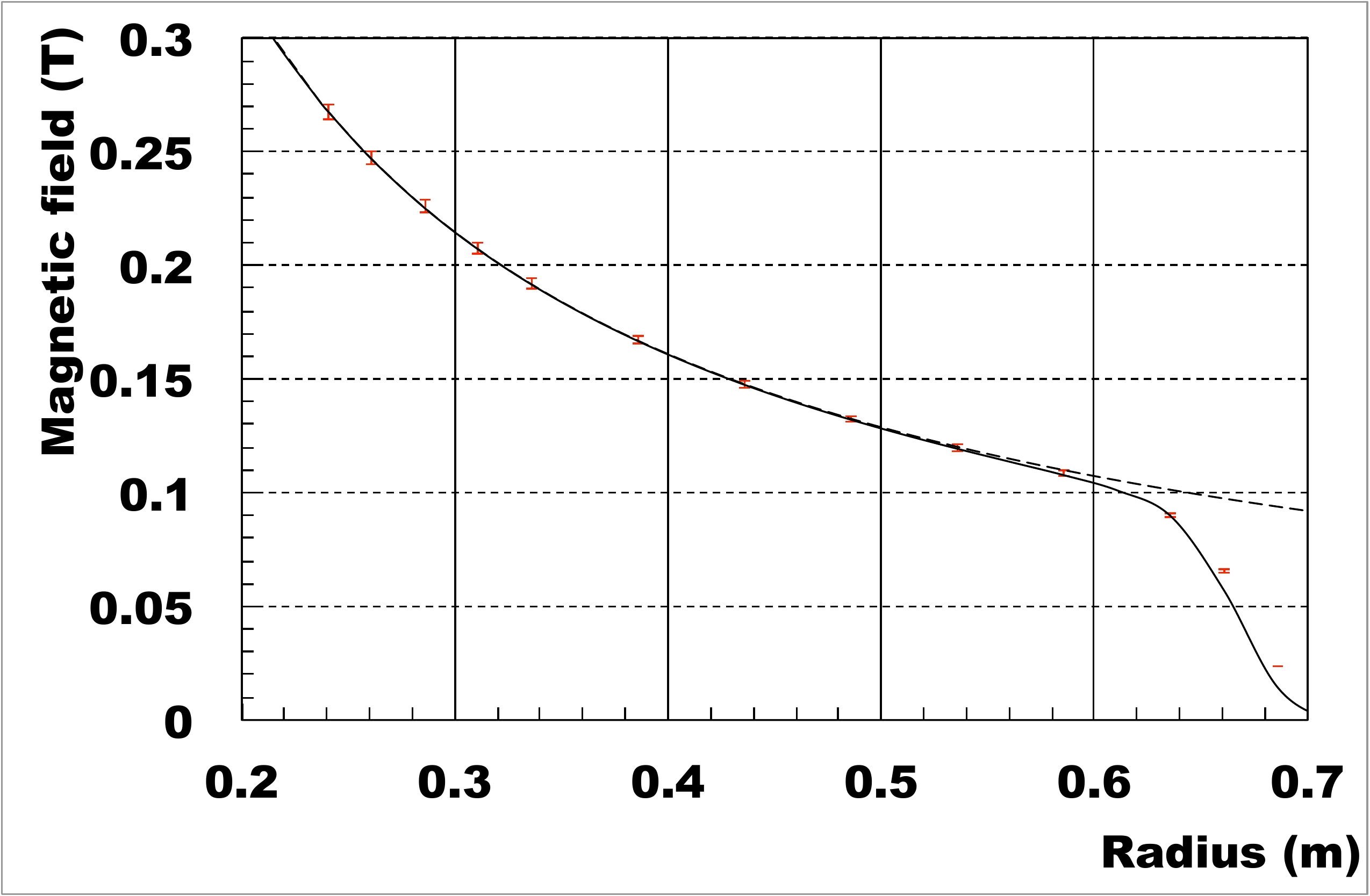}
\caption{Measured magnetic fields as functions of radial distance from center axis for horn-1 (top),
horn-2 (middle), and horn-3 (bottom). The solid lines represent the expected field by taking into account a fringe effect.
The dashed lines represent the expected magnetic fields calculated by
Equation (\ref{eq:Bfield}). The curves represent the magnetic fields expected from the operation currents of 320 kA (horn-1),
250 kA (horn-2), 320 kA (horn-3).}
\label{fig:FieldMeasurements}
\end{figure}

\begin{figure}
\centering
\includegraphics[clip,width=0.4\textwidth,bb=15 5 730 475]{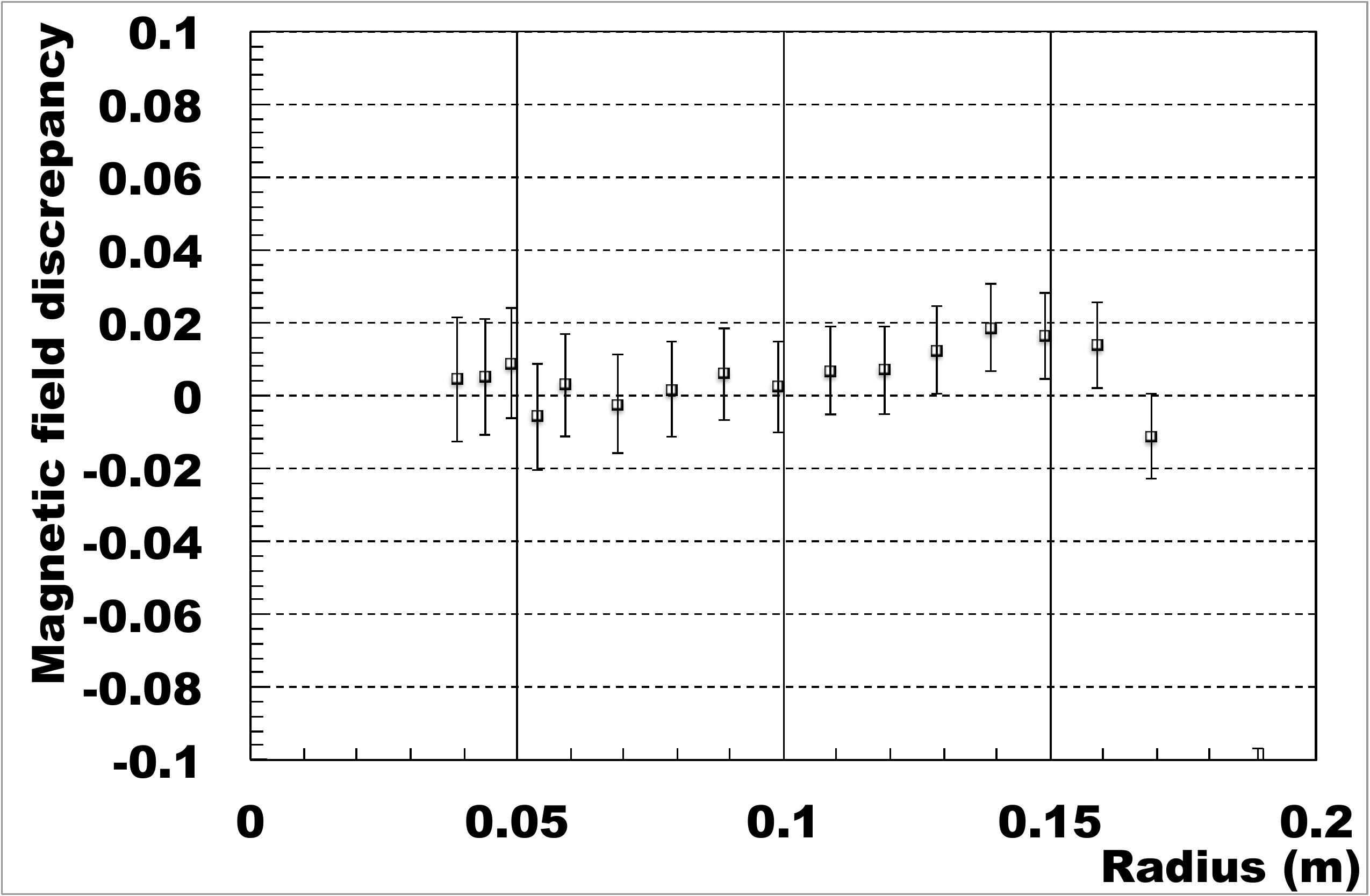}
\includegraphics[clip,width=0.4\textwidth,bb=15 5 730 475]{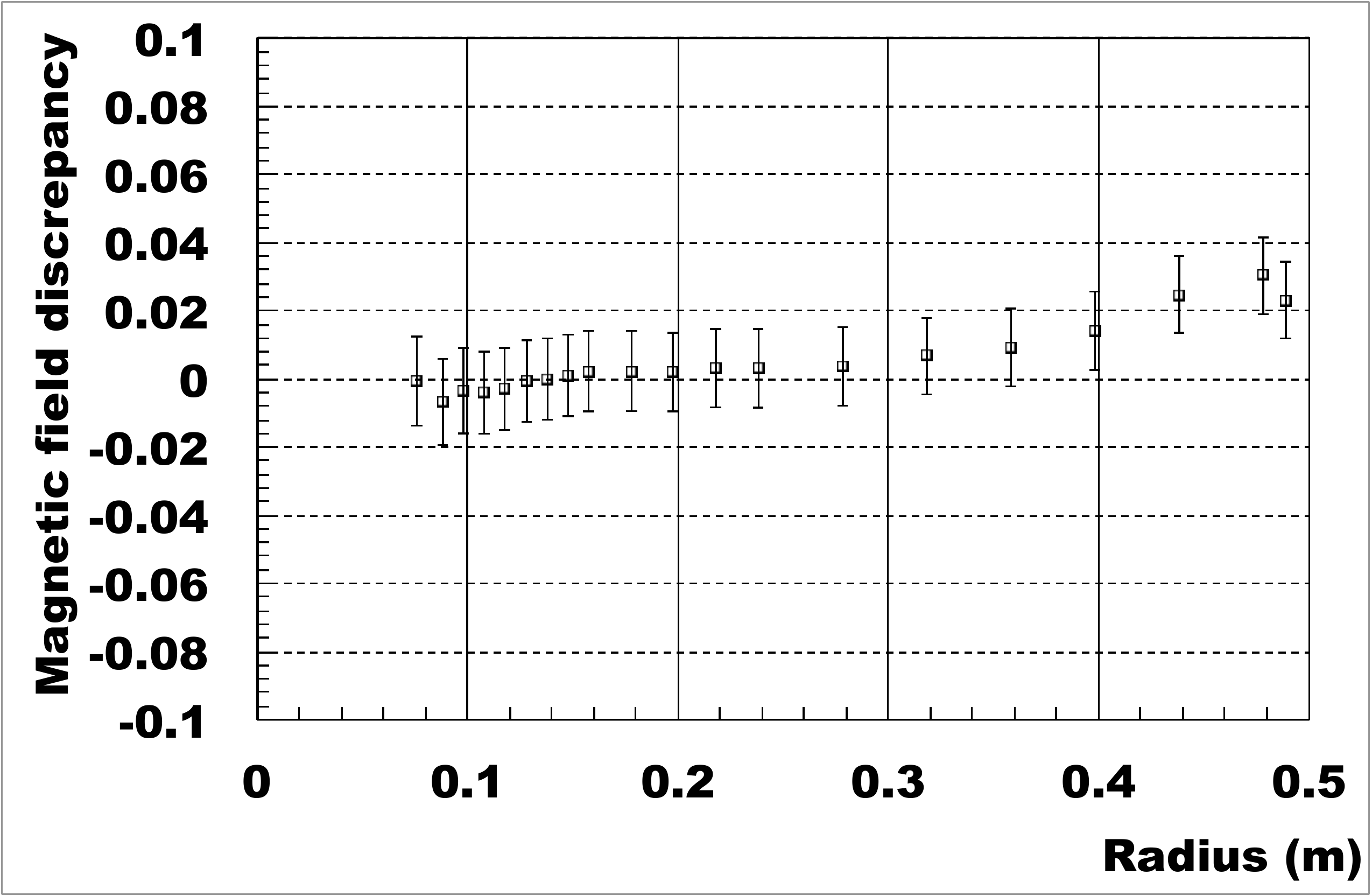}
\includegraphics[clip,width=0.4\textwidth,bb=15 5 730 475]{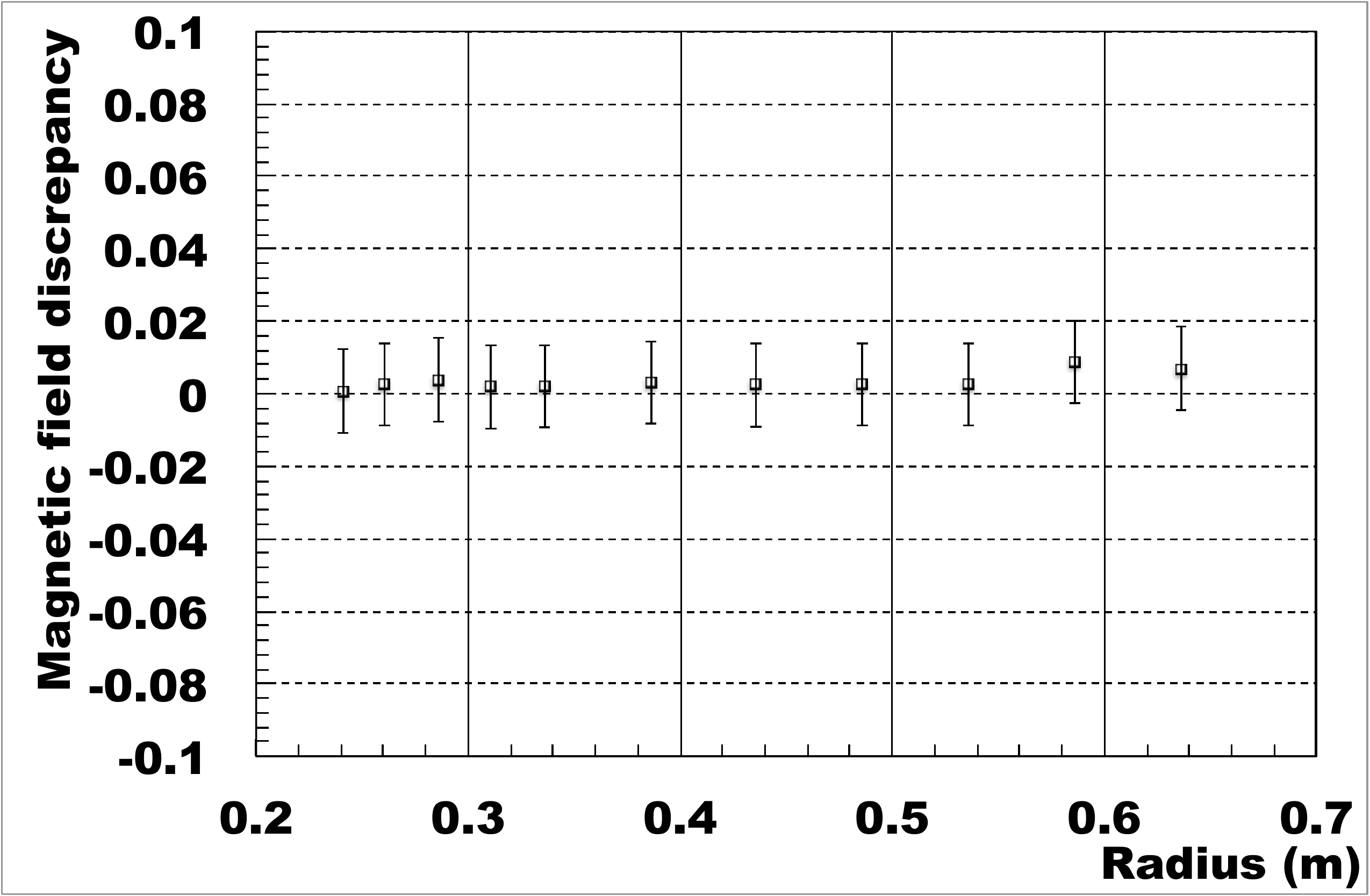}
\caption{The discrepancies between the measured and expected magnetic fields, $(B_{meas.}-B_{exp.})/B_{exp.}$,
as functions of radial distance from the center axis for horn-1 (top), horn-2 (middle), and horn-3 (bottom).
The expected magnetic fields are taken into account the fringe effect.}
\label{fig:FieldMeasurements-2}
\end{figure}

The magnetic field measurements for all three magnetic horns were performed at a current-testing stand.
The measured magnetic fields and their discrepancies from the expected value
are shown in Figs. \ref{fig:FieldMeasurements} and \ref{fig:FieldMeasurements-2} as a function of radial position.
As measurement errors, the calibration errors (0.2\%), radial position precision (0.5 mm), intrinsic measurement
precision of the field probe (0.5\%), and current measurement precision (1\%) were allocated.
Smaller magnetic fields than predicted were measured near the outer conductor inner surface (at 0.18 m for horn-1,
0.49 m for horn-2, and 0.65 m for horn-3). A leakage of magnetic flux 
from the holes through which the field probe was inserted caused a reduction in the magnetic field. 
The expected magnetic fields are in good agreement by taking into account the fringe field correction.
These measurements indicate that the measured magnetic field values are consistent with 
those expected based on the measured current values, within a measurement precision of 1-2\%.


%
%
\section{Cooling performance}
\label{sec:cooling}

\begin{figure}
\centering
\includegraphics[clip,width=0.45\textwidth,bb=0 0 732 478]{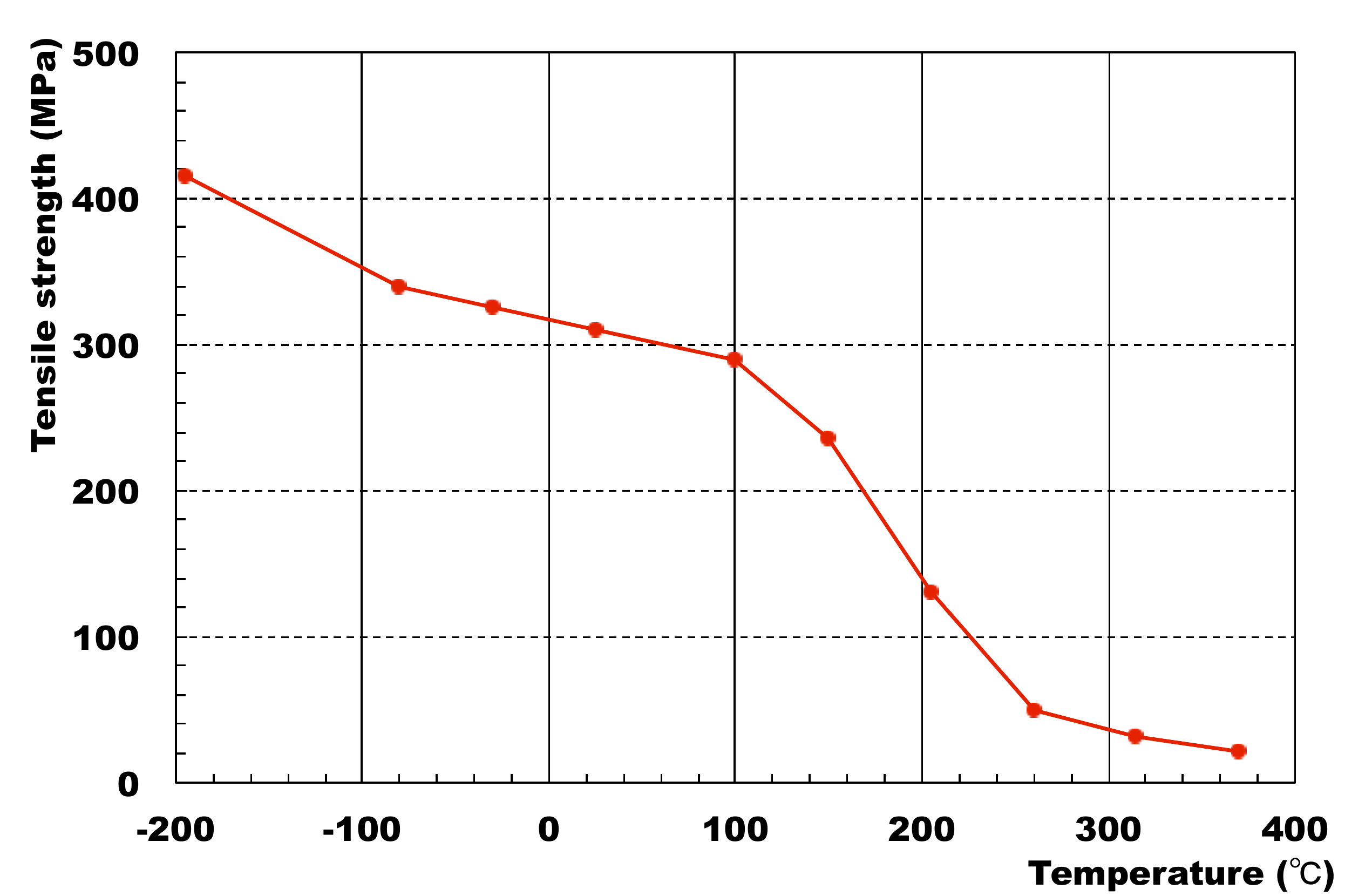}
\caption{Temperature dependence of aluminum alloy A6061-T6 tensile strength. 
Data points are taken from \cite{Alhandbook}.}
\label{fig:Al_strength_temp}
\end{figure}

The technical cooling issues of the T2K magnetic horns and the cooling performance are described in this section.
The tensile strength of A6061-T6 decreases as the temperature increases, 
as shown in Fig. \ref{fig:Al_strength_temp}.
It can be seen that the strength drastically decreases above 100$^{\circ}$C.
Therefore, the maximum permitted temperature of the T2K magnetic horns is set to 80$^{\circ}$C,
to prevent conductor strength degradation.
Cooling of the T2K magnetic horns is conducted using two methods: one is water cooling, used for the inner
conductors and vertical frames, and the other is cooling of the striplines using a forced helium flow.

\subsection{Horn-conductor water cooling}

\begin{figure}
\centering
\includegraphics[clip,width=0.4\textwidth,bb=100 0 500 300]{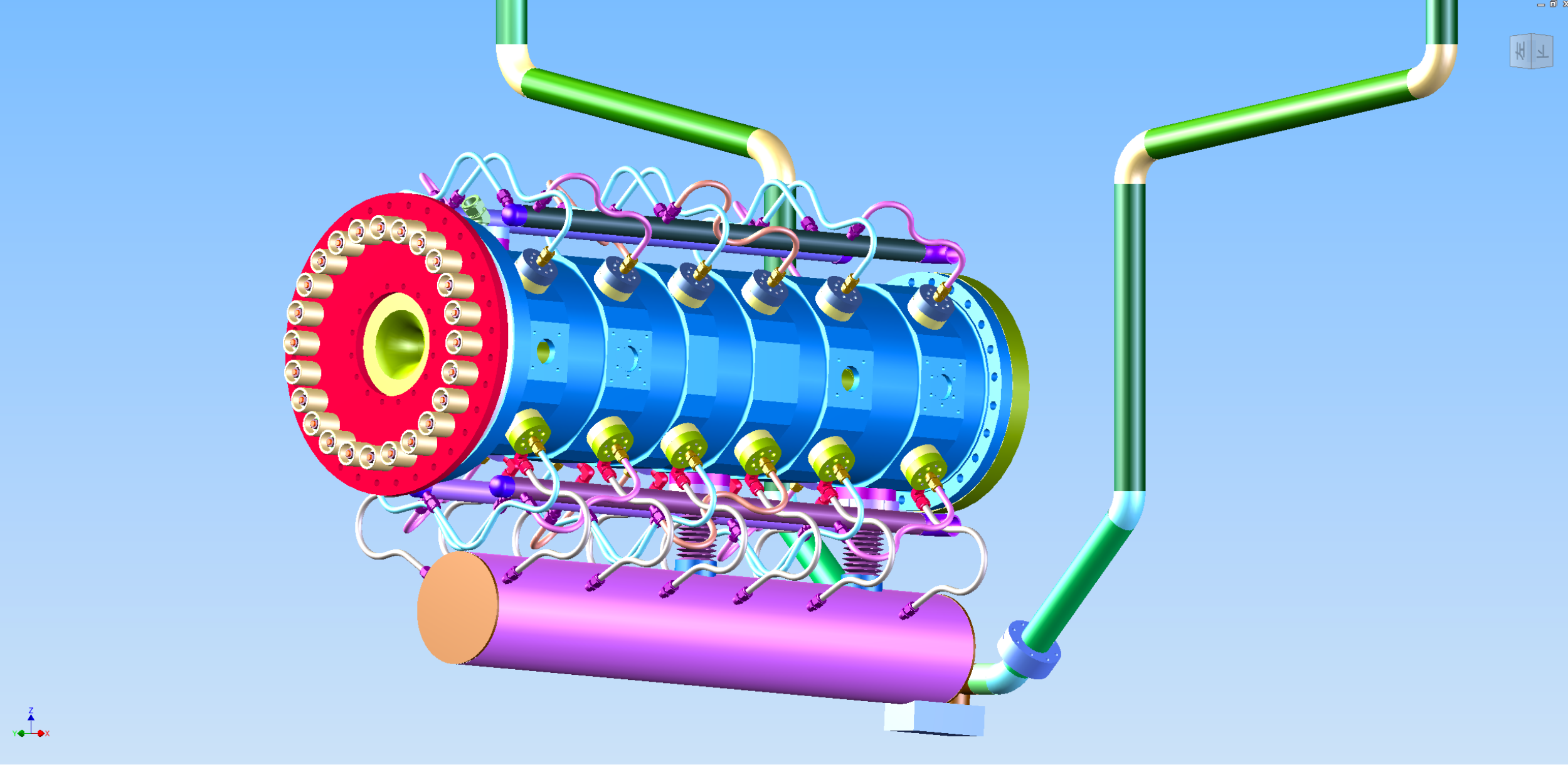}
\includegraphics[clip,width=0.4\textwidth,bb=100 0 500 300]{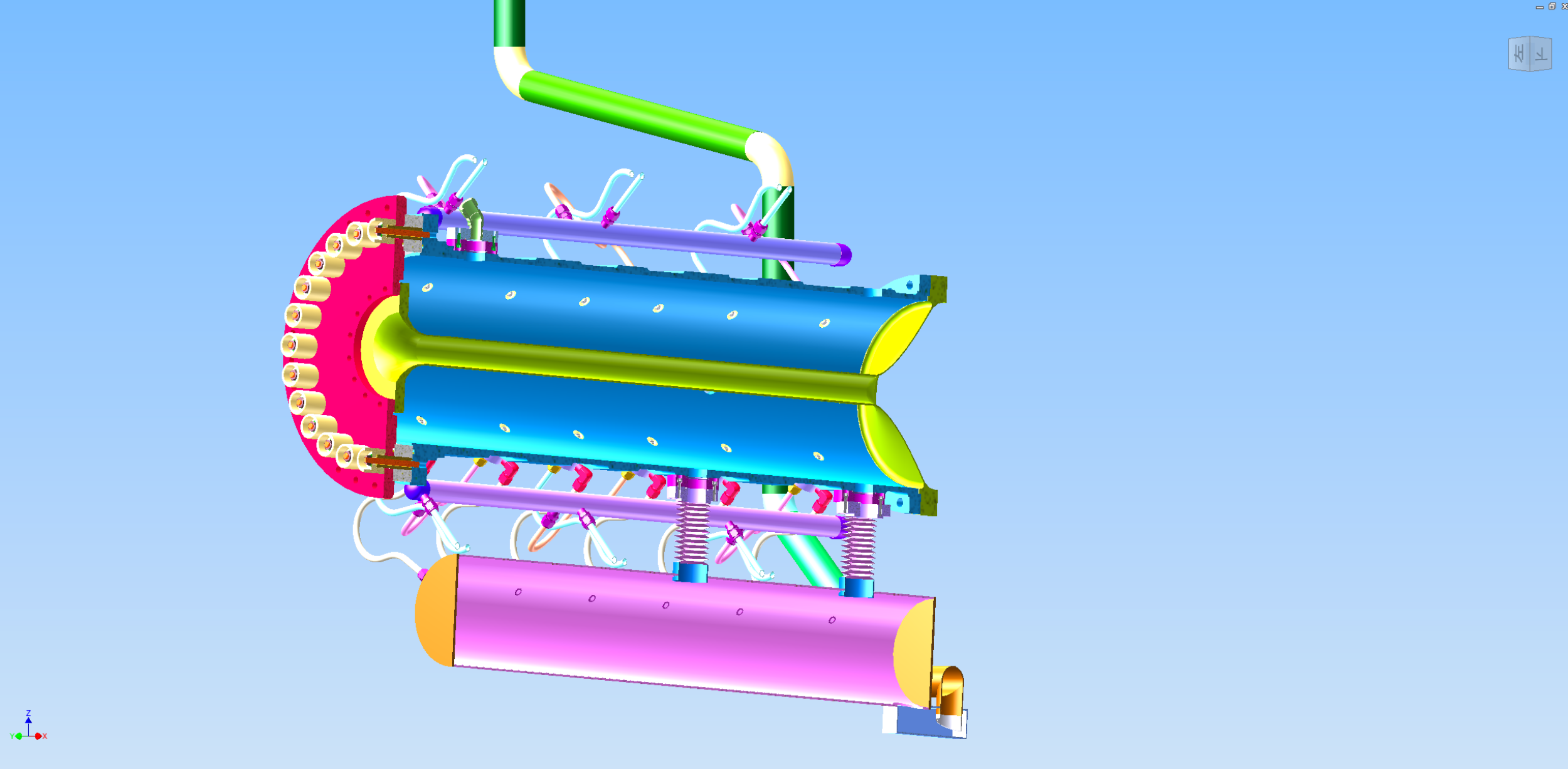}
\caption{3D model of horn-1 showing water nozzles attached to outer conductor (top),
and its cross-section (bottom).}
\label{fig:horn_water}
\end{figure}

\begin{figure}[tbh]
\centering
\includegraphics[clip,width=0.45\textwidth,bb=20 0 820 400]{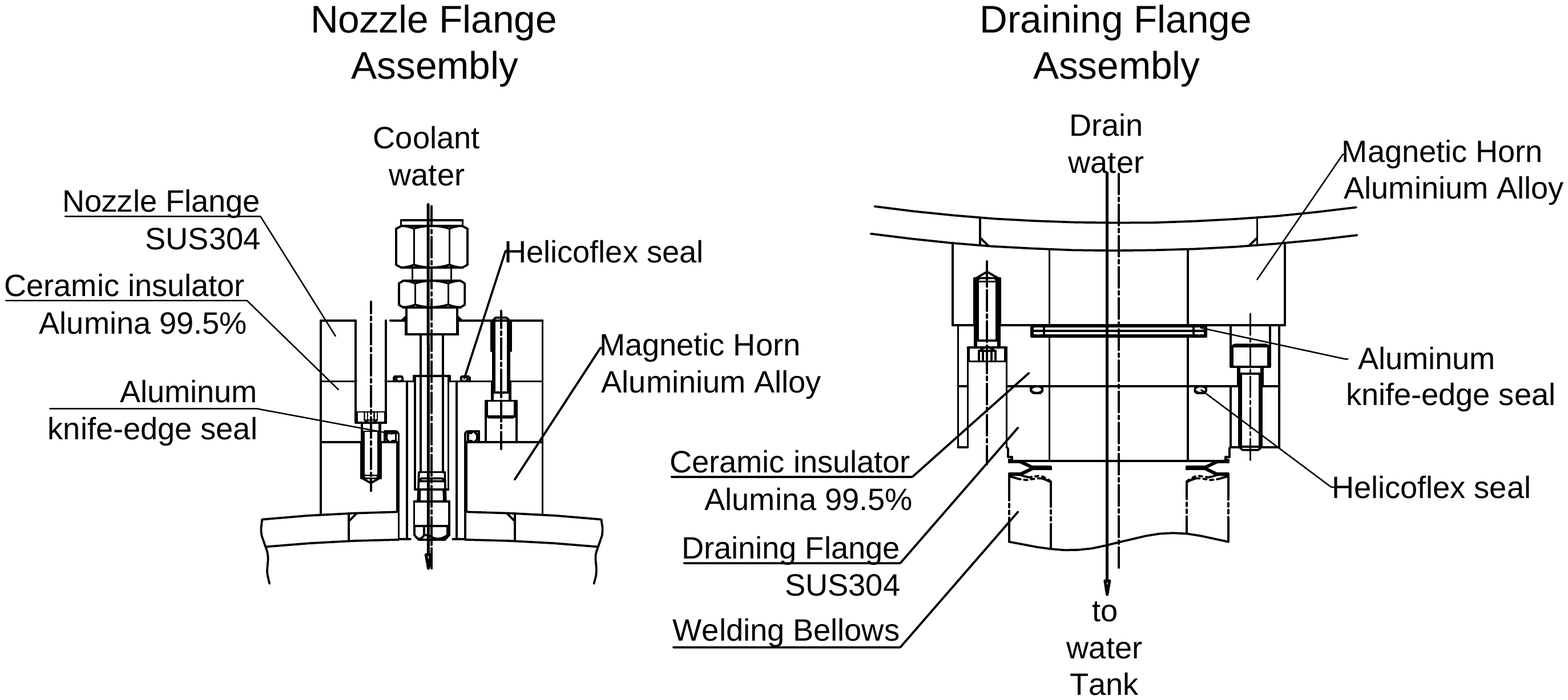}
\caption{Detailed pictures of water nozzle ports (left) and water drain ports (right).}
\label{fig:nozzle_port}
\end{figure}

\begin{figure}
\centering
\includegraphics[clip,width=0.4\textwidth,bb=0 25 554 486]{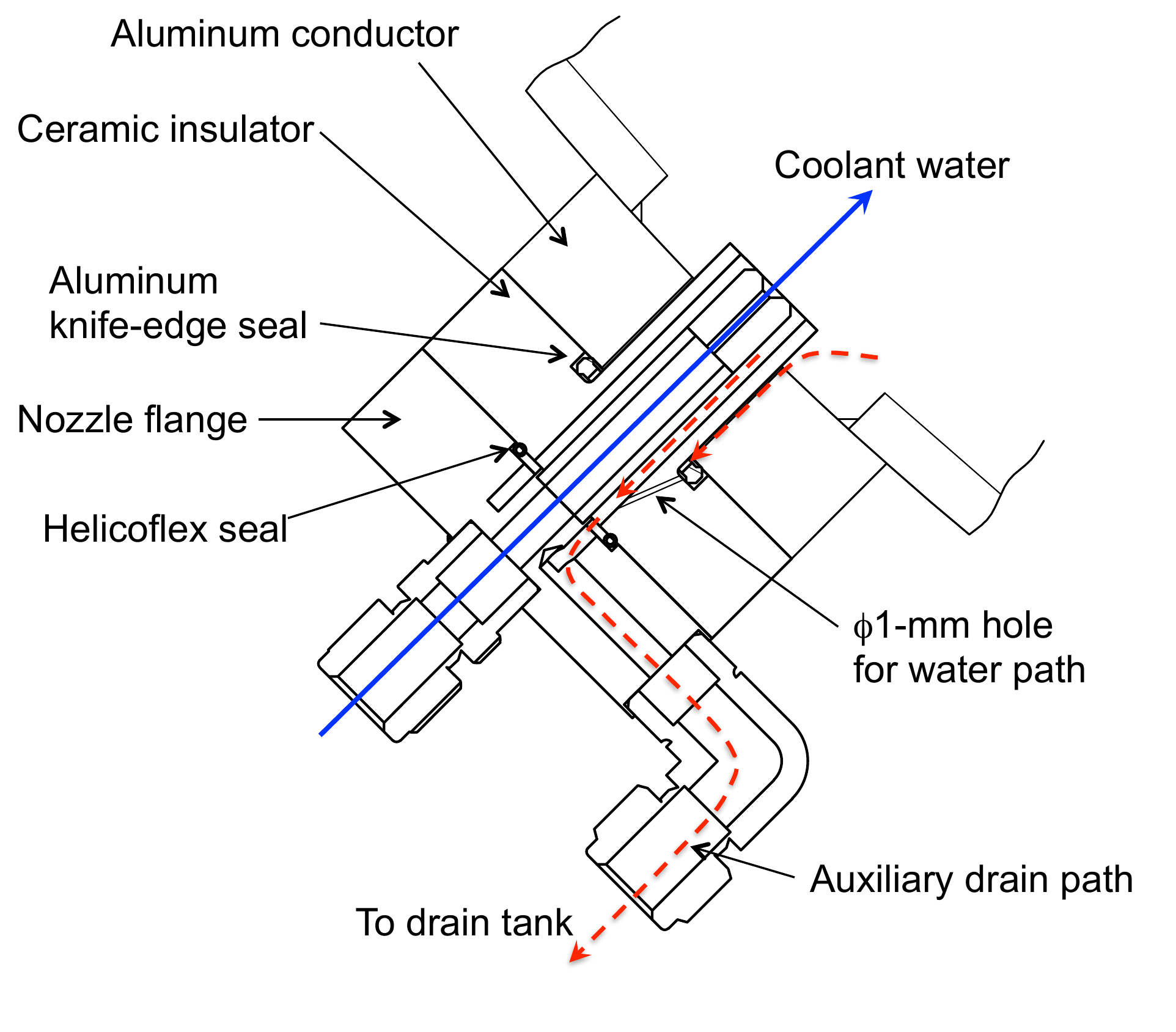}
\caption{Detailed figure of auxiliary drain port attached to bottom nozzle flange. The coolant-water path is represented by
a solid line and the auxiliary drain paths are shown using dashed lines.}
\label{fig:AuxiliaryDrain}
\end{figure}

Water cooling of the inner conductors is performed by spraying water from spray nozzles
onto the inner conductors. The water nozzles are attached to the outer conductors and face
inward, as shown in Fig. \ref{fig:horn_water}.
Horn-1 and horn-2 have 24 spray nozzles, with four (six) nozzles oriented in the
azimuthal (longitudinal) direction, while horn-3 has 16 nozzles, with four (four) in the azimuthal (longitudinal) direction.
The nozzles are connected to four manifolds surrounding the outer conductors with
bent tubes. This bent shape plays an important role in absorbing vibration
from the outer conductors and preventing the tubes from experiencing a mechanical break.
The sprayed water is collected into a drain tank located below the outer conductors
and pumped upwards by suction pumps.
The horn-1 nozzles have a flat spray pattern to match the cylindrical inner conductor,
while horn-2 and horn-3 nozzles have a cone spray pattern to disperse the cooling water widely.
The water nozzles, made of stainless steel, are insulated from the outer conductors to prevent both
a high voltage breakdown and Galvanic corrosion due to contact between different metals (e.g.
aluminum and stainless steel). 
Two seals are used in the nozzle ports to prevent Galvanic corrosion: one is the aluminum knife-edge seal,
which is used between the ceramic parts and aluminum conductors, and the other is a Helicoflex seal with a silver jacket,
which is used between the ceramic parts and the stainless steel flanges.
Insulation is achieved using alumina ceramic parts, and  the same scheme is also applied to the water drain ports.
Details of the water nozzle ports and water drain ports are shown in Fig. \ref{fig:nozzle_port}.
Also, this insulation scheme utilizes bolted joints instead of brazed joints.
A brazed joint between ceramic parts and metals is commonly used for the sealing of water plumbing,
however, the brazing joint often fails in an environment that is affected by vibration, which causes a water leak.
The bolted fixation used here is mechanically strong and can withstand vibration.
Another concern regarding the potential loosening of such small bolts then exists, but this can be solved by 
taking appropriate measures (e.g., welding the bolts to prevent loosening).
Since tiny gaps exist between the ceramic insulators and outer conductors, water can be trapped
within them, especially at the bottom nozzle flanges, and the trapped water can corrode metal seals.
The bottom flanges have auxiliary drain ports connected to the drain tank in order to prevent water from
accumulating. Figure \ref{fig:AuxiliaryDrain} shows the auxiliary drain port attached to the bottom nozzle flange in detail.

\begin{figure}
\centering
\includegraphics[clip,width=0.5\textwidth,bb=50 0 720 345]{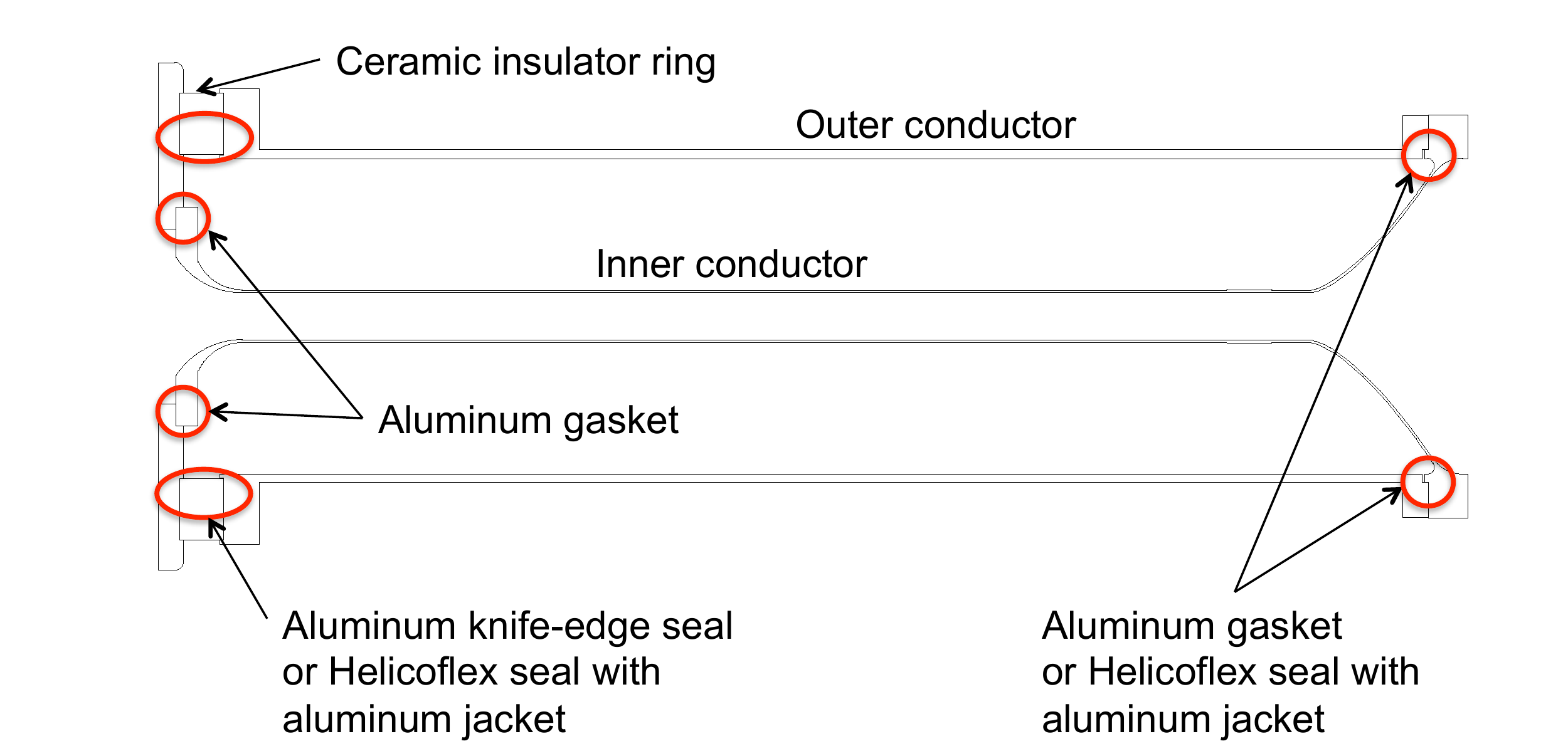}
\caption{Schematic figure of the horn conductor assembly showing several water seals.}
\label{fig:seals}
\end{figure}

For the horn conductor assemblies, several types of water seals are embedded, as shown in Fig. \ref{fig:seals}.
At the upstream part, aluminum gaskets are used between the inner conductor and upstream endplate.
At both ends of the ceramic insulator ring, Aluminum knife-edge seals are used for horn-1 and Helicoflex
seals with aluminum jackets are used for horn-2 and horn-3. At the downstream end, an aluminum gasket
is used for horn-2 and a Helicoflex seal with an aluminum jacket is used for horn-3, however, no seal is embedded 
in horn-1. This is because the tight fit of the downstream flanges between the inner and outer conductors works well as a water seal.
No leak has been observed at the downstream end of horn-1 for four years of beam operation.
However, a water leak was observed at the upstream end of this horn, details of which are given in Section \ref{sec:leak}.

\begin{figure}
\centering
\includegraphics[clip,width=0.4\textwidth,bb=10 10 725 470]{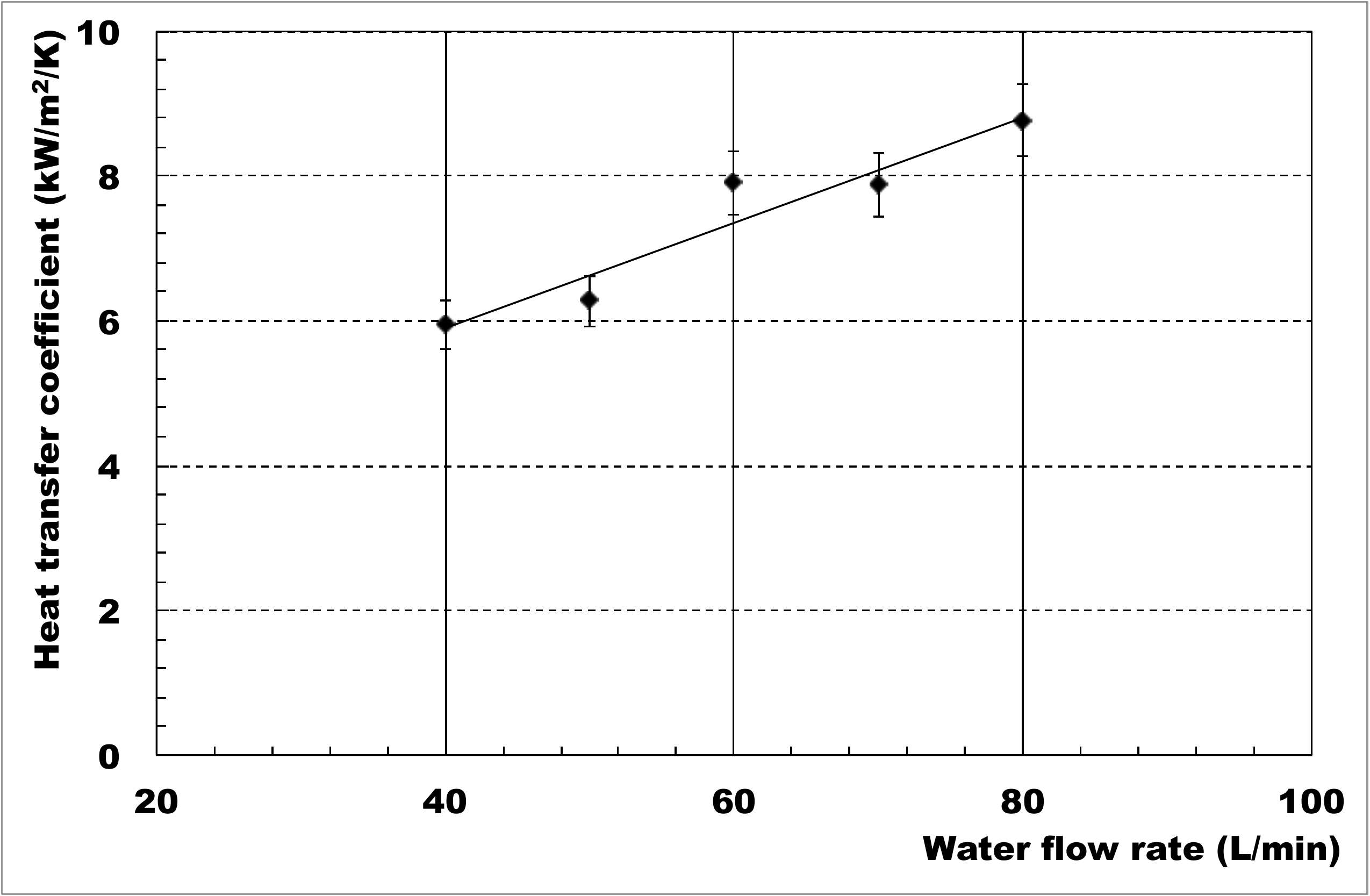}
\caption{Measured heat transfer coefficients for horn-1 as a function of water flow rate.
The data points are fitted with a linear function.}
\label{fig:cooling_performance}
\end{figure}

The cooling performance of the T2K magnetic horns was measured during current testing at the test stand.
Temperatures were measured with infrared thermometers at several inner conductor positions in order 
to evaluate the heat transfer coefficients.
Figure \ref{fig:cooling_performance} shows the measured heat transfer coefficients for horn-1 as a function of the water flow rate.
For a nominal water flow rate of 60 L/min, a heat transfer coefficient of 7.9 kW/m$^2\cdot$K is achieved.
The heat transfer coefficients obtained during current testing are summarized in Table \ref{tab:heat_coeff}. 
\begin{table}[htb]
\centering
\caption{Summary of measured heat transfer coefficients for horn-1, horn-2, and horn-3 in several
water flow conditions and at several positions. The origin of the longitudinal position is the center of the horns.}
\begin{tabular}{l|c|c|c}
\hline
 & Flow rate & Longitudinal & Heat transfer \\
 & (L/min.)& position & coefficient \\
 & & (cm) & (kW/m$^2\cdot$K) \\ \hline
horn-1 & 40 & 50 & 5.9 \\
           & 60 & 50 & 7.9 \\
           & 80 & 50 & 8.8 \\ \hline
horn-2 & 40 & 20 & 1.0 \\
            & 40 & 70 & 0.2 \\ \hline
horn-3 & 40 & 10 & 1.3 \\
            & 40 & 75 & 0.1 \\
            & 50 & 10 & 2.2 \\
            & 50 & 75 & 0.2 \\ \hline
\end{tabular}
\label{tab:heat_coeff}
\end{table}
From the measured heat transfer coefficients, the maximum temperature for 750-kW operation
is estimated to be 49.0 (horn-1), 50.3 (horn-2), and 29.6$^{\circ}$C (horn-3), respectively, 
as summarized in Table \ref{tab:est_max_temp}.
\begin{table}[htb]
\centering
\caption{Estimation of maximum temperatures at horn-1, horn-2, and horn-3 for 750-kW operation.}
\small
\begin{tabular}{lrrr}
\hline
& horn-1 & horn-2 & horn-3 \\ \hline
Instantaneous temperature rise & 18.0 $^{\circ}$C & 5.9 $^{\circ}$C & 1.8 $^{\circ}$C \\
~~~(beam exposure)  & 11.0 $^{\circ}$C & 1.1 $^{\circ}$C & 0.1 $^{\circ}$C \\
~~~(Joule heating)  & 7.0 $^{\circ}$C & 4.8 $^{\circ}$C & 1.7 $^{\circ}$C \\
Steady state temperature rise & 6.0 $^{\circ}$C & 19.4 $^{\circ}$C & 2.8 $^{\circ}$C \\
Coolant water temperature & 25.0 $^{\circ}$C & 25.0 $^{\circ}$C & 25.0 $^{\circ}$C \\ \hline
Maximum temperature & 49.0 $^{\circ}$C & 50.3 $^{\circ}$C & 29.6 $^{\circ}$C \\ \hline
\end{tabular}
\label{tab:est_max_temp}
\end{table}
The estimated maximum temperatures are much lower than the allowable maximum temperature of 80$^{\circ}$C.
Therefore, a cooling performance appropriate for 750-kW operation was achieved.

\begin{figure}[bt]
\centering
\includegraphics[clip,width=0.5\textwidth,bb=50 50 720 530]{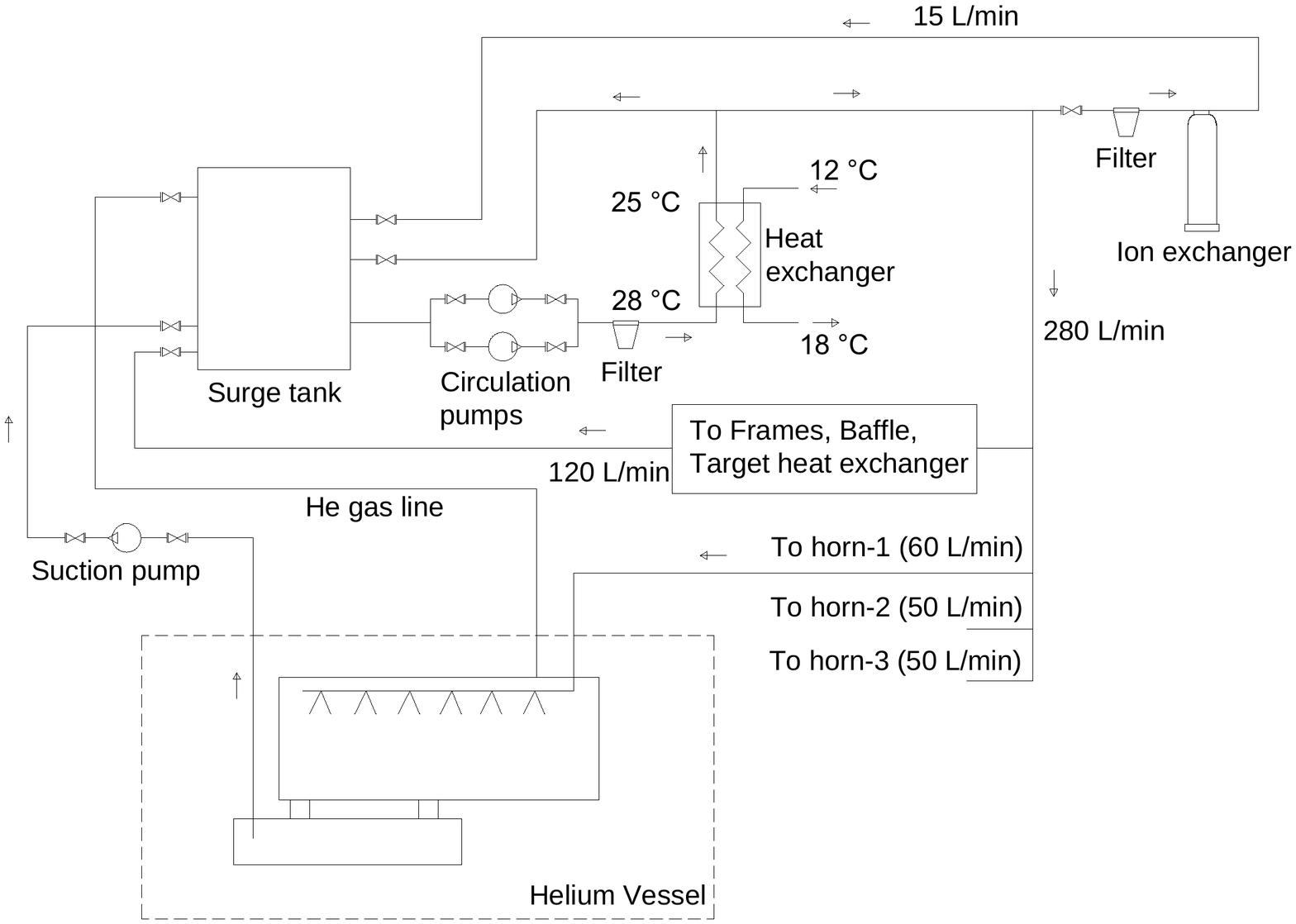}
\caption{Schematic diagram of water circulation system of T2K magnetic horns. The temperatures in this diagram
are design values.}
\label{fig:water_circ}
\end{figure}

The cooling water flow rates of horn-1, horn-2, and horn-3 are 60, 50, and 50 L/min, respectively.
Water circulation is achieved as follows:
a circulation pump supplies coolant water to each horn and the water is sprayed on the inner conductors;
the sprayed water is collected by a drain tank below the outer conductor and then pumped upwards 
by another suction pump located 6-8 m above the drain tank; the pumped water then
returns to a surge tank in a machine room. Each horn has its own suction pump.
A schematic diagram of the water circulation system of the T2K magnetic horns is shown in Fig. \ref{fig:water_circ}.
The use of suction pumps to lift the drained water to an 8-m height and
also the use of two independent pumps (a circulation and a suction pump) for circulation are considerable challenges.
The most important consideration in this system is to balance the flow rates of the supply and suction water 
so as to avoid emptying or filling the magnetic horns.
This raising of the water by 8 m using a suction pump was evaluated with a mockup and finally tested on
the actual horn configuration, which was suspended by the support modules at the test stand. 
It was found that, provided the flow rate of the return water is first adjusted, the flow rate is then automatically 
balanced in such a way that it is reduced by air bubble contamination if too much water is pumped upwards and 
the drain tank is then emptied.
All of the tests were successfully completed and this scheme was then adopted as the circulation system.  
Thereafter, no problems related to coolant water circulation have occurred during several years of operation of this system.

\subsection{Vertical frame water-cooling}
\begin{figure}
\centering
\includegraphics[clip,width=0.45\textwidth,bb=50 50 800 842]{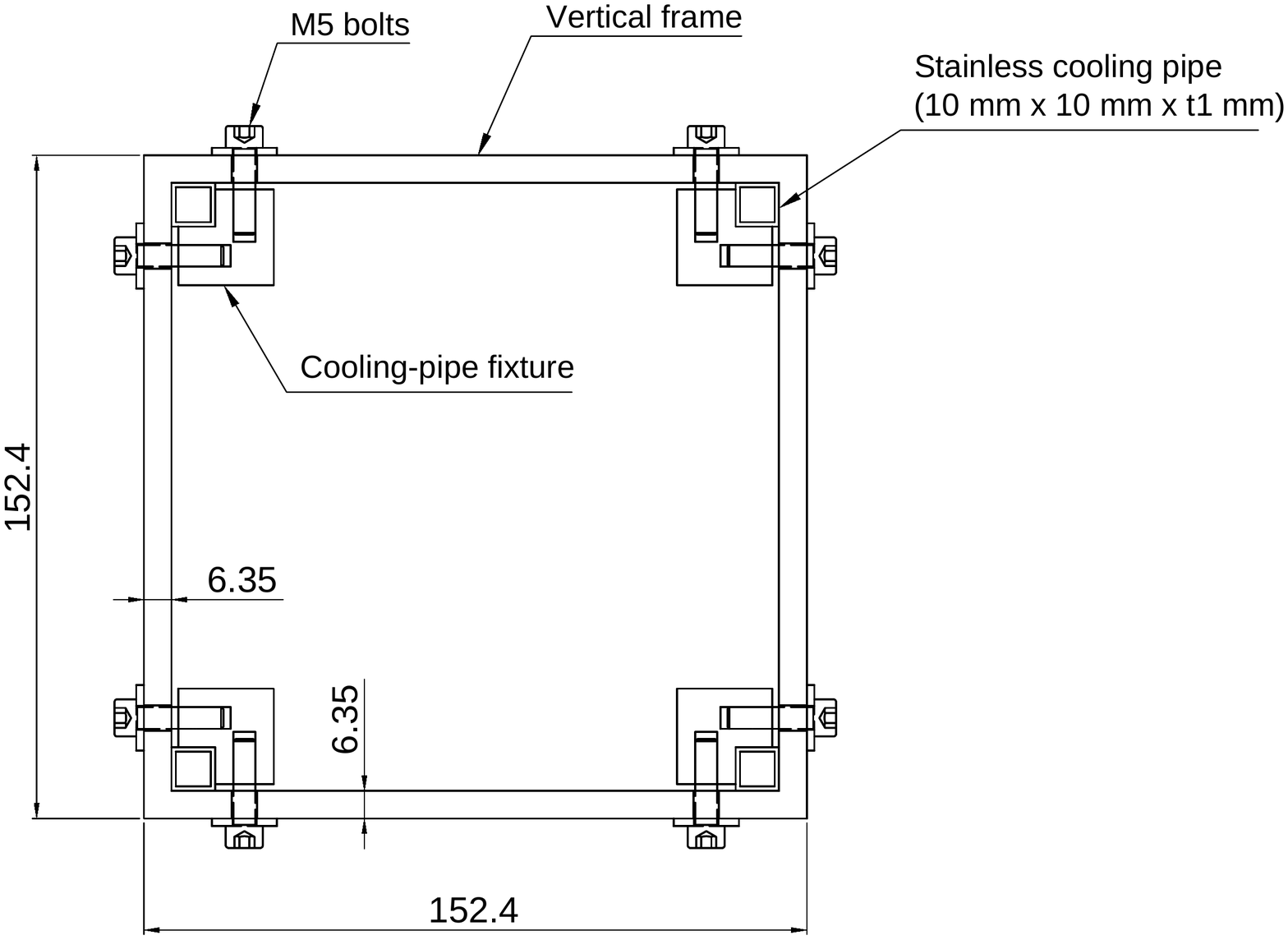}
\caption{Cross-section of vertical frame with cooling channels.}
\label{fig:frame_cooling}
\end{figure}

Since four vertical support frames are located adjacent to the outer conductor, the frames have
non-negligible heat deposits. The heat deposits at the frames are estimated to be at most 0.15 (horn-1),
0.07 (horn-2), and 0.01 W/g (horn-3). For cooling of the frames, four rectangular stainless steel
pipes are attached to the inner corners of each frame and are clamped at 10-cm intervals. This yields good contact
pressure, as shown in Fig. \ref{fig:frame_cooling}.
A mockup test for cooling of the vertical frame resulted in a heat transfer coefficient of
at least 1.1 kW/m$^2\cdot$K, which is larger than the value required for 750-kW operation by a factor of 3.8.
 
\subsection{Stripline cooling by helium}

\begin{figure}
\centering
\includegraphics[clip,width=0.45\textwidth,bb=80 100 750 510]{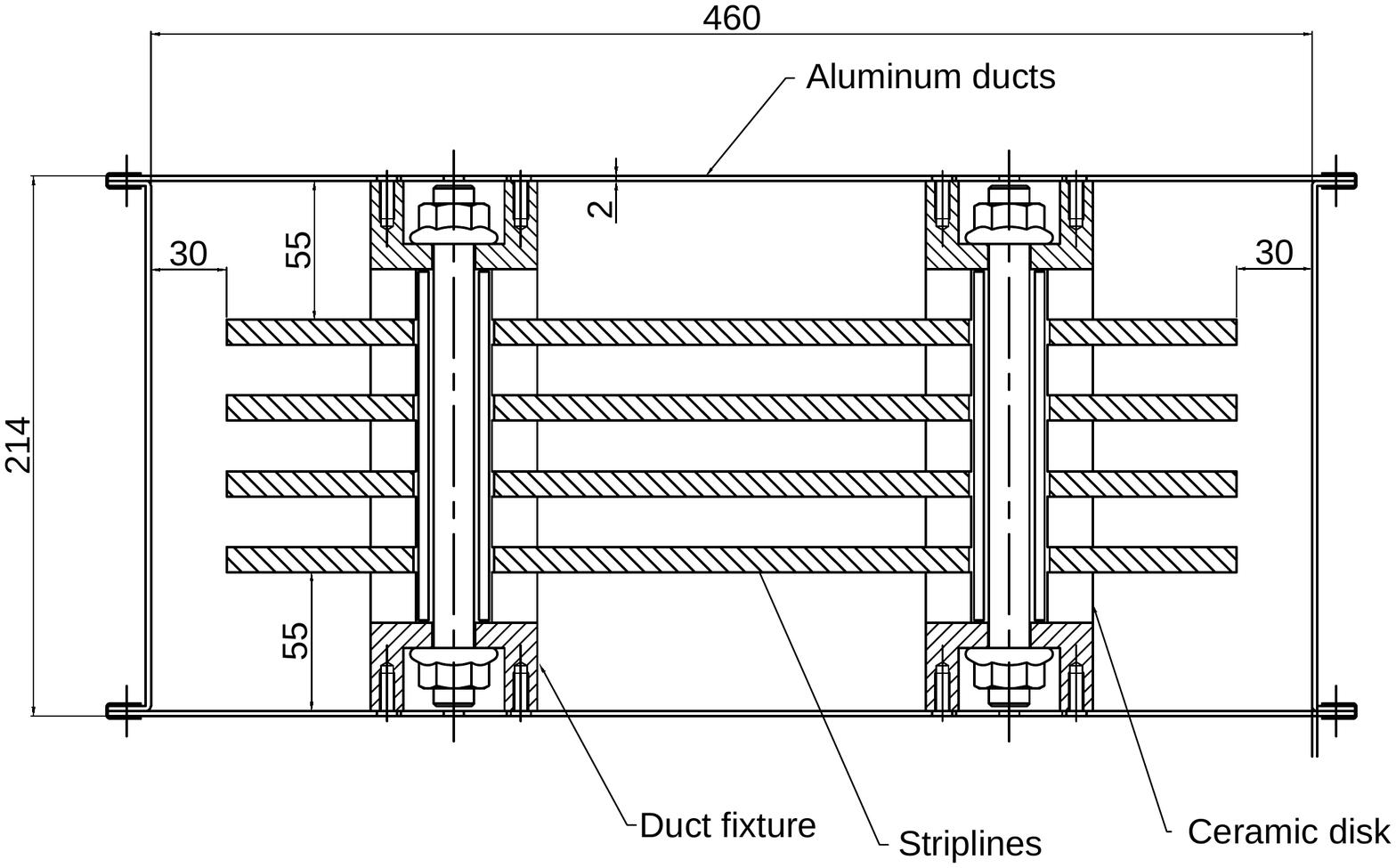}
\caption{Cross sections of striplines covered by ducts. The duct fixtures are attached
to some of the clamps.}
\label{fig:ducts}
\end{figure}

\begin{figure}[htb]
\centering
\includegraphics[clip, width=0.4\textwidth,bb=10 10 724 470]{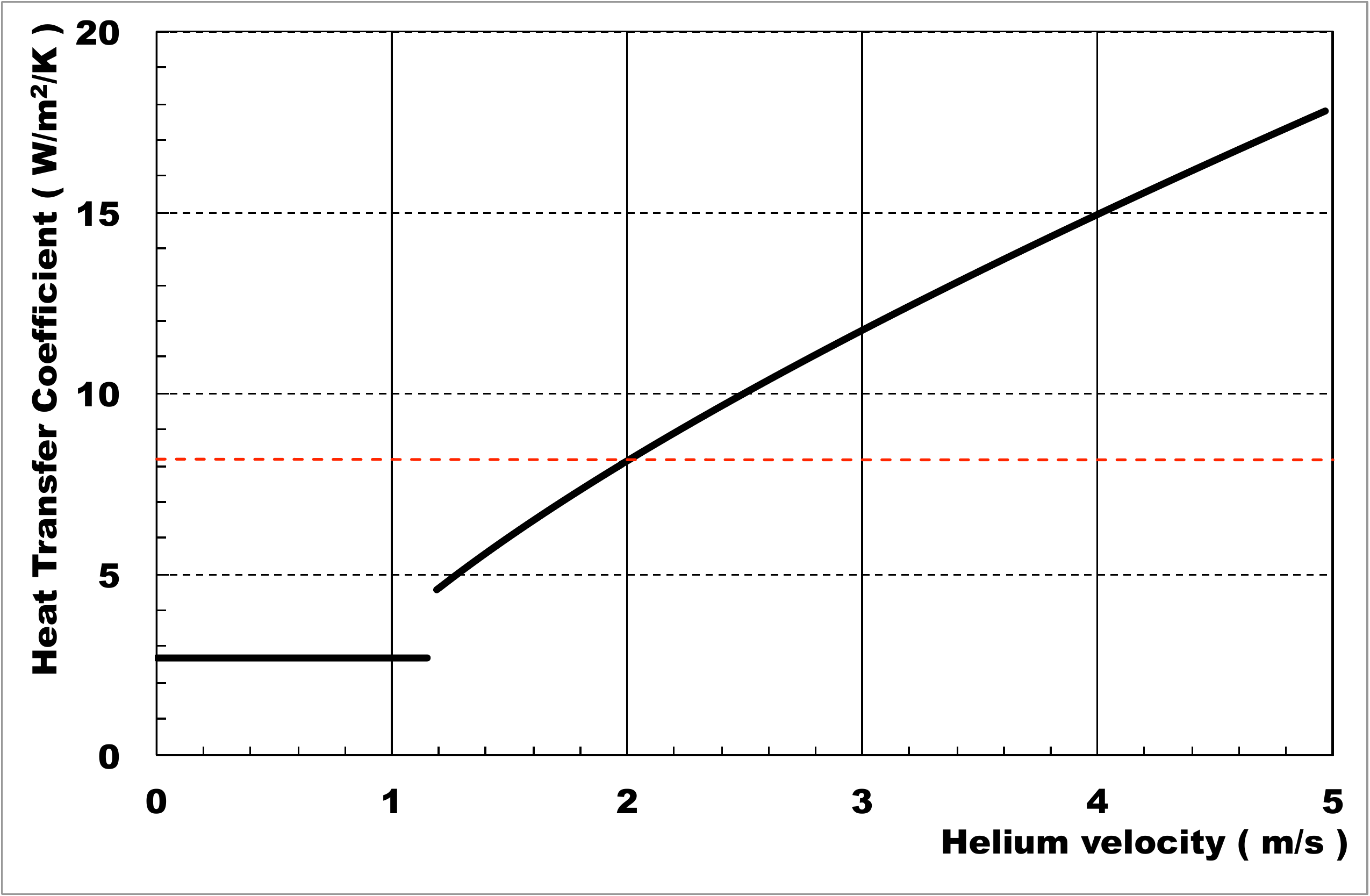}
\caption{Estimated heat transfer coefficients as a function of helium velocity.
The velocity of 1.2 m/s is the critical velocity from laminar to turbulent flow.
The heat transfer coefficient is constant in a laminar flow and,
in a turbulent flow, it is calculated using Gnielinski's formula.
A dashed horizontal line shows the required heat transfer coefficient for 750-kW operation.}
\label{fig:heat_transfer_coeff}
\end{figure}

Non-negligible heat deposits occur at the striplines, mostly due to defocused negatively charged pions.
Many negative pions defocused by horn-1 pass through horn-2 striplines and hit the walls of the HV
beside horn-2. The horn-2 striplines have the largest heat deposits among the three horns.
In addition, Joule heating also contributes to the total heat deposits at the striplines.
The maximum heat fluxes at the striplines are summarized in Table \ref{tab:stripline_heat},
\begin{table}[htb]
\centering
\caption{Summary of heat fluxes at striplines near horn conductors.}
\begin{tabular}{lrrr}
\hline
 & horn-1 & horn-2 & horn-3 \\ \hline
 Beam (J/m$^2$) & 140.5 & 895.0 & 106.0 \\
 Joule heating (J/m$^2$) & 96.0 & 47.7 & 32.5 \\ \hline
 Total (J/m$^2$) & 236.5 & 942.7 & 138.5 \\ \hline
 \end{tabular}
 \label{tab:stripline_heat}
 \end{table}
and occur at the striplines near the horn conductors.
Cooling of the striplines is performed using a forced helium flow.
The striplines inside the HV, especially those around the magnetic horns and support
modules, are enclosed within aluminum ducts through which helium gas flows.
Figure \ref{fig:ducts} shows a cross section of stripline surrounded by a duct.
The heat transfer coefficient calculated according to the fluid dynamics formula is shown
as a function of helium velocity in Fig. \ref{fig:heat_transfer_coeff}.
At 750-kW operation, the heat flux for horn-2 is estimated to be 942.7 J/m$^2$.
Assuming that the coolant helium temperature is 25$^{\circ}$C,
the heat transfer coefficient for horn-2 is required to be over 8.2 W/m$^2\cdot$K
in order to maintain a maximum stripline temperature of below 80$^{\circ}$C. 
A velocity of 2.0 m/s is required to achieve the above heat transfer coefficient.

\begin{figure}
\centering
\includegraphics[clip,width=0.5\textwidth,bb=0 0 685 540]{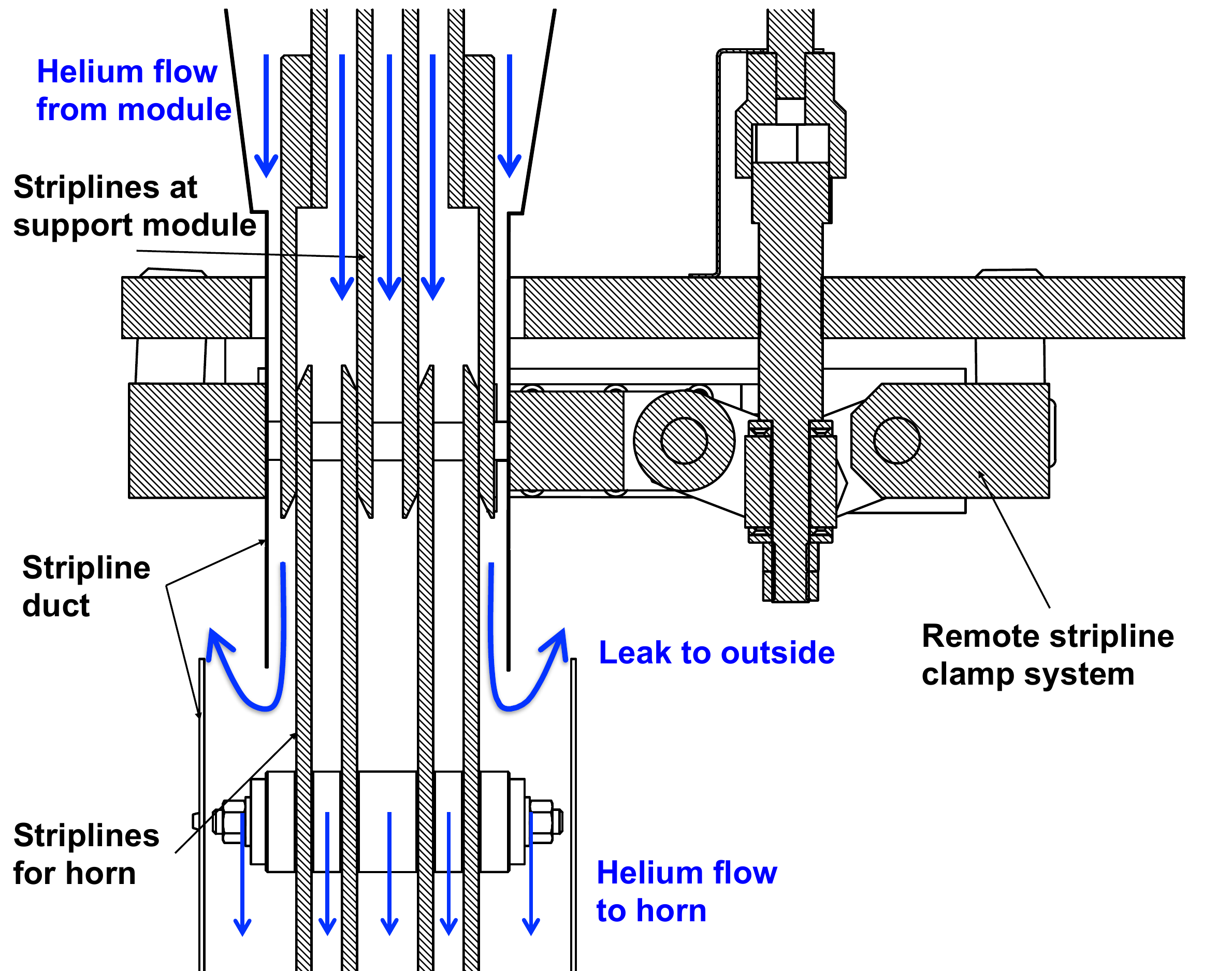}
\caption{Helium flow around remote stripline clamp system.}
\label{fig:helium_leak}
\end{figure}

Before installation of the magnetic horns into the HV, an air flow test was performed using the full stripline
ducts from the support modules to the horns. Since the ducts around the remote stripline clamp system
could not be tightly connected, because the structure around the stripline remote connection was very complicated (as
shown in Fig. \ref{fig:helium_leak}),
a significant amount of air exited the duct connection gaps (more than 60 \%). Hence, the expected flow rate
was not achieved at the duct exits. Although the expected stripline cooling performance was not achieved, 
no modification of the duct connection was made, because the necessary design modification was not completed 
by the date of installation.

After installation into the HV, an in-situ helium flow measurement was performed.
The measured flow rates at the inlet of the HV were 17.0 (horn-1), 15.6 (horn-2), 
and 16.0 m$^3$/min (horn-3), which corresponded to exit velocities of 1.10 (horn-1), 1.07 (horn-2), 
and 0.73 m/s (horn-3), respectively.
However, the actual helium flow rate at the exit could not be estimated correctly because of the helium leak 
at the stripline remote connection. 
Therefore, the actual heat transfer coefficient is estimated to be 2.69 W/m$^2\cdot$K.
This must be improved by a factor of 3 in the future, and
several modifications to the next magnetic horn production design are necessary.
The duct shapes, especially the duct cross sections, should be optimized to achieve an even helium velocity distribution
inside the ducts. Another important modification is a leak-tight duct connection in the remote stripline clamp system.

\subsection{Production of radioactivity, hydrogen, and nitrogen oxide}

Since secondary particles are also exposed to the cooling water, many kinds of radioactive nuclei are produced in it \cite{rad_water}. 
Dominant contributors to radioactivity in the coolant water are $^7$Be and $^3$H (Tritium or T), which have
half-lives of  53.3 d and  12.3 y, respectively. The production rates of $^7$Be and $^3$H are estimated to be 300 and
150 GBq, respectively, for 1 year of operation at 750 kW, which is equivalent to $1.56\times 10^{21}$ protons on target (POT). 
Since $^7$Be exists as cations in the water, its particles can be removed using an ion exchanger.
First, radioactive water is circulated through the ion exchanger and also the horns.
It is found that 1\% of the $^7$Be cannot be removed by ion exchangers because the majority of the remaining $^7$Be cations
are absorbed by metal-oxide colloids and are electrically neutral. In the first stage, the reduction efficiency is measured as 99\%
\cite{water_radioactivity}.
In order to further remove $^7$Be, another water tank (called a buffer tank) was constructed for radioactivity reduction,
after beam operation for physics data-taking began in 2010.
In this system, after the first ion-exchange operation, the radioactive water is moved to the buffer tank and 
sulfuric acids are added, so that the metal-oxide colloids with $^7$Be cations are 
degraded under strong acidity conditions. Then, the water is circulated through the
ion exchanger so that the remaining $^7$Be cations are removed. 
A total reduction efficiency of 99.9\% is achieved by two stages of ion-exchange operation.
For drainage of the processed water, hydrochloric acids are added to neutralize the water.
Details of the dual-stage $^7$Be reduction system are given in \cite{Oyama_NBI2012}. 
On the other hand, $^3$H (T) exists as molecules of HTO in the water, therefore, 
it is not removed by an ion exchanger; instead, it is simply diluted below the regulation level and disposed of, so as to reduce
the amount of tritiated water in the water circulation system. Additional information on the
observation and estimation of radioactivity in the horn cooling water is given in \cite{water_radioactivity}.

The beam exposure to the coolant water also creates hydrogen gas through water radiolysis.
Its production rate is approximately 40 L/d at 750-kW operation and it is primarily produced at horn-1. 
The inner volumes of the magnetic horns and cover gas region in the surge tank are filled with helium gas 
at 1 atm., and both regions are connected by the single-path plumbing for each horn, as shown in Fig. \ref{fig:water_circ}.
The total volume of the helium gas region is approximately 5.5 m$^3$. The produced hydrogen gas is removed by
a hydrogen recombination catalyst that is adapted to the water circulation system in order to prevent a hydrogen gas
explosion. This system was adopted after beam operation for physics data-taking was initiated, and is being
improved to achieve superior removal. Details of the hydrogen reduction are given below, in Section \ref{sec:operation}.

Beam exposure to nitrogen in the air produces nitrogen oxide. Once these particles are dissolved in the coolant water,
nitric acids are produced and the water is acidized \cite{NitricAcid}, following
\begin{equation}
2\textrm{NO}_2 + 2\textrm{H}_2\textrm{O} \rightarrow 2\textrm{HNO}_3 + \textrm{H}_2.
\end{equation}
For aluminum, such acidic conditions are very dangerous,
because aluminum corrosion proceeds dramatically in strongly acidic conditions (e.g. pH $<$ 4) \cite{Alhandbook}.  
The helium environment inside the magnetic horns significantly reduces the production of nitrogen oxides.
The aluminum corrosion occurring during beam operation will be described below, in Section \ref{sec:operation}.


%
%
\section{Mechanical properties}
\label{sec:mechanical}

The mechanical properties of the T2K magnetic horns are described in this section.
The horn conductors experience both a Lorentz force and a thermal shock due to
beam exposure. Design optimization was performed based on stress analyses
conducted using the finite element method (FEM).

The tensile strength of A6061-T6 is 310 MPa and the yield strength is 275 MPa. 
The T2K magnetic horns are expected to be used for over five years and the
accumulated number of pulses during the five-year operation is designed to be 10$^8$ or more.
Fatigue due to repetitive forces must be considered in the mechanical design of magnetic horns.
The fatigue strength of A6061-T6 after 5$\times$10$^8$ cycle of the repetitive forces is 95 MPa \cite{Alhandbook}.
The magnetic horns are water-cooled, thus, the inner surfaces of the inner and outer conductors
make contact with the coolant water. 
The aluminum strength is reduced by constant exposure to water. An empirical factor of 0.43 was used
to reduce the fatigue strength of the wet aluminum in the horns, and the allowable stress was further reduced in order to
25 MPa to increase the confidence level for fatigue failure from 50\% to 97.5\%, and to account for the effect
of the fatigue stress ratio.

\subsection{Stress analyses}
Design optimization was first performed using static stress analyses and confirmations were
made based on dynamic stress analyses.
For a thermal stress analysis, energy deposits in the conductors are calculated using the MARS \cite{MARS} simulation code.
For analysis of the stress due to the Lorentz force, the pressure on the inner conductors is calculated from
\begin{eqnarray}
i(r) & = & \frac{I}{\pi (r_2^2 - r_1^2)}, \\
B(r) & = & \frac{\mu_0}{2\pi}\frac{I}{r_2^2 - r_1^2}\frac{r^2 - r_1^2}{r}, \\
p & = & \frac{1}{2\pi r_2}\int_0^{2\pi} \int_{r_1}^{r_2} i(r)\times B(r) dr(rd\phi) \nonumber \\
   & = & \frac{1}{r_2}\int_{r_1}^{r_2}\frac{\mu_0}{2\pi}\frac{I^2}{\pi (r^2_2 - r^2_1)^2}\frac{r^2 - r^2_1}{r} rdr  \nonumber \\
  & = & \frac{\mu_0}{2\pi} I^2 \frac{1}{3\pi}\frac{r_2+2r_1}{r_2(r_2+r_1)^2}, \label{eq:press}
\end{eqnarray}
where $r$ is the radial position, $r_1$ and $r_2$ are the inner and outer radii of the inner conductors, respectively,
$i(r)$ and $B(r)$ are the current density and magnetic field inside the inner conductors at $r$, respectively,
and $p$ is the pressure on the inner conductors.
For $I$ = 320 kA, Equation (\ref{eq:press}) gives a simple approximation formula
\begin{equation}
p \approx \frac{1690}{r_2 \textrm{[mm]}^2} \textrm{[MPa]}, \label{eq:simple_press}
\end{equation}
where $t\equiv r_2 - r_1$ = 3 mm. In later stress analyses for Lorentz forces, Equation (\ref{eq:simple_press})
is used to calculate the pressures on the conductors.

\begin{figure}
\centering
\includegraphics[clip,width=0.45\textwidth,bb=0 0 720 448]{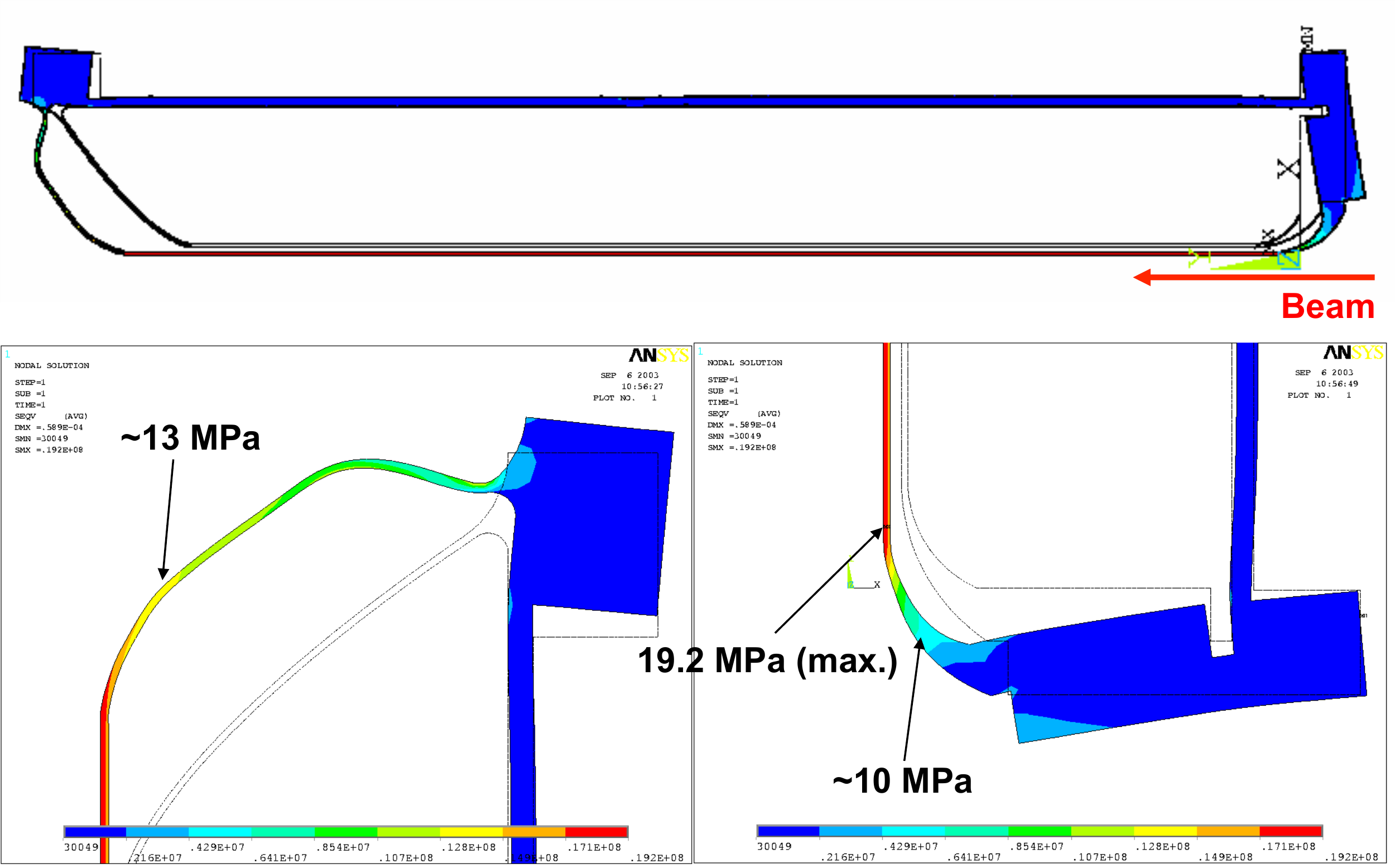}
\caption{Simulation results of static stress analysis showing Von Mises equivalent stresses due to Lorentz force for horn-1.}
\label{fig:stress_ana_Lorentz_h1}
\end{figure}

\begin{figure}
\centering
\includegraphics[clip,width=0.45\textwidth,bb=10 0 720 411]{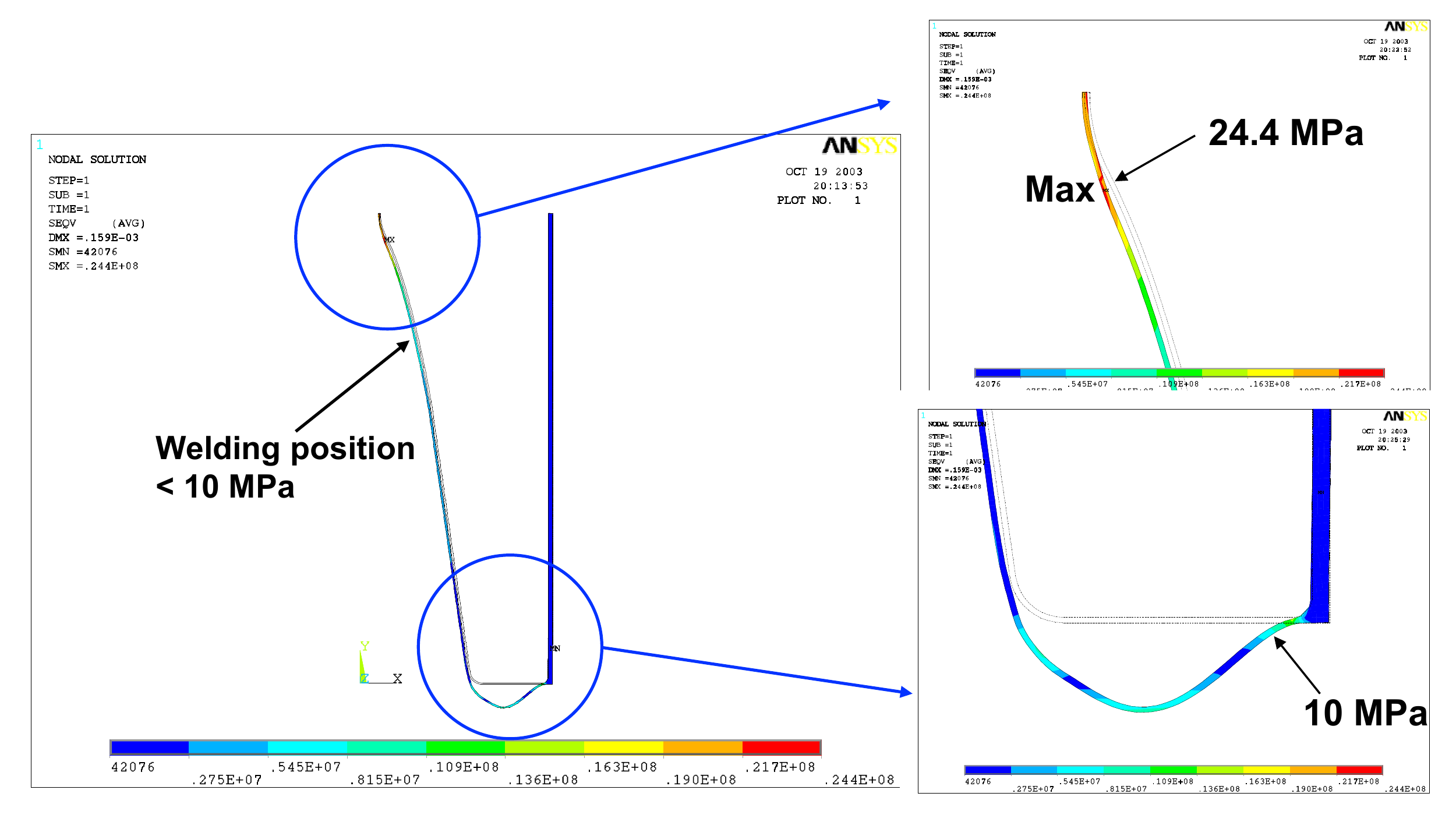}
\caption{Simulation results of static stress analysis showing Von Mises equivalent stresses due to Lorentz force for horn-2.}
\label{fig:stress_ana_Lorentz_h2}
\end{figure}

\begin{figure}
\centering
\includegraphics[clip,width=0.45\textwidth,bb=0 0 720 419]{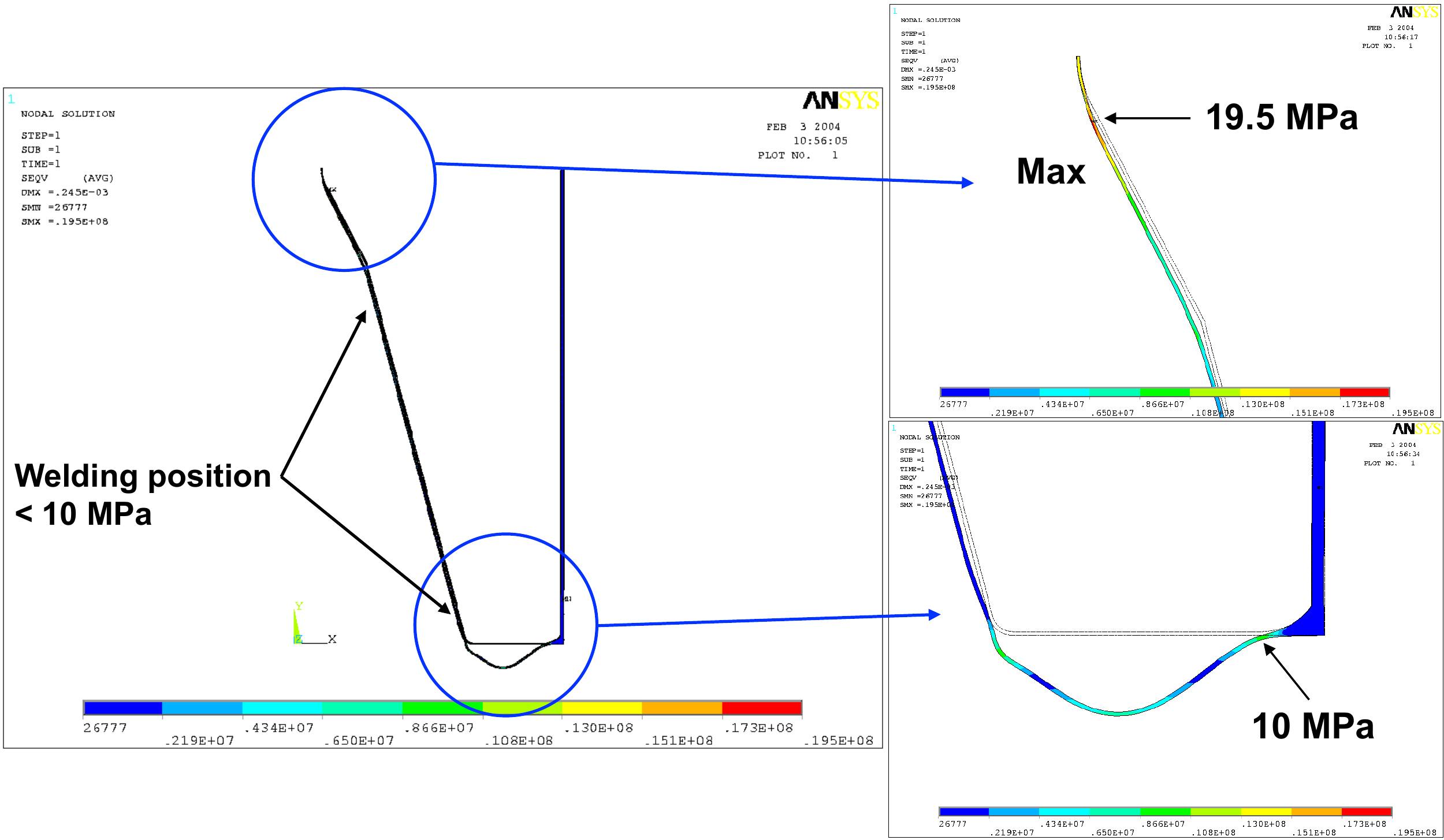}
\caption{Simulation results of static stress analysis showing Von Mises equivalent stresses due to Lorentz force for horn-3.}
\label{fig:stress_ana_Lorentz_h3}
\end{figure}

The static stress analyses for the Lorentz forces were performed using the ANSYS \cite{ANSYS} simulation code.
Figures \ref{fig:stress_ana_Lorentz_h1}-\ref{fig:stress_ana_Lorentz_h3} 
show the results of the static stress analyses for horn-1, horn-2, and horn-3, respectively.
The values for the Von Mises equivalent stresses obtained from the static stress analyses are summarized in 
Table \ref{tab:stress_ana_Lorentz},
\begin{table}[htb]
\centering
\caption{Summary of maximum Von Mises equivalent stress from static stress analyses considering Lorentz force.}
\begin{tabular}{cccc}
\hline
& horn-1 & horn-2 & horn-3 \\ \hline
Maximum pressure (MPa) & 1.88 & 0.91 & 0.32 \\
Maximum stress (MPa) & 19.2 & 24.4 & 19.5 \\
Maximum displacement (mm) & 0.06 & 0.16 & 0.25 \\
Position at maximum stress & cylinder & neck & neck \\ \hline
\end{tabular}
\label{tab:stress_ana_Lorentz}
\end{table}
along with the maximum pressures calculated from Equation (\ref{eq:simple_press}).
For horn-1, the maximum stress of 19.2 MPa arises at the straight cylindrical section.
For both horn-2 and horn-3, the maximum stress of 24.4\footnote{For the horn-2 inner conductor,
the conductor shape was slightly modified in the manufacturing design. After horn-2 production,
stress analysis was performed based on the actual manufacturing drawing, and it was found that
the maximum stress reached 30 MPa.} and 19.5 MPa arise at the``neck" section, where
their diameters are at a minimum.

\begin{figure}[bh]
\centering
\includegraphics[clip,width=0.45\textwidth,bb=0 0 721 430]{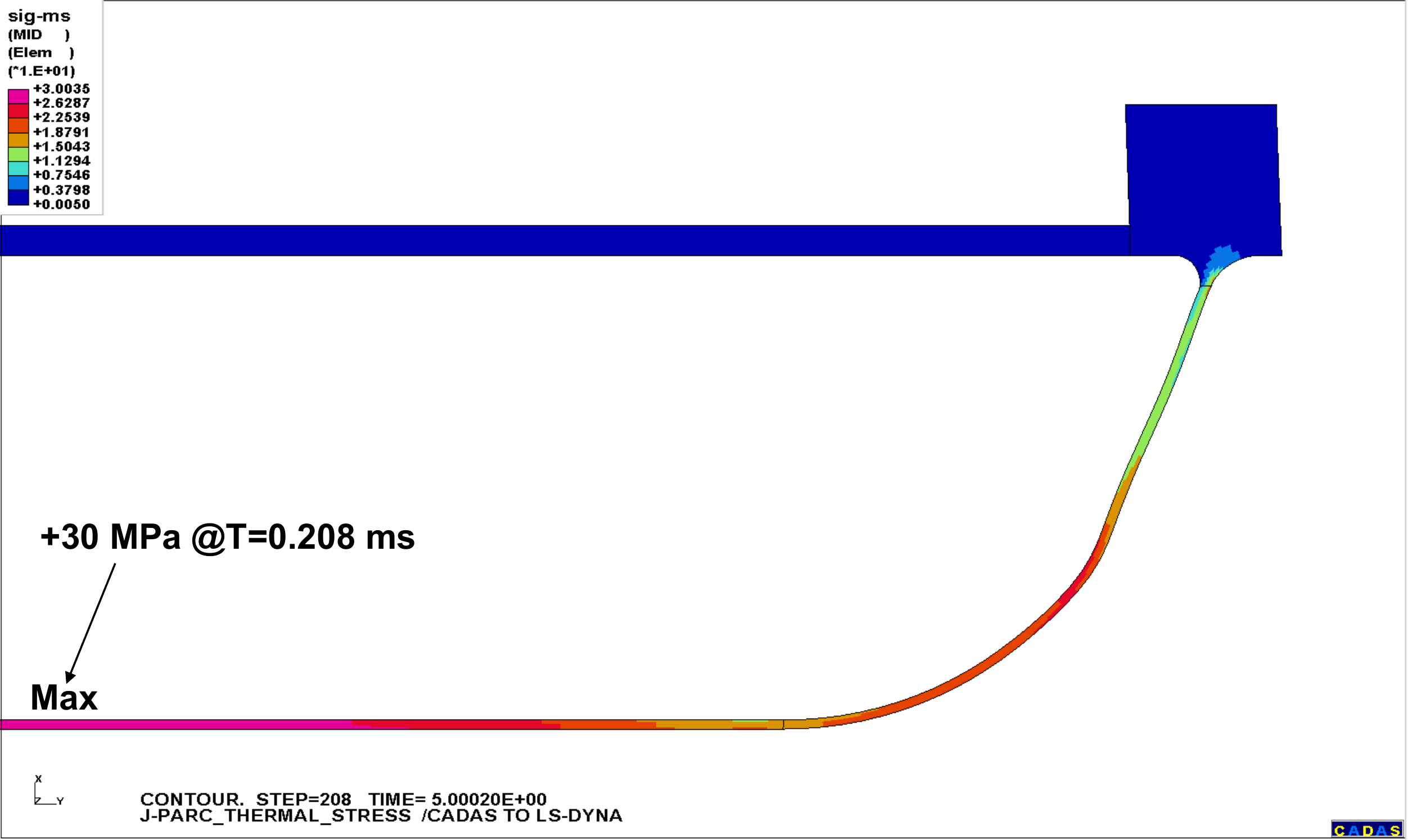}
\includegraphics[clip,width=0.45\textwidth,bb=0 0 724 220]{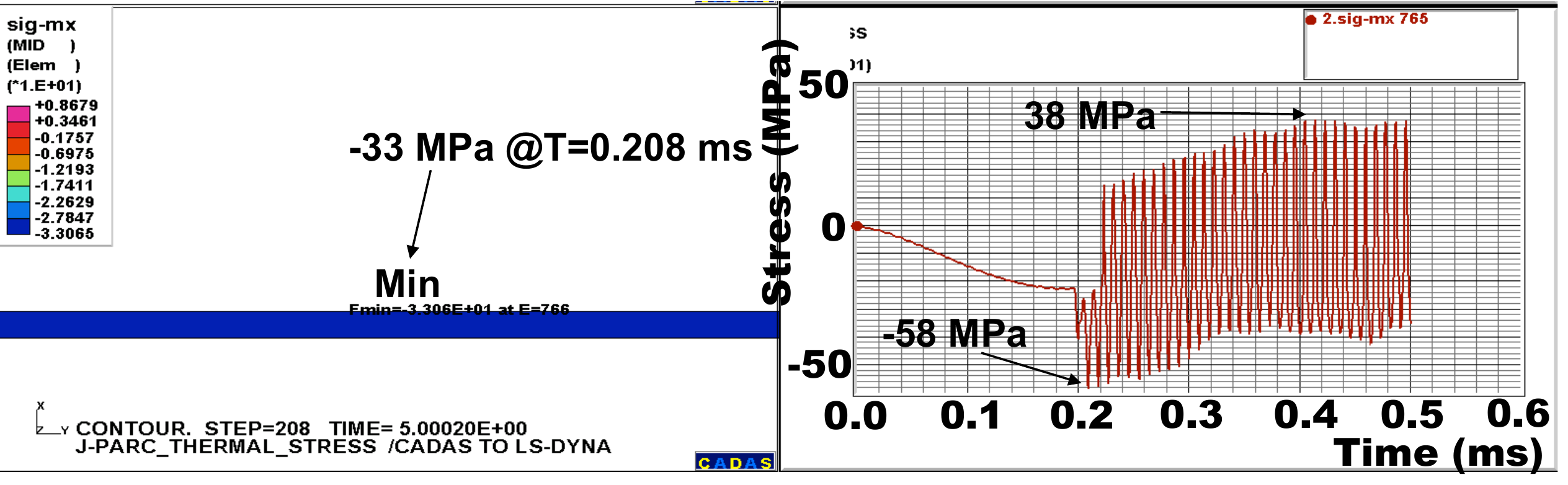}
\caption{Results of dynamic stress analyses for horn-1. Top figure: Von Mises equivalent stress at
horn-1 conductors immediately after beam exposure. Bottom figures: principal stress perpendicular to longitudinal
direction around cylindrical part immediately after beam exposure (left) and time variation of principal stress at 
a point on the cylindrical part, in units of MPa (vertical) and ms (horizontal), respectively (right).}
\label{fig:stress_dynamic1}
\end{figure}

\begin{figure}
\centering
\includegraphics[clip,width=0.45\textwidth,bb=0 0 723 459]{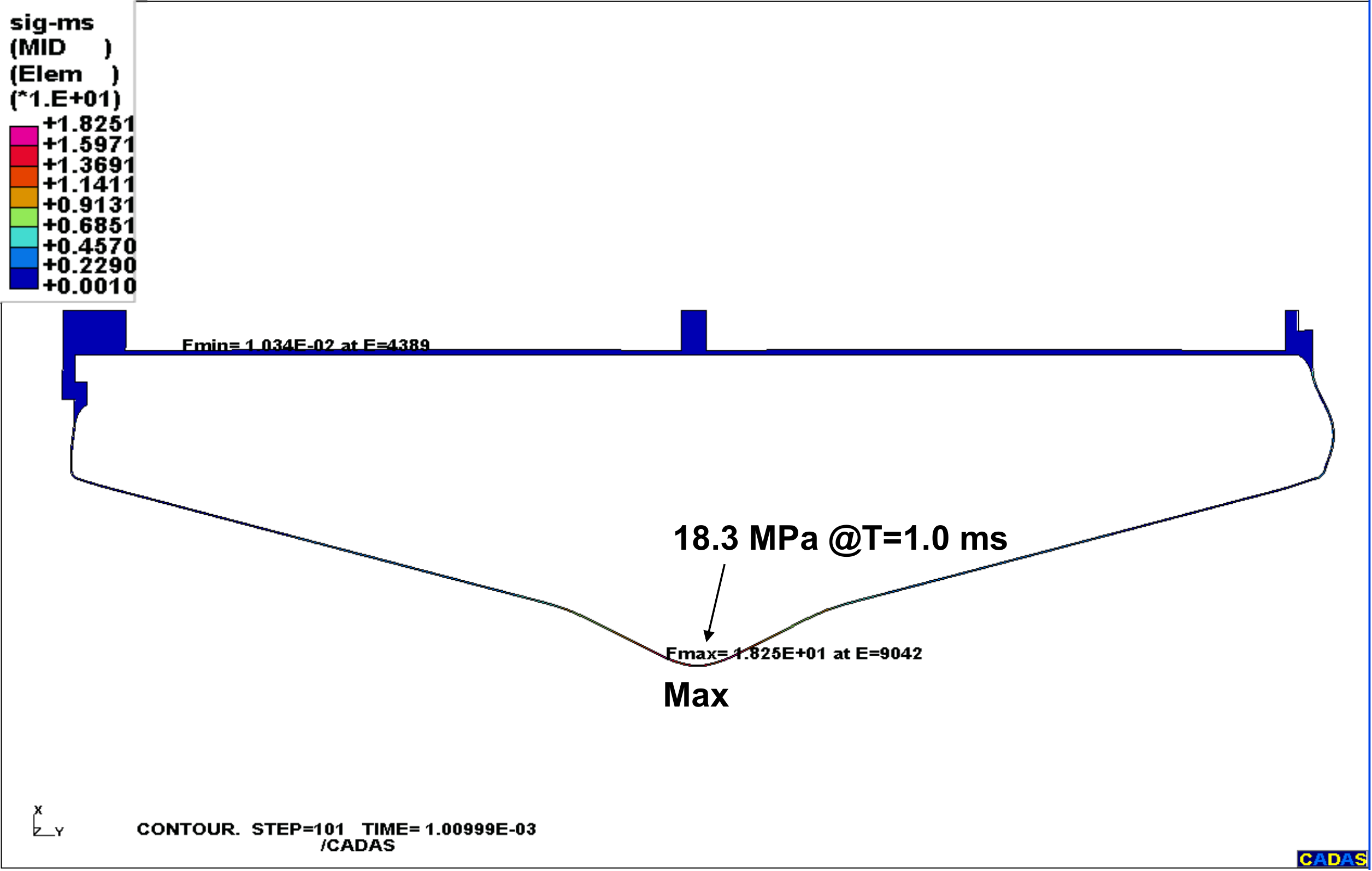}
\includegraphics[clip,width=0.45\textwidth,bb=0 0 721 458]{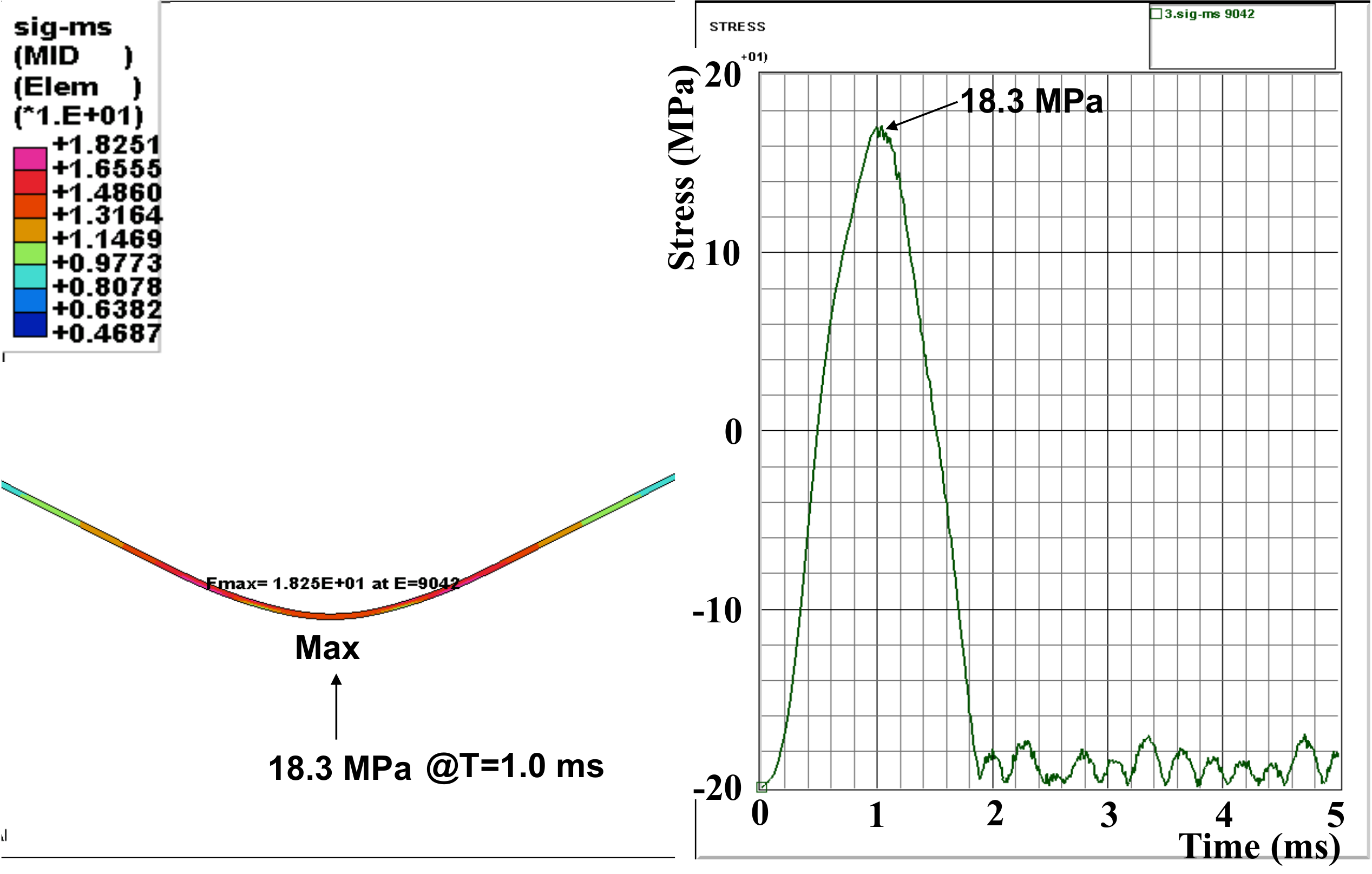}
\includegraphics[clip,width=0.45\textwidth,bb=0 0 721 458]{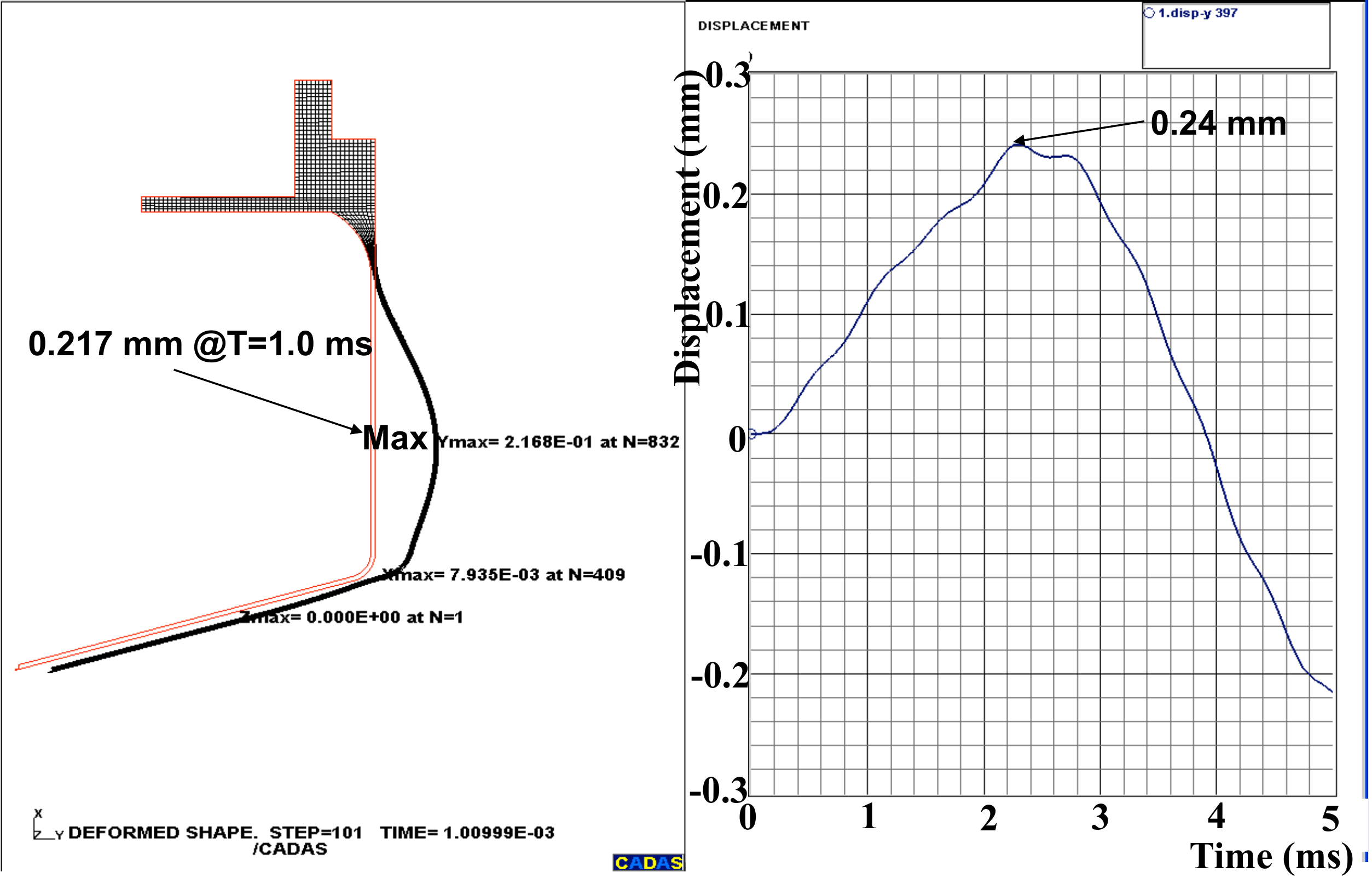}
\caption{Results from dynamic stress analyses for horn-3. Top figure: Von Mises equivalent stress
for full horn-3 conductors. Middle figures: Magnification of top figure around neck part (left) and time variation
of Von Mises equivalent stress around neck part in units of MPa (vertical) and ms (horizontal), respectively.
Bottom figures: displacement in longitudinal direction around downstream endplate (left) and time variation
of displacement around endplate in units of mm (vertical) and ms (horizontal), respectively.}
\label{fig:stress_dynamic3}
\end{figure}

Dynamic stress analyses were performed for horn-1 and horn-3 by Hitachi Engineering Co., 
Ltd.\footnote{The company name has now been changed to Hitachi Power Solutions Co., Ltd.}
The dynamic analyses take into account both thermal stresses due to beam exposure and Joule heating and 
stresses due to the Lorentz force. For horn-1, the dynamic stress analyses were performed with old conductor shapes only,
which were not optimized, and the value of the Von Mises equivalent stress was slightly larger than 25 MPa,
which is consistent with the static analysis result when the same conductor shape is used.
The results of the dynamic stress analyses are shown in Figs. \ref{fig:stress_dynamic1} and \ref{fig:stress_dynamic3}
and summarized in Table \ref{tab:stress_dynamic}.
\begin{table}[htb]
\centering
\caption{Summary of dynamic stress analyses for horn-1 and horn-3. The stresses shown in this table are
Von Mises equivalent stresses.}
\begin{tabular}{lrr}
\hline
 & horn-1 & horn-3 \\ \hline
Time duration (ms) & 0.5 & 5.0 \\
Time step (ms) & 0.001 & 0.01 \\
Pulse width (ms) & 0.4 & 2.0 \\
Time at beam exposure (ms) & 0.2 & 1.0 \\  
Maximum stress (MPa) & 30.0 & 18.3 \\
Position at maximum & cylinder & neck \\
Time at maximum (ms) & 0.208 & 1.00 \\ \hline
\end{tabular}
\label{tab:stress_dynamic}
\end{table}
Although the pulse widths of the pulsed current in these analyses are not the same as in the actual conditions,
it is expected that the maximum stress does not change, because the results from the dynamic stress analyses 
are consistent with those from the static analyses.

\subsection{Vibration measurements}
\begin{figure}
\centering
\begin{tabular}{r}
\includegraphics[clip,width=0.45\textwidth,bb=5 20 730 470]{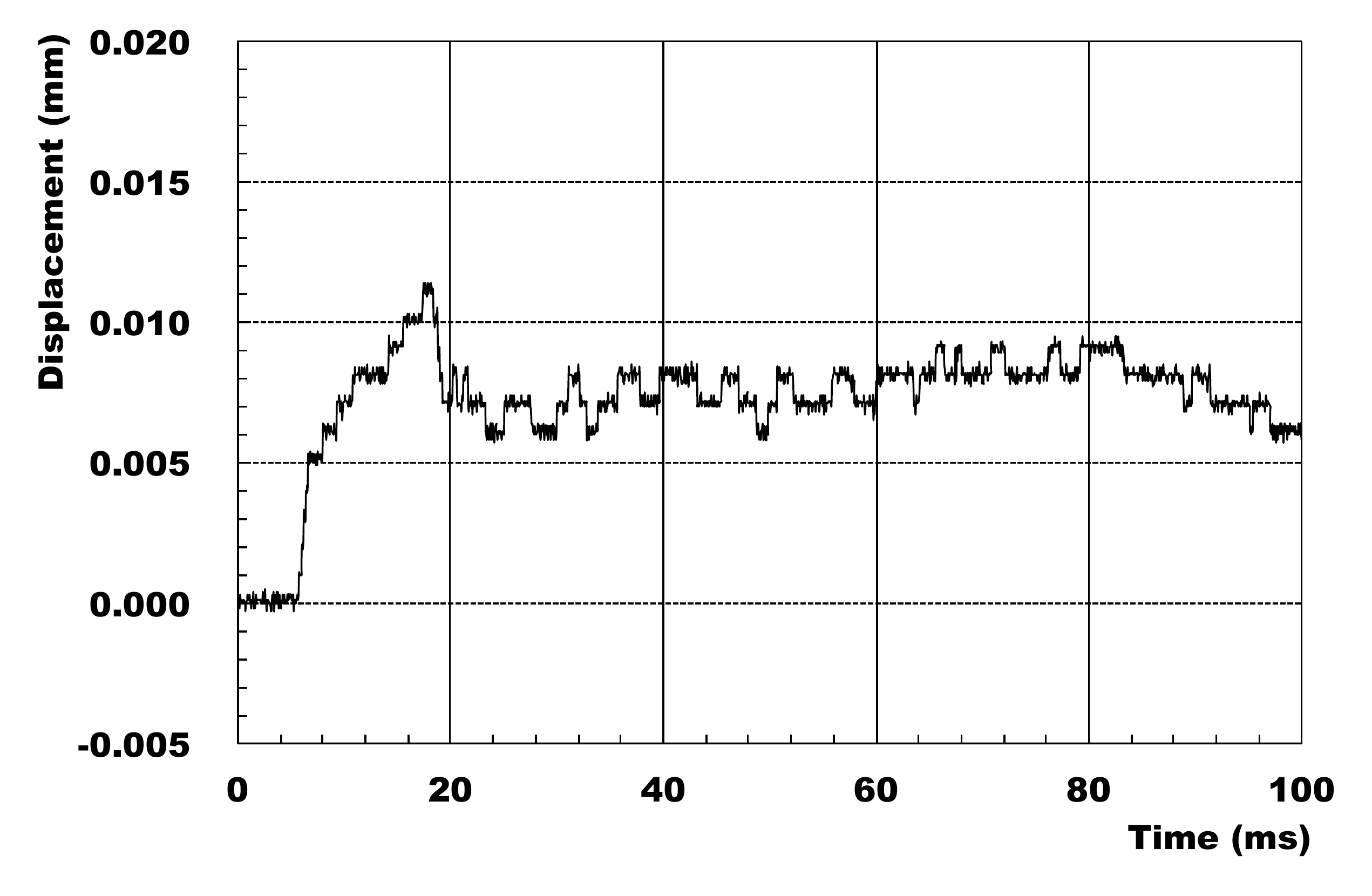} \\
\includegraphics[clip,width=0.45\textwidth,bb=5 10 730 475]{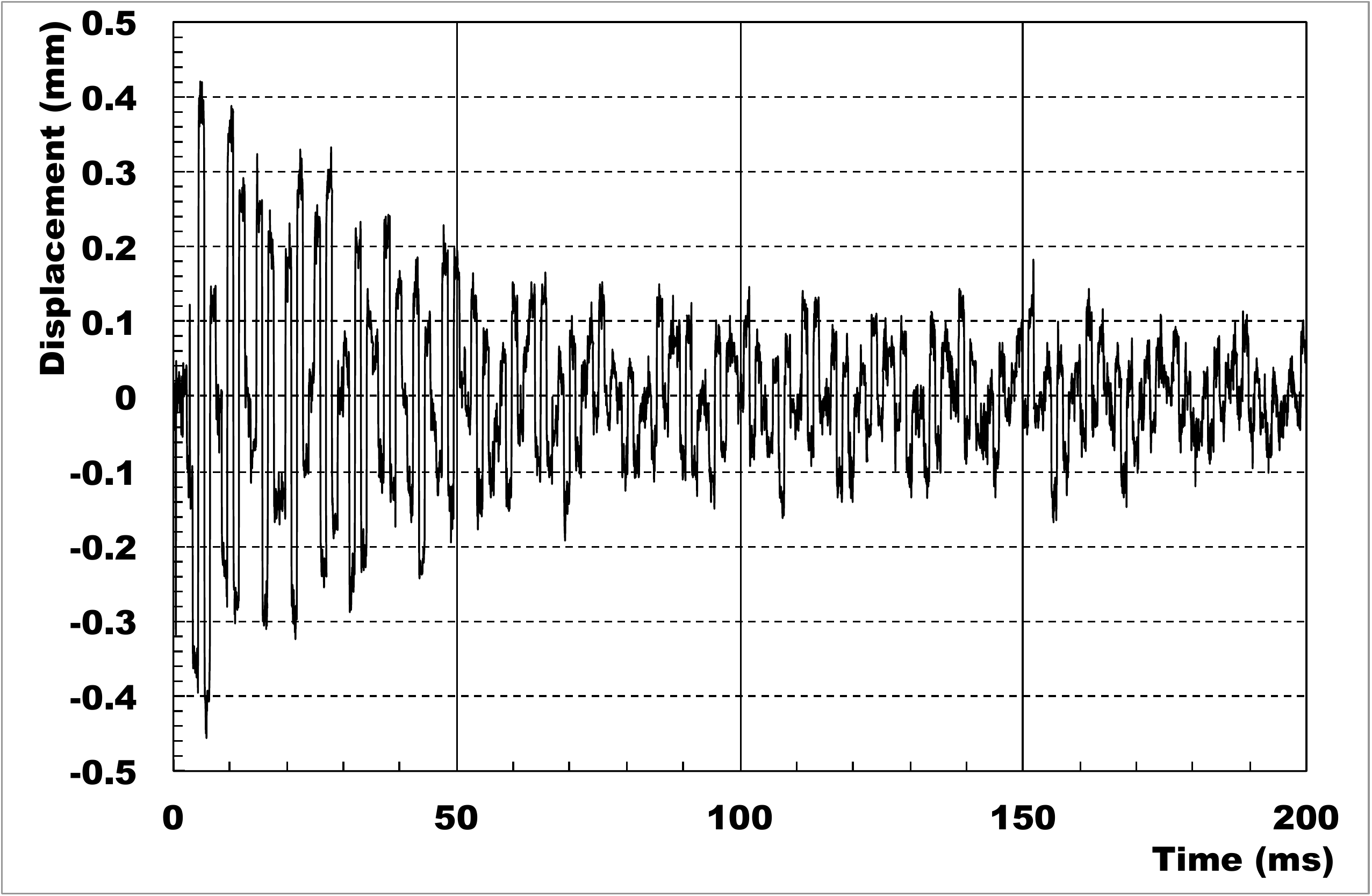} \\
\includegraphics[clip,width=0.45\textwidth,bb=5 5 730 468]{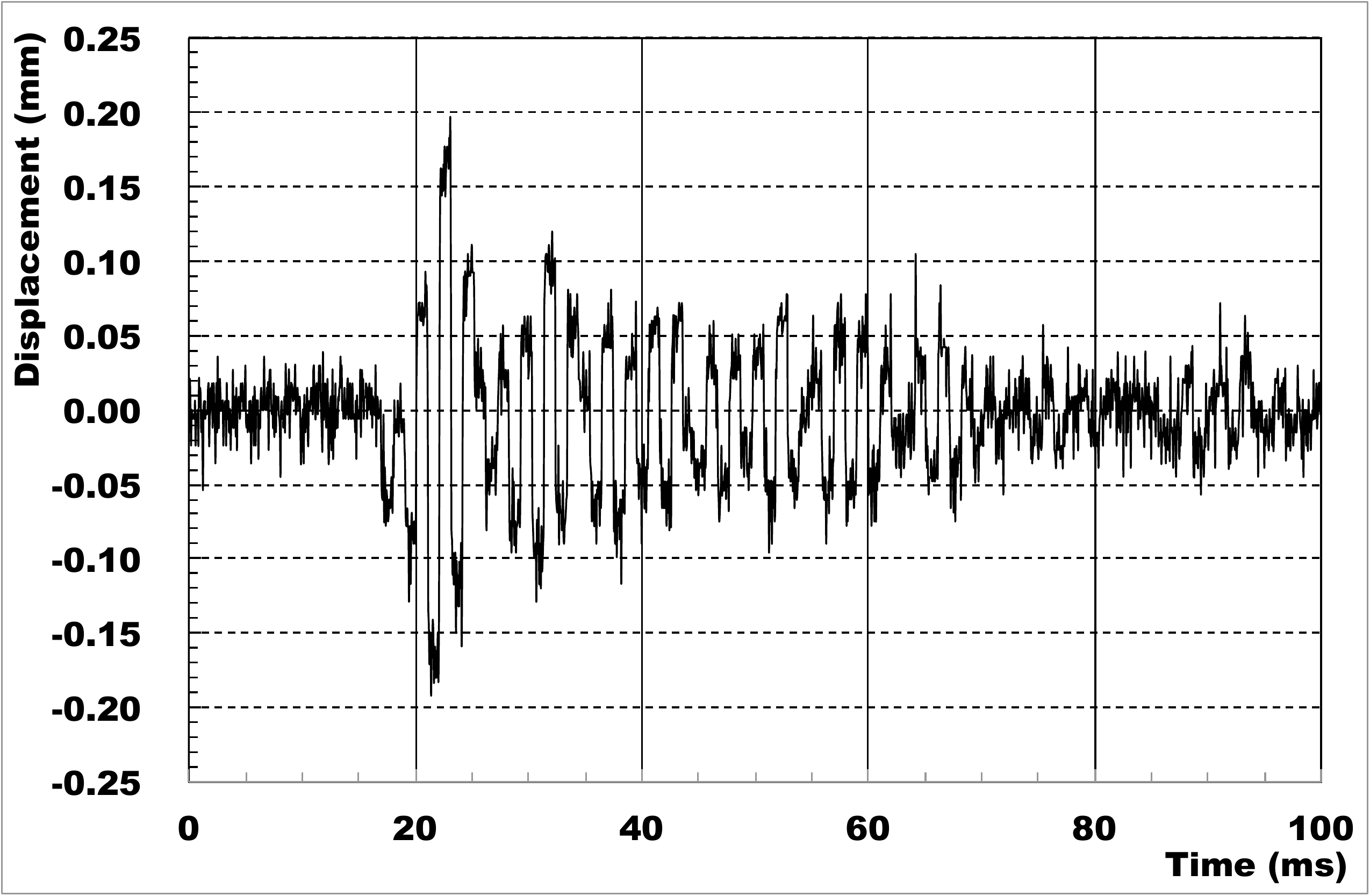}
\end{tabular}
\caption{Time variations of measured displacements at downstream parts of inner conductors
for horn-1 (top), horn-2 (middle), and horn-3 (bottom). The operation currents for these measurements
are 320 (horn-1), 250 (horn-2), and 266 kA (horn-3).}
\label{fig:displacement}
\end{figure}

Vibrations at the horn conductors during pulsed current operations were measured during current testing before
installation. The vibration measurements were performed with laser displacement meters made by 
Keyence Cooperation, and the operation currents for the vibration measurements were 320 (horn-1),
250 (horn-2), and 266 kA (horn-3). The vibrations at several inner-conductor positions, 
for example, the upstream endplates, conductor centers, downstream endplates, and so on, were measured, and
the measured displacements are summarized in Table \ref{tab:vib_meas}.
\begin{table}[htb]
\centering
\caption{Summary of vibration measurements. The expected value for horn-2 (horn-3) at 250-kA (266-kA)
operation was obtained by scaling against the current value. No measurement was conducted for the center of horn-1
since the laser can not reach inside the cylinder. Larger measurement errors for horn-2 and horn-3 
are due to small signal sizes (10-20 mV) compared with electrical noise ($\sim$5 mV).}
\small
\begin{tabular}{lrrr}
\hline
 & horn-1 & horn-2 & horn-3 \\ \hline
Laser displacement meter  & LK-G400 & LK-80 & LK-80 \\
Measurement range (mm)  & 400 $\pm$ 100 & 80 $\pm$ 15 & 80 $\pm$ 15 \\
Intrinsic precision ($\mu$m)  & $\pm$2 & $\pm$3 & $\pm$3 \\
Sampling rate (kHz) & 50 & 1 & 1 \\
Operation current (kA)  & 320 & 250 & 266 \\ \hline
Upstream: meas. ($\mu$m) & 35 $\pm$ 2 & 33 $\pm$ 18 & 21 $\pm$ 18 \\
Upstream: exp. ($\mu$m)  & 45 & 31 & 69 \\
Middle: meas. ($\mu$m)            & -     & 54$ \pm$ 18 & 28 $\pm$ 18 \\
Middle: exp. ($\mu$m)              & -    &  60 & 58 \\
Downstream: meas. ($\mu$m) & 12 $\pm$ 2 & 440 $\pm$ 18 & 180 $\pm$ 18 \\
Downstream: exp. ($\mu$m) & 50 & 427 & 166 \\ \hline
\end{tabular}
\label{tab:vib_meas}
\end{table}
Figure \ref{fig:displacement} shows the time variations of the measured displacements at the downstream endplates of
the inner conductors.
Although the measurement errors are large for horn-2 and horn-3, mainly due to small signal sizes,
the measured displacements are consistent with the estimations from the FEM simulations.
Therefore, no large discrepancy between the designed and actually produced horn conductors is expected
as regards mechanical properties.
At the downstream endplates of the inner conductors, especially for horn-2, a very large displacement
was observed, as expected from the simulation. 
Ribs will be added to the upstream and downstream endplates in the next version of horn-2 in order to reduce the large displacement.

\subsection{Manufacturing of conductors}

Since the inner conductors are very thin in spite of their large dimensions, the manufacturing process of these 
components is very important, as the designed conductor shapes must be achieved. 
A brief description of the conductor manufacturing process is given in this subsection.
 
Each of the horn-1 inner and outer conductors was machined from a forged A6061-T6 block.
The horn-2 inner conductor was made from three components:
a central neck section and upstream and downstream taper and endplate sections.
Each section was machined from a forged A6061-T6 block and welded by
friction stir welding (FSW).
The horn-3 inner conductor was composed of five pieces: a central neck section,
upstream and downstream taper sections, and upstream and downstream
endplate sections. The taper sections were made from rolled A6061-T6 plate,
while each of the other sections was machined from a forged A6061-T6 block.
All the sections were welded with TIG welding. For both horn-2 and horn-3 inner conductors,
the welding positions were chosen so that the mechanical stresses at the positions were small.
The horn-2 and horn-3 outer conductors were made from rolled A6061-T6 plates and assembled
with TIG welding. All the manufacturing of the conductors was successful and the full 
magnetic horn assemblies were completed in 2008.


%
%
\section{Remote maintenance issues}
\label{sec:remote}

Details of the remote connection system used for the T2K magnetic horns are given in this section.

\begin{figure}
\centering
\includegraphics[clip,width=0.45\textwidth,bb=0 0 715 478]{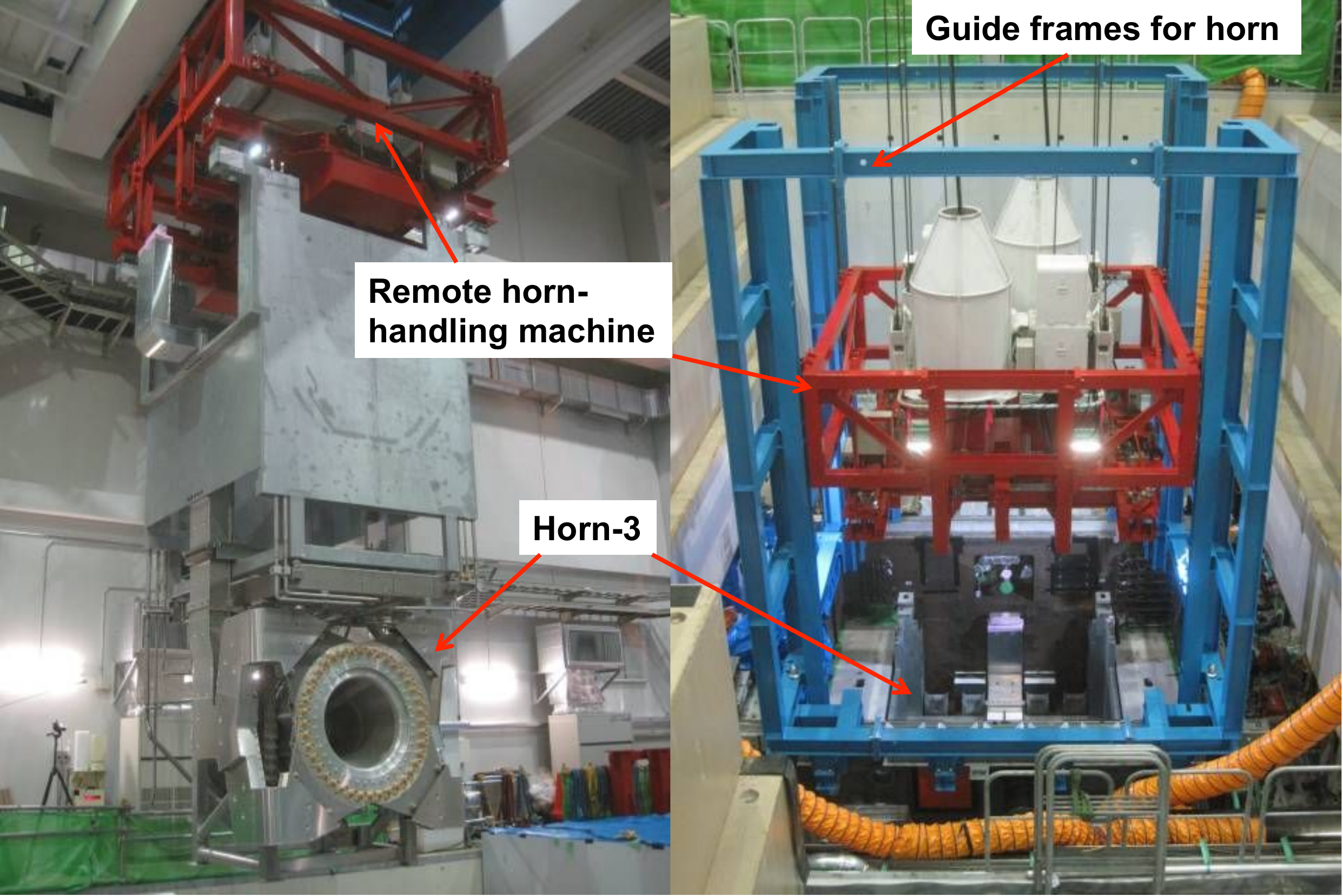}
\caption{Picture of horn-3, which was moved to the maintenance area using a remote horn-handling machine (left).
The picture shows the remote horn-handling machine and its guide frames attached on the top of the HV.}
\label{fig:horn3remote}
\end{figure}

Irradiated magnetic horns or other instruments inside the HV can be removed
from the vessel using a numerically-controlled overhead crane in the Target Station (TS), as shown in Fig. \ref{fig:horn3remote}. 
In the TS building, there is an area where radiated instruments can be handled (maintenance area).
The maintenance area has two sections; one is a human-access space, with an area of 4$\times$1.5 m$^2$, 
and another is a 4$\times$4-m$^2$ space, where radioactive equipment are set.
The two sections are separated by a 1-m-thick concrete wall, while
a lead glass window located in the wall gives clear views of the radioactive equipment from the human-access area.
Two manipulators are fabricated in the maintenance area and used for the handling of radioactive materials, other than
the horns\footnote{Remote handling of the horns is not performed using the manipulators but by manual operation from the top
of the maintenance area.}.

If one of the magnetic horns is found to be defective, it is moved to the maintenance area.
Then, the horn itself is disconnected from its support module and stored 
in a storage area located inside the TS. Thereafter, a new horn is set up in the maintenance area
and connected to the support module. The support modules can be reused even if the horns are broken.
After connection, the new horn is moved to the HV.

\begin{figure}
\centering
\includegraphics[clip,width=0.5\textwidth,bb=0 0 600 540]{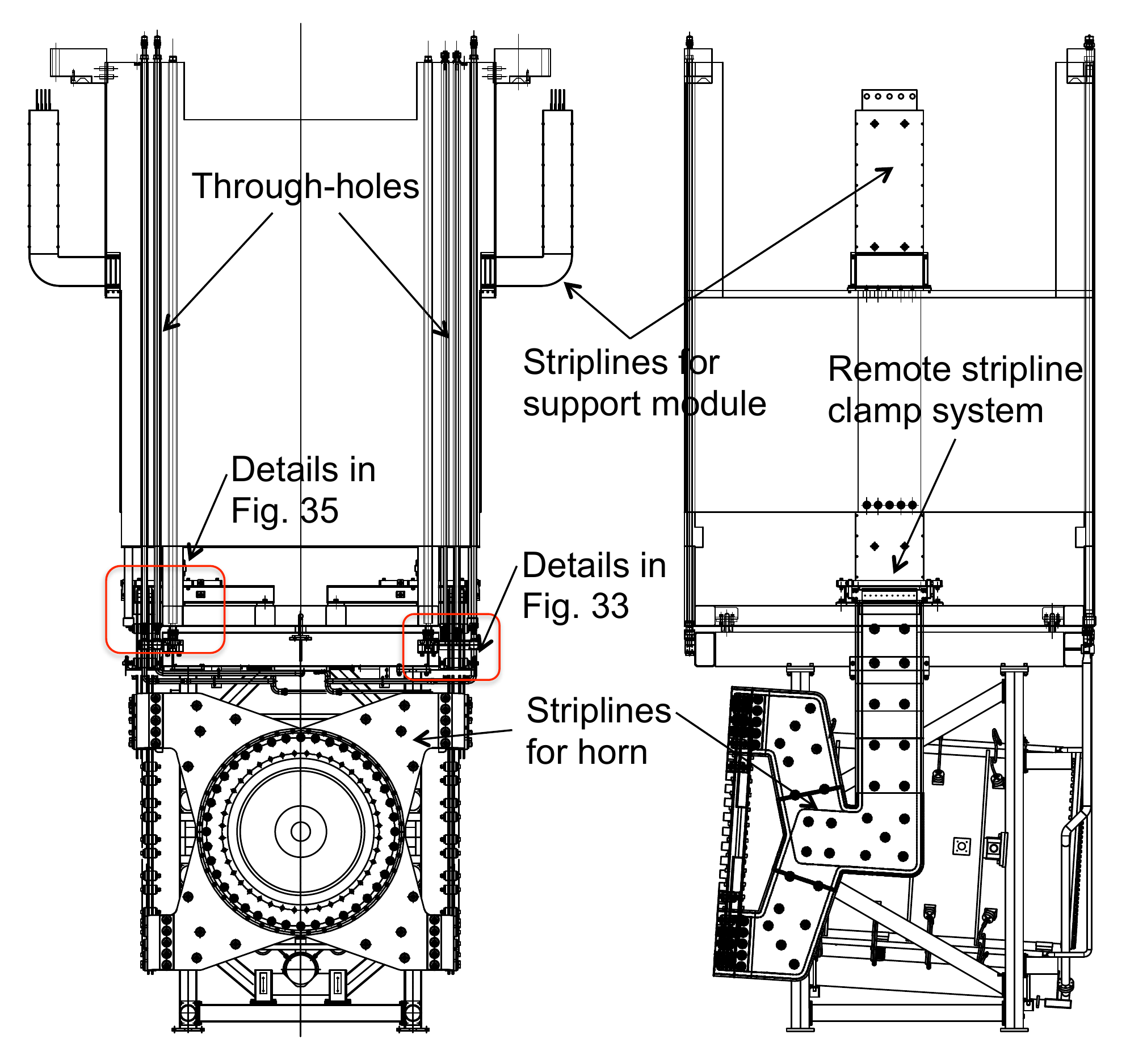}
\caption{Horn-3 attached to its support modules: front view (left) and side view (right).}
\label{fig:horn3module_fig}
\end{figure}

\begin{figure}[H]
\centering
\includegraphics[clip,width=0.5\textwidth,bb=0 50 725 450]{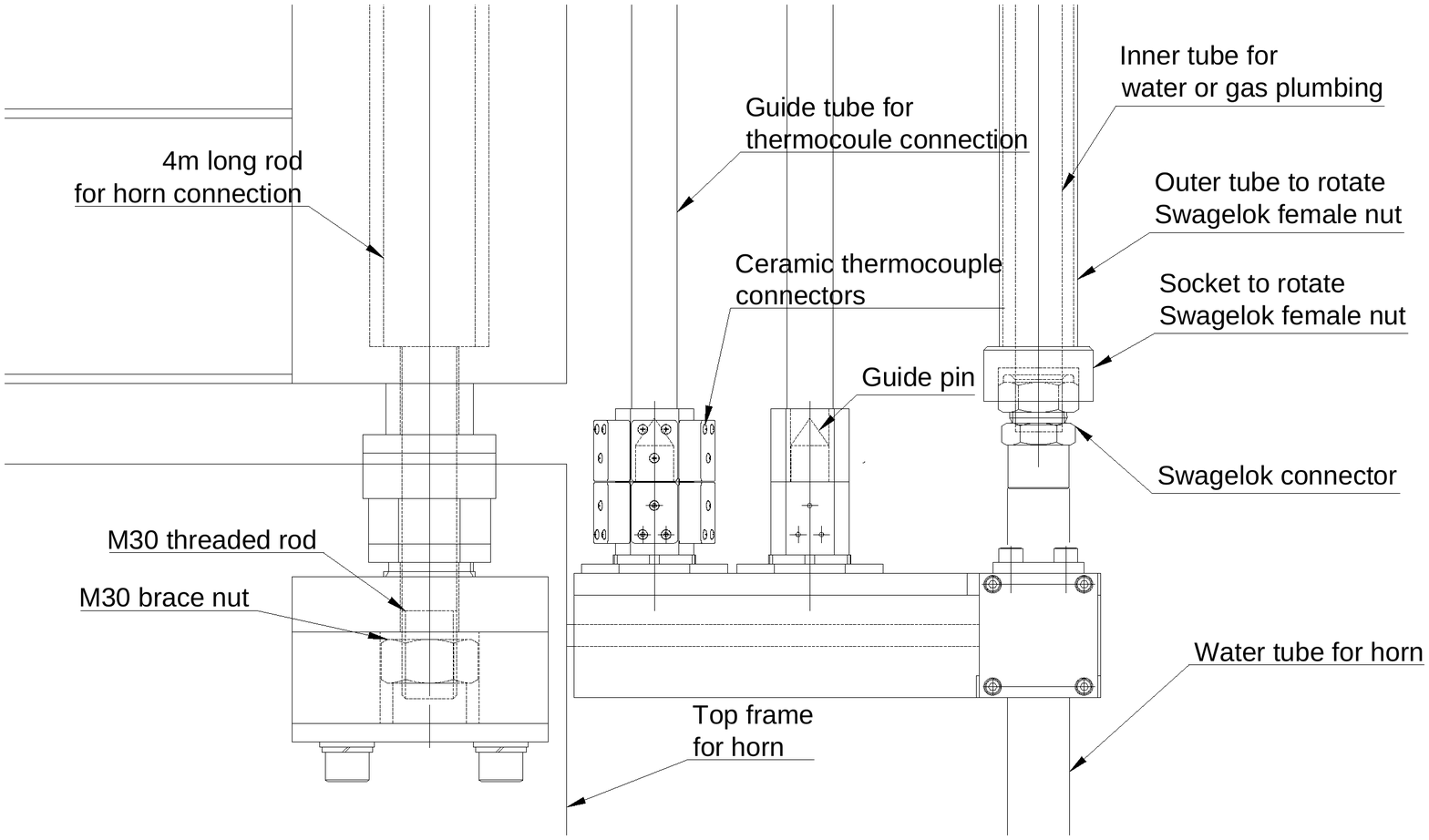}
\caption{Detailed schematic of remote connections.}
\label{fig:remote_coupling_tubes}
\end{figure}

\begin{figure}[tbh]
\centering
\includegraphics[clip,width=0.3\textwidth,bb=0 0 358 268]{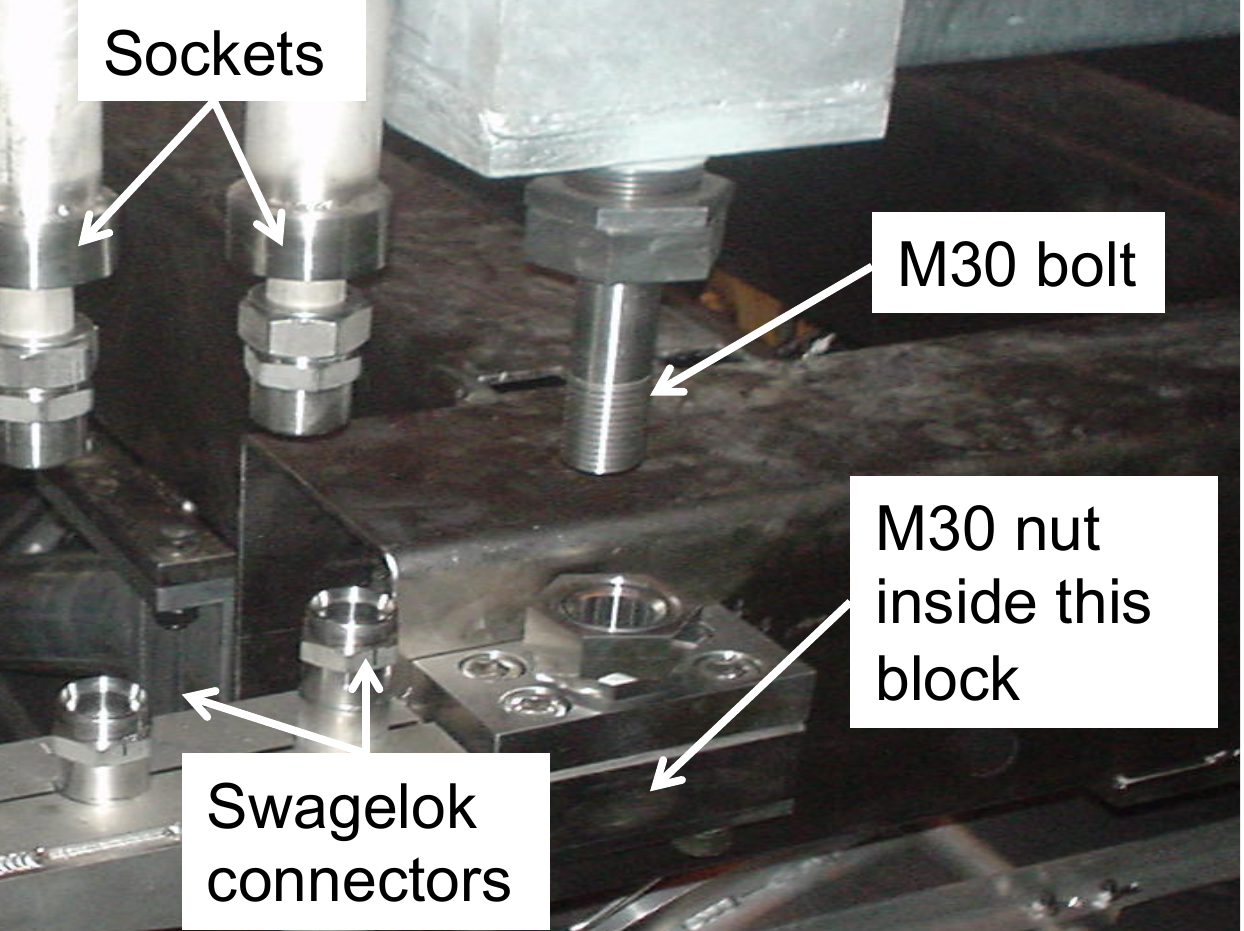}
\includegraphics[clip,width=0.3\textwidth,bb=0 0 356 269]{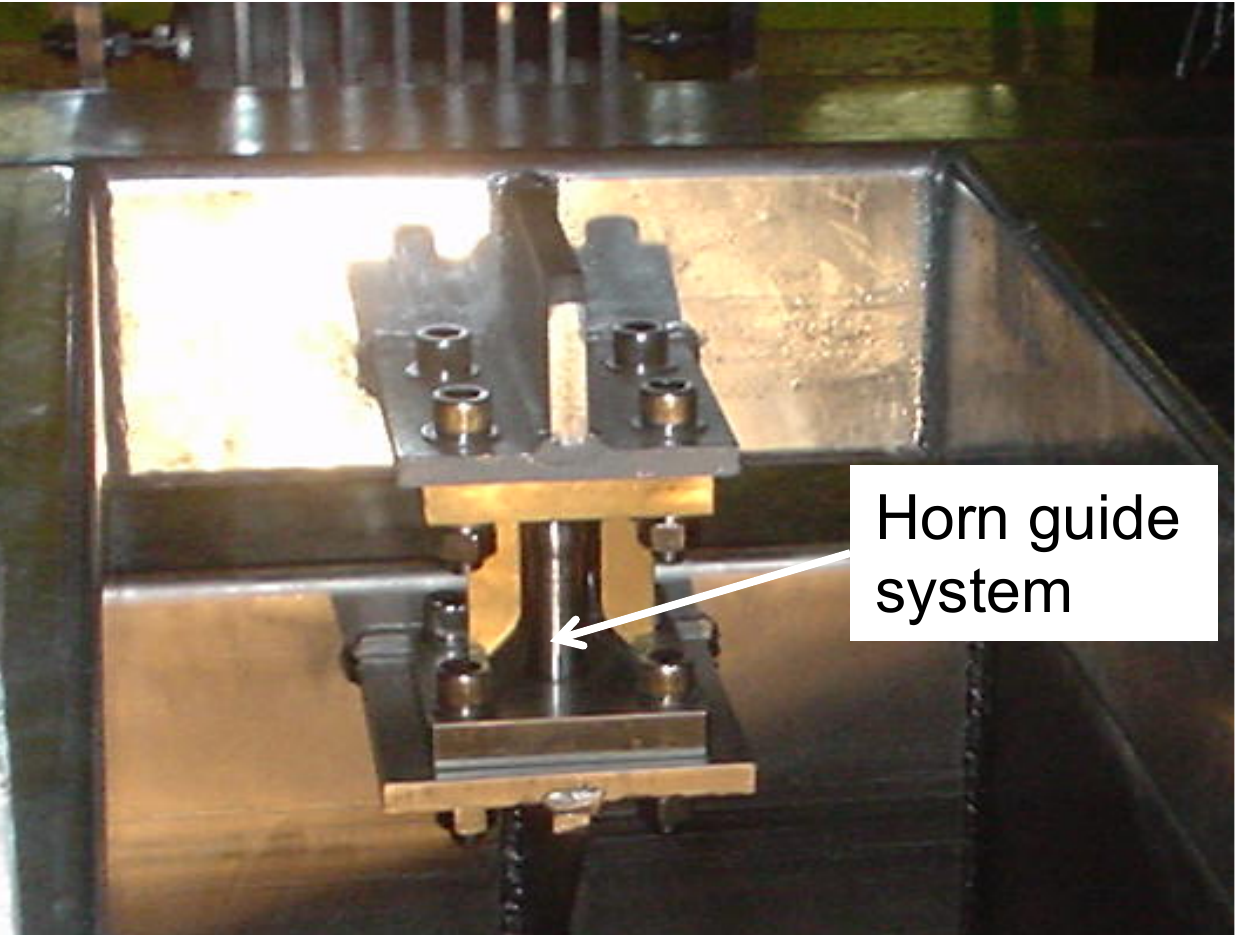}
\includegraphics[clip,width=0.3\textwidth,bb=0 0 358 269]{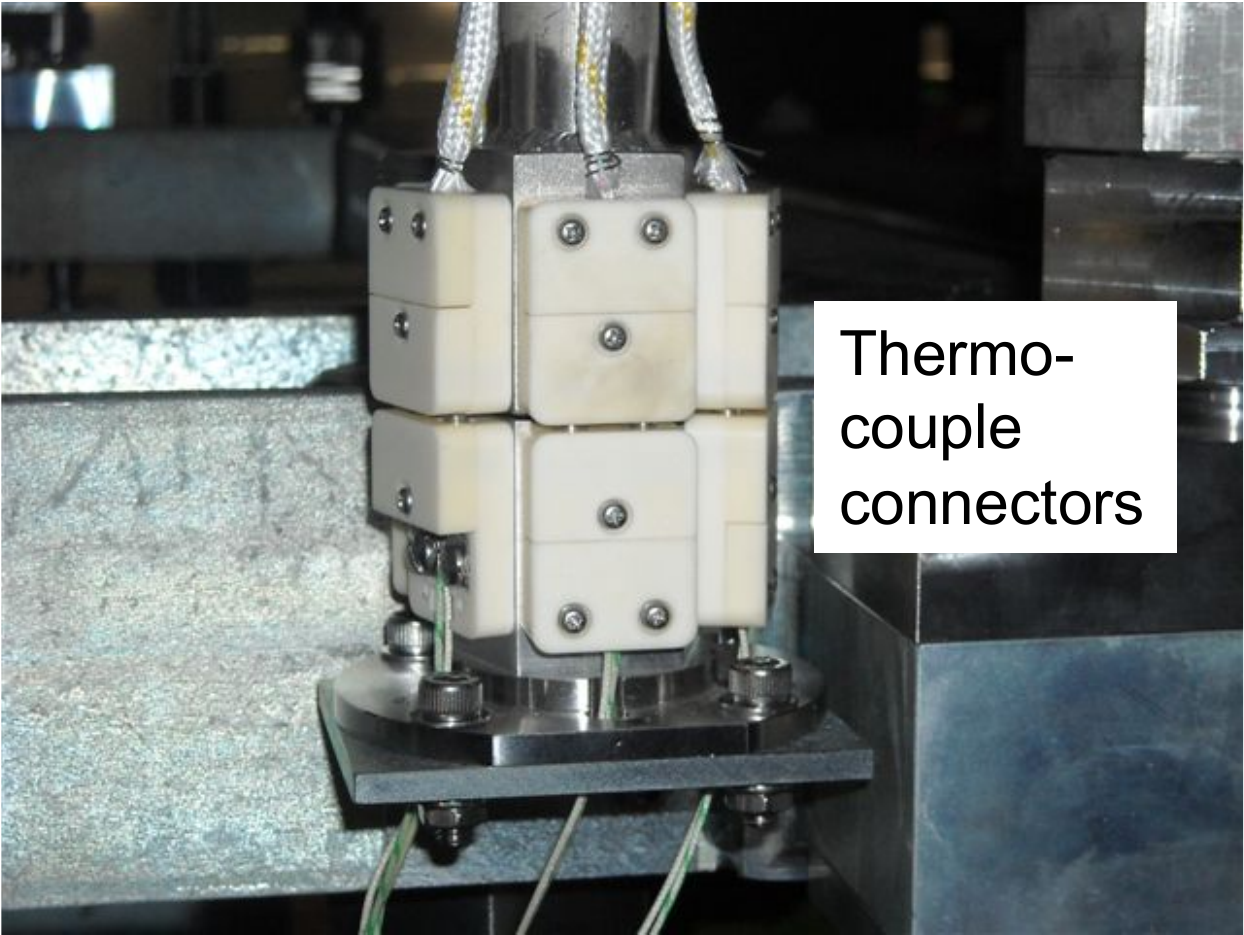}
\caption{Pictures of remote connections for horn and water pipes (top left), guide system (top right), and thermocouples (bottom).}
\label{fig:remote_pic}
\end{figure}

The support modules, as shown in Figs. \ref{fig:ModulePic} and \ref{fig:horn3module_fig}, have a box-shaped structure.
Both the upstream and downstream walls are made of 75-mm-thick steel plates and while the side walls are 9-mm-thick steel plates.
The thick steel plates have several through-holes with diameters of $\phi$50-65 mm.
For the connection of the horns and their support modules, two guide pins define a relative position between the horns and 
support modules. A reproducibility of 0.3 mm for the relative position can be achieved by the guide pins.
Once the horns are aligned to the support modules, they are connected with bolted joints.
Four 4-m-long stainless rods with M30 threads at the bottom end penetrate
the through-holes, and the four M30 brace nuts at the top frame of the horns can be tightened by turning the rods.  
Water and gas plumbing components also penetrate the through-holes and are connected to the horn plumbing via Swagelok connectors.
The penetrating plumbing has a double-tube structure, the inner tubes of which function as water and gas plumbing and
the outer tubes of which, with a hex-socket shape at the bottom, are used to turn the female nuts of the Swagelok connectors. 
Because of this double tube
structure, the Swagelok connectors at the bottom of the support modules can be loosened and tightened from the top of the
support modules. Thermocouple wires are drawn with two guide tubes penetrating the through-holes and
connected to the horn thermocouples via ceramic thermocouple connectors. Alignment of the thermocouple connectors
is achieved using guide pins and guide holes.
Figures \ref{fig:remote_coupling_tubes} and \ref{fig:remote_pic} show these connections in detail.

\begin{figure}
\centering
\includegraphics[clip,width=0.4\textwidth,bb=0 10 685 530]{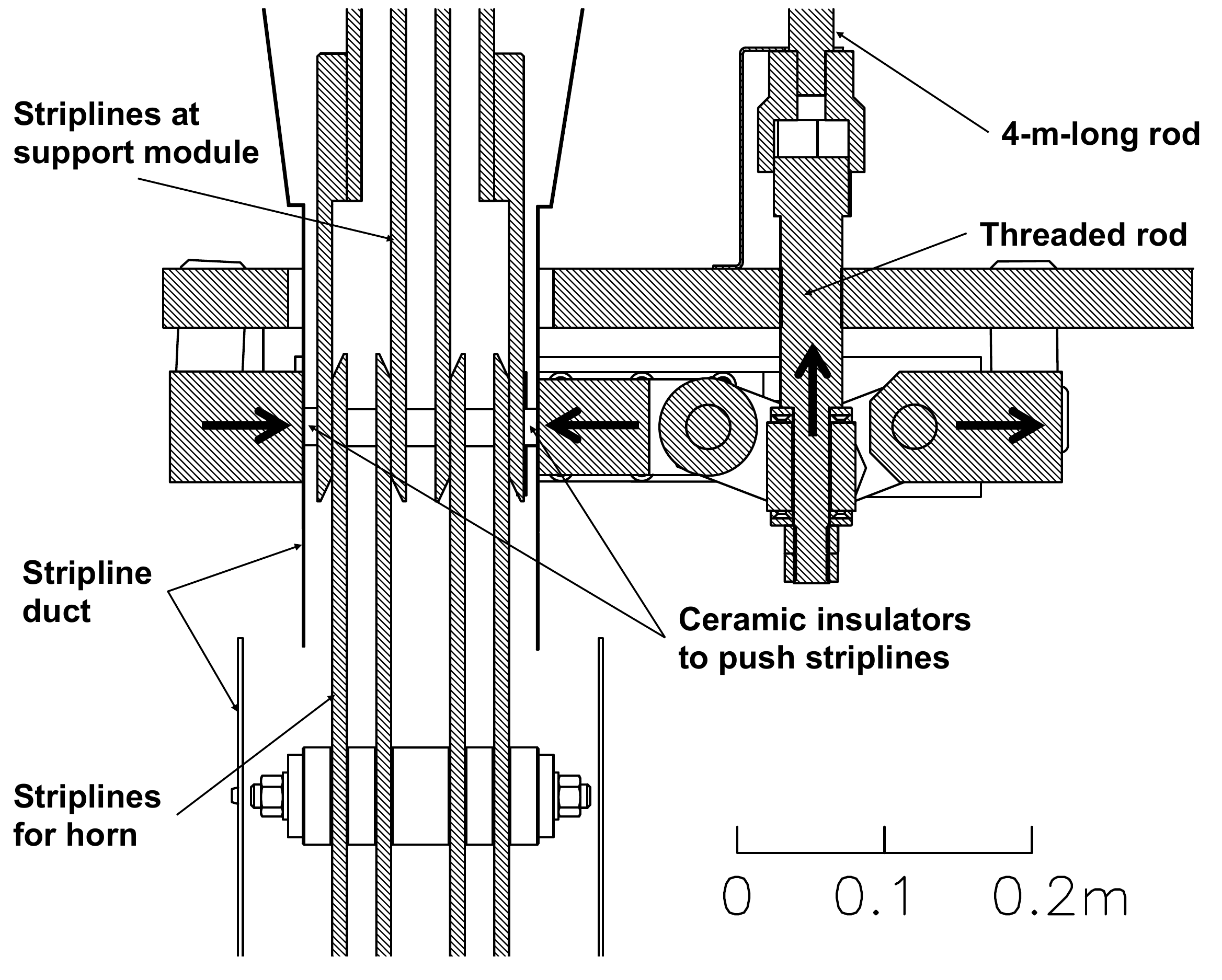}
\caption{Detailed drawing of remote stripline clamp system.}
\label{fig:stripline_remote_clamp}
\end{figure}

The striplines are also detachable at the bottom of the support modules. Figure \ref{fig:stripline_remote_clamp} shows
the remote stripline clamp system located at the bottom frames of the support 
modules\footnote{The remote stripline clamp system of the NuMI magnetic horns at FNAL \cite{NuMI} is used as a reference 
for this system by courtesy of NuMI beamline group.}.
The mechanism of this system is similar to a car jack, in that the rotation of a threaded rod changes torque to a horizontal force.
The remote clamp can load 15 tons of clamping force, equivalent to 5 MPa pressure, onto the stripline joints.
By rotating the threaded rod with 4-m-long rod, with a box socket at the end on top of the support module,
the stripline clamping force is released, which allows remote disconnection of the striplines.

During connection and disconnection work at the maintenance area, a concrete shielding block is inserted inside the support module
to provide radiation protection. Workers can stand on the concrete block, where the radiation level is a few tens of $\mu$Sv/h or less.


%
%
\section{Operation experiences}
\label{sec:operation}

Operational experience involving use of the T2K magnetic horn system is described in this section.

\subsection{Development period}

Development of the T2K magnetic horn system began in 2003, when design optimization of the horn conductor shapes
and several component tests were performed \cite{T2KhornDesign}.

The first production of the prototype magnetic horn (horn-1) was completed in March 2006 and
the first pulsing of the prototype horn-1 at 320-kA was achieved in June 2006.
Various performance tests such as a field measurement, cooling test, and vibration test
were performed on the prototype. A long-term operation test (three months) was
conducted in 2007 and 0.85 million pulses were successfully applied to the prototype horn-1.
During the long-term operation test, loosening of small bolts at the water nozzle flanges occurred
at approximately 0.5 million pulses. After measures were taken to prevent this, no further bolts were loosened. 
Cracks in the ceramic insulator disks used for stripline fixation were also found during the long-term test.
Alumina ceramic disks with 96\% purity were used for the test stand and were replaced with alumina ceramic disks
with higher purity (over 99.5\%), which have slightly higher flexural strength (380 MPa for $>$99.5\% and 350 MPa for 96\%
\cite{Ceramics}). 
The other modification was to improve the conditions of contact between the ceramic disks and fixation parts. 
Both of the modifications were effective in preventing cracking of the ceramic disks and, thereafter, no cracks were found.
Prototype horn-3 was produced in March 2007. Then, a pulsing test was established with a series configuration of horn-1 and horn-3
in order to evaluate the series operation of dual magnetic horns. Performance tests for horn-3 were also successfully
completed. A long-term test with dual horn operation was also performed and 0.45 million pulses were successfully
applied to both prototype horns. No bolt loosening occurred during the long-term test.

In the final stage of the development period, from October 2007 to May 2008, a mockup test was conducted for a full 
configuration of prototype horn-3 suspended by its support module.
A 10-meter-high large test stand was constructed for this purpose. Using the mockup,
a water-circulation test to confirm 8-m-high pumping of the drain water, validation of the remote connection system, development
of the magnetic horn handling system, and current testing with a full configuration, were performed.
During the current testing, vibrations were measured at various positions on the horn. Even when supported from an 8-m height,
similar to a pendulum, the measured oscillation of the entire horn was, at most, 5 $\mu$m, which is acceptable. 

Performance tests for horn-2 were performed with a production version in 2008 and adequate performance was
confirmed. 

\subsection{Installation}

\begin{figure}
\begin{tabular}{ll}
\begin{minipage}{0.23\textwidth}
\includegraphics[clip,width=\textwidth,bb=120 20 750 550]{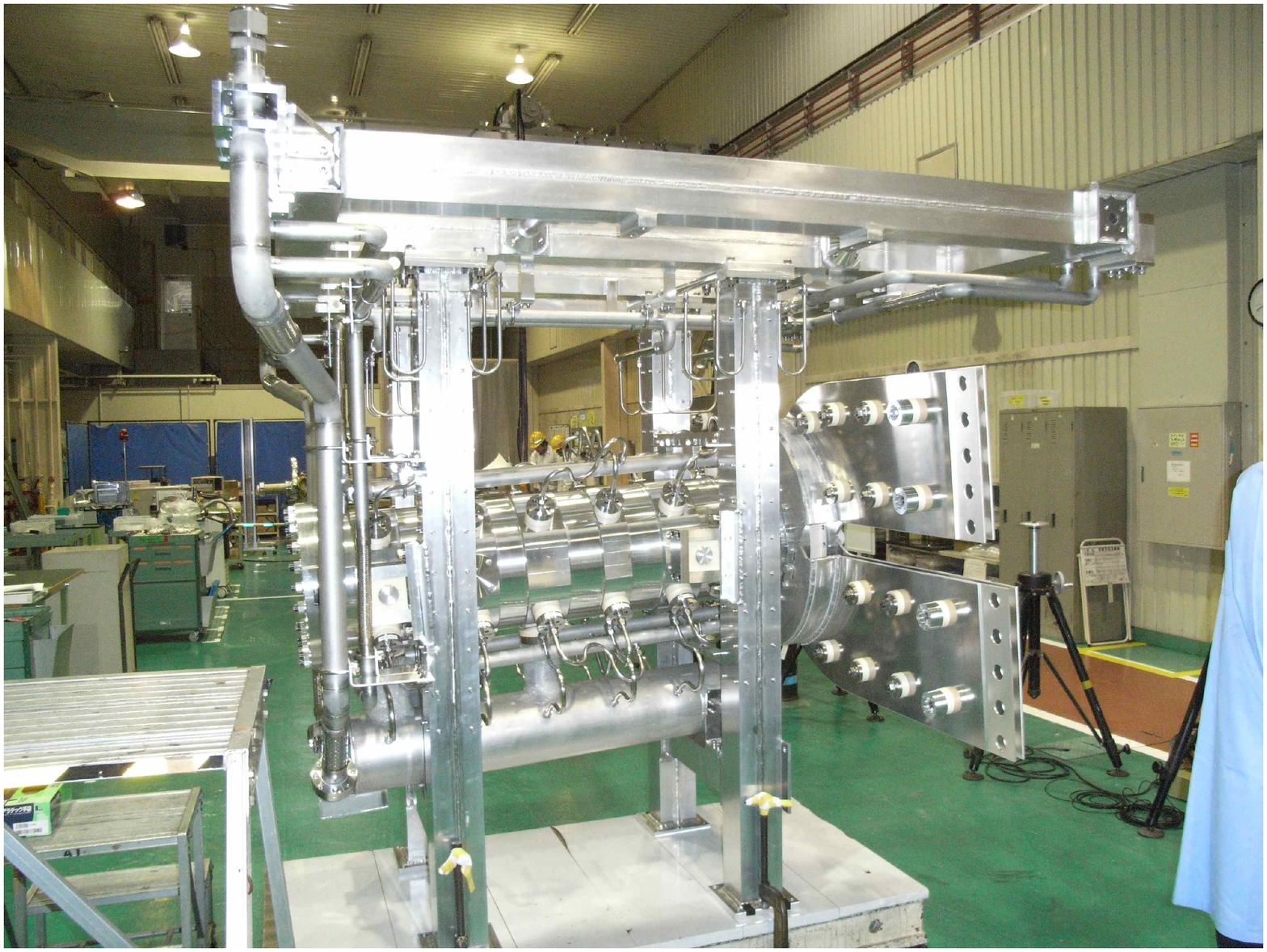}
\end{minipage}
\begin{minipage}{0.23\textwidth}
\includegraphics[clip,width=\textwidth,bb=120 20 750 550]{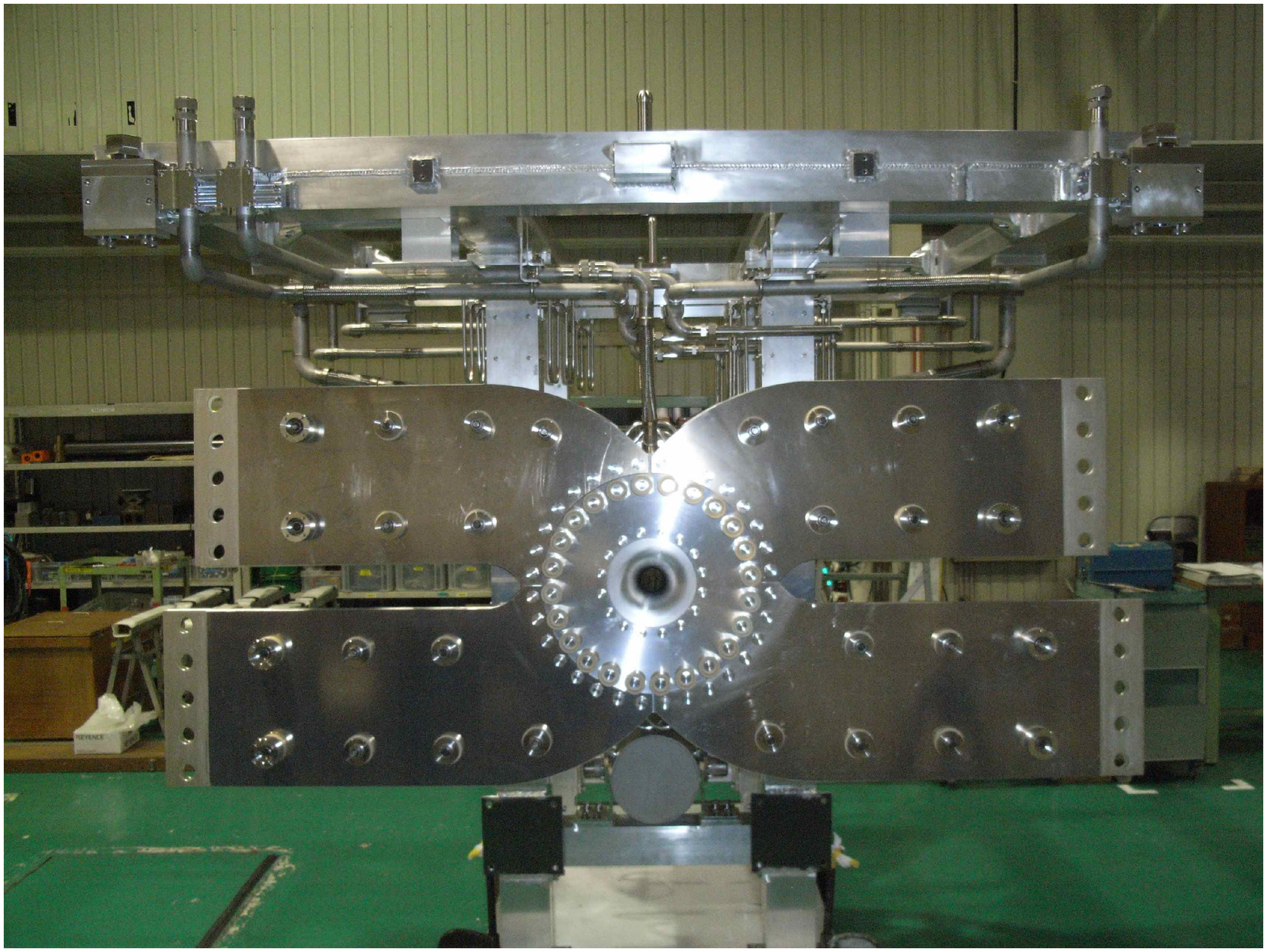}
\end{minipage} \\
\begin{minipage}{0.23\textwidth}
\includegraphics[clip,width=\textwidth,bb=100 70 700 550]{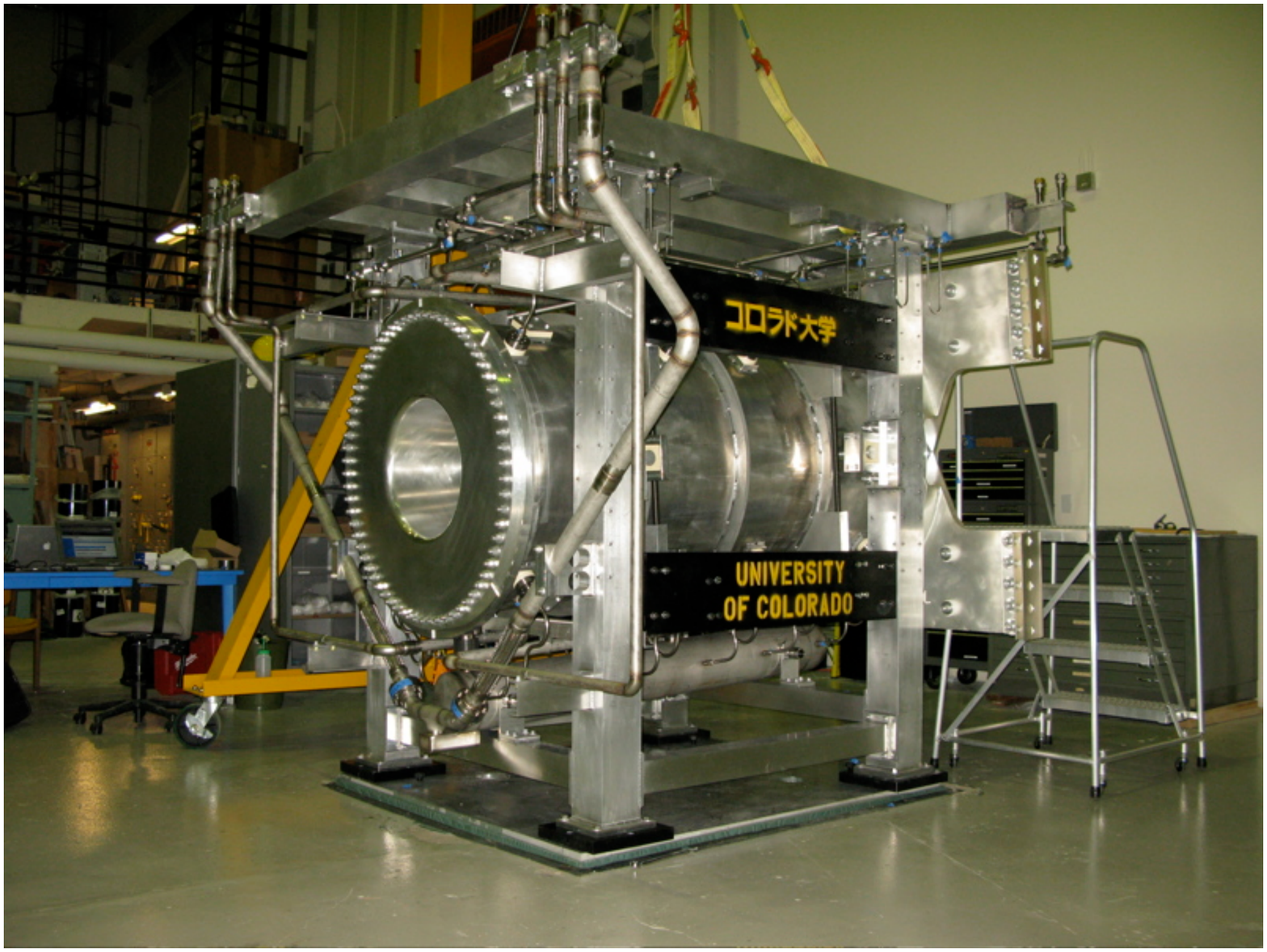}
\end{minipage}
\begin{minipage}{0.23\textwidth}
\includegraphics[clip,width=\textwidth,bb=40 10 760 590]{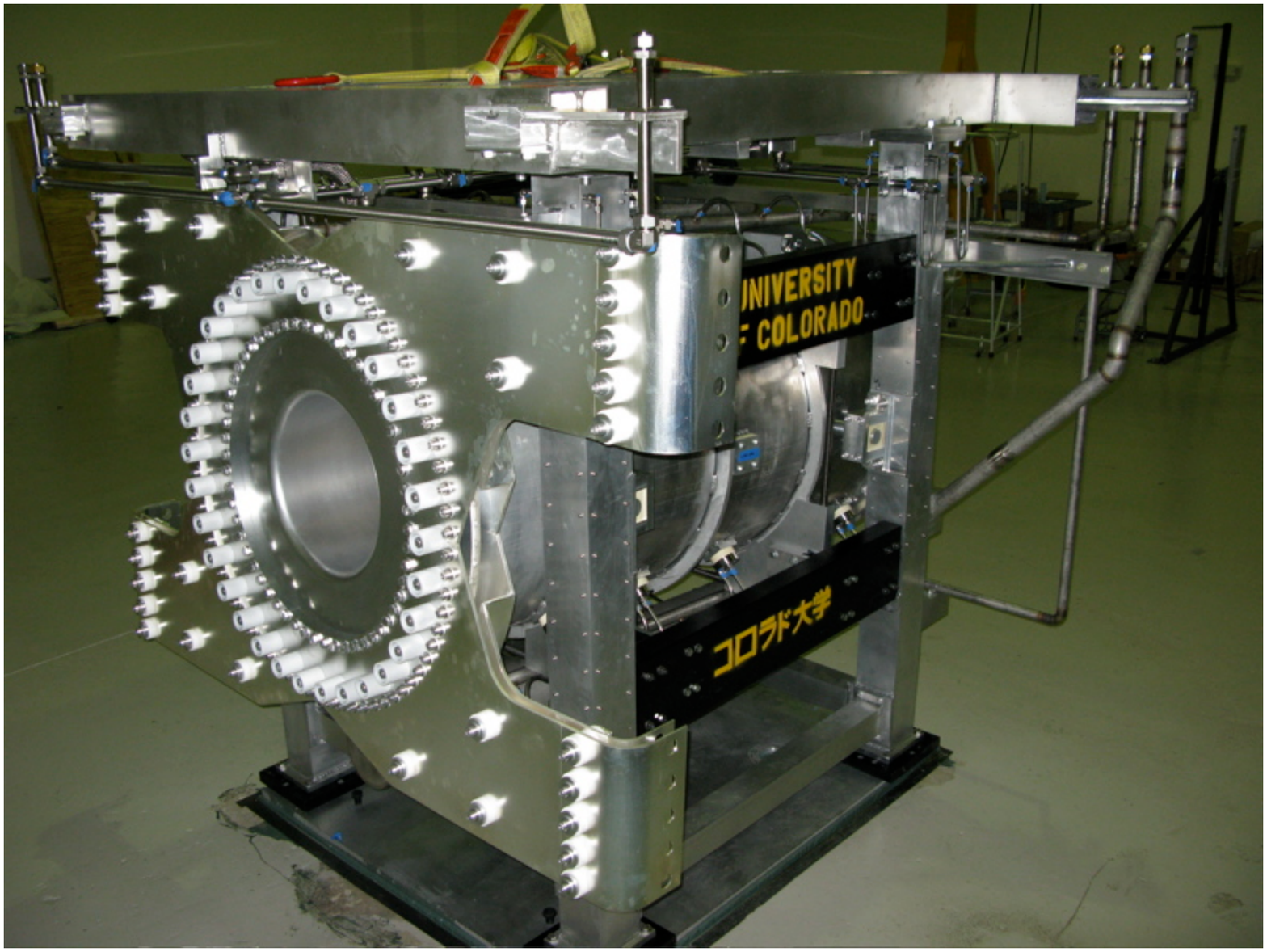} 
\end{minipage} \\
\begin{minipage}{0.23\textwidth}
\includegraphics[clip,width=\textwidth,bb=0 30 265 310]{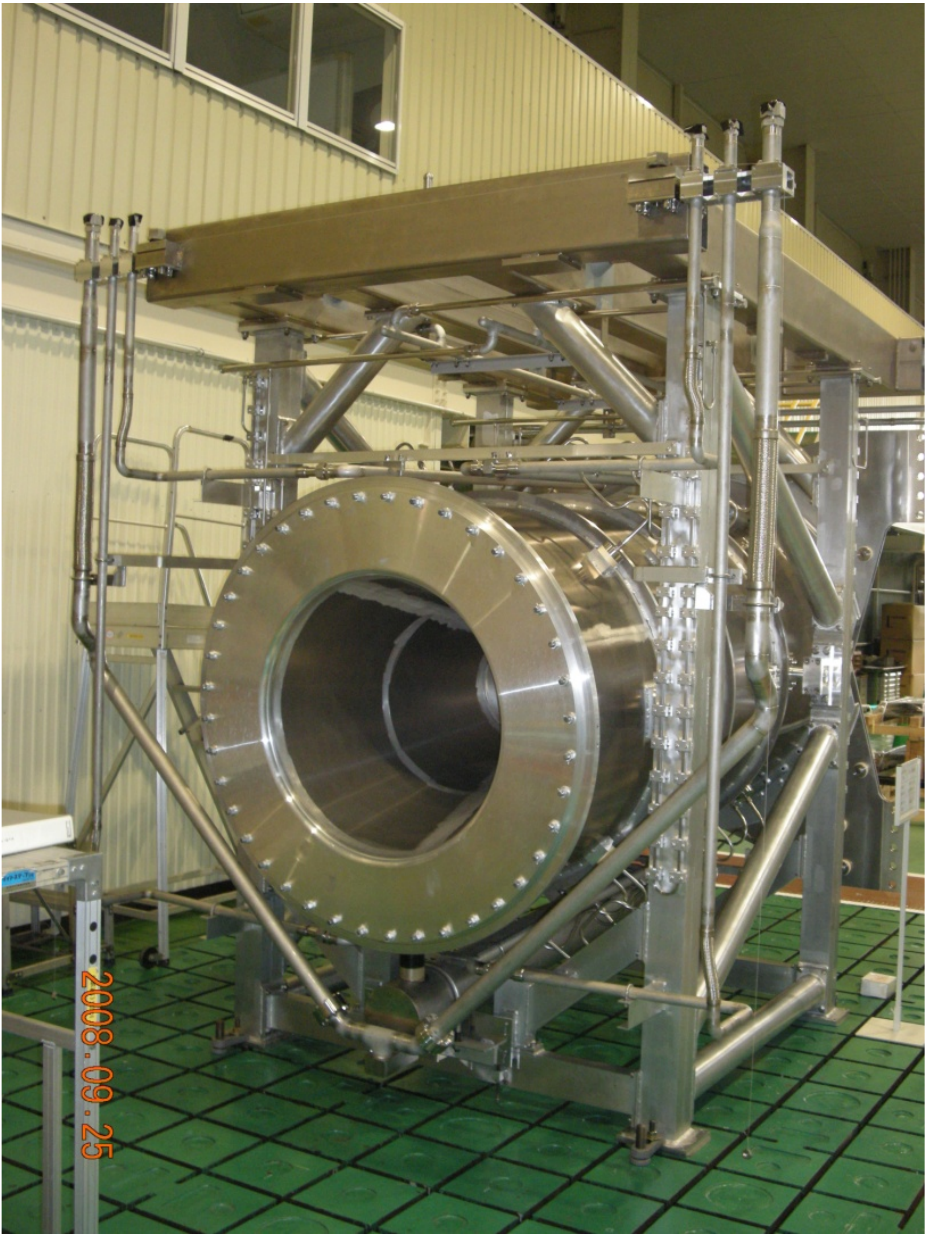}
\end{minipage}
\begin{minipage}{0.23\textwidth}
\includegraphics[clip,width=\textwidth,bb=0 30 265 310]{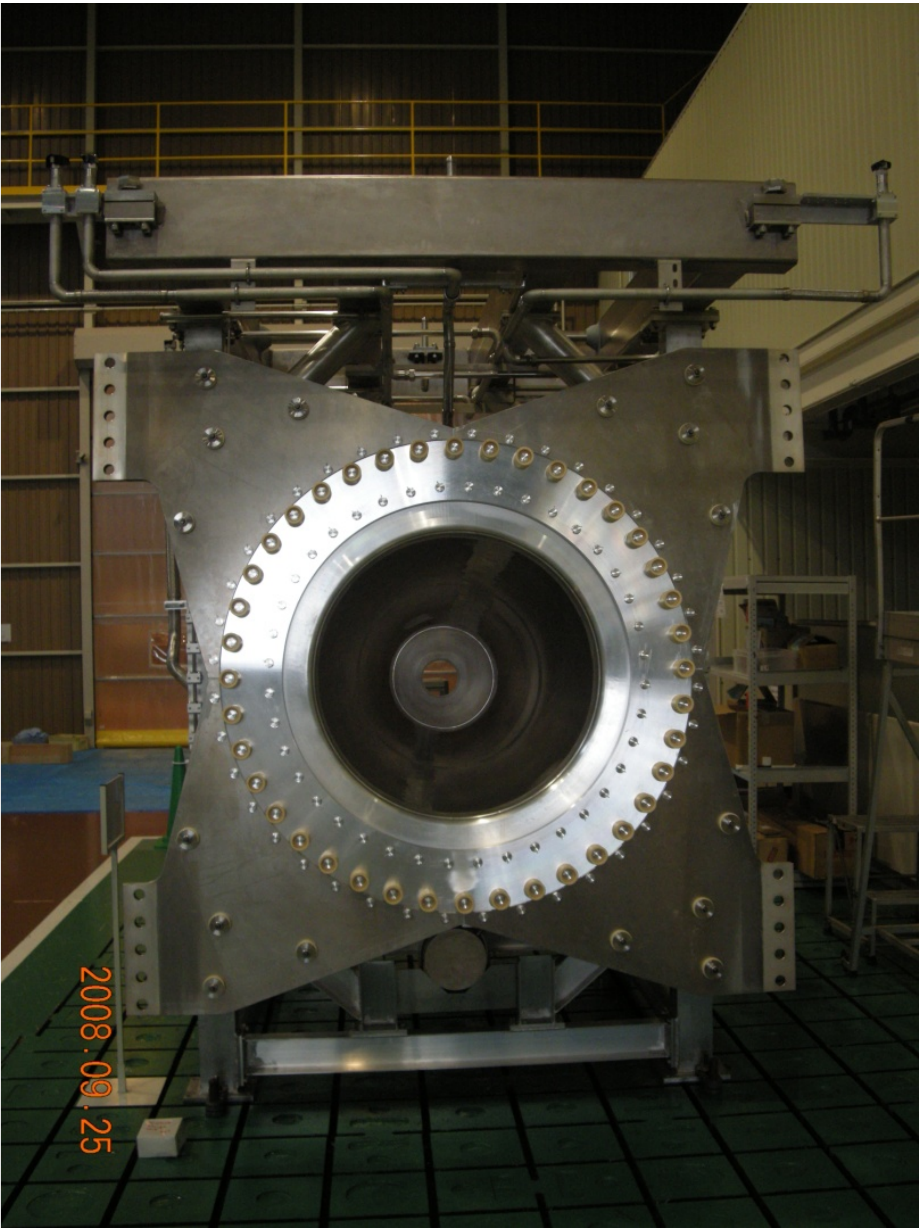}
\end{minipage}
\end{tabular}
\caption{Pictures of T2K magnetic horns: horn-1 side view (top left), horn-1 upstream end view (top right), horn-2 side view (middle left),
horn-2 upstream end view (middle right), horn-3 downstream end view (bottom left), and horn-3 upstream end view (bottom right).}
\label{fig:fig_horns}
\end{figure}

The production version of the T2K magnetic horns was completed in
2008\footnote{With collaboration between KEK and T2K colleagues in the United States 
(the main efforts were made by a group at The University of Colorado at Boulder)}. 
Horn-1 and horn-3 were produced in Japan, while horn-2 was manufactured in the U.S.
and then shipped to Japan. Pictures of the production version of the T2K magnetic horns can be seen in Fig. \ref{fig:fig_horns}.

The installation work began in August 2008. All the infrastructure used for the current testing in KEK was
transported to J-PARC and set up in the TS building. The horns and support modules
were assembled and aligned at the test stand placed on the ground floor of the TS.
Then, current testing was performed with the horns connected to the support modules at the test stand.
Horn-1 was installed in the HV in January 2009. After water connection, stripline connection, and
cabling work between the power supply and transformer, the first in-situ operation of horn-1 was successfully conducted in April 2009.
The first beam commissioning with horn-1 only was achieved on April 23rd, 2009 and the first neutrino beam
was produced at the neutrino beamline of J-PARC.
After beam commissioning, installation of horn-2 and horn-3 was accomplished by August 2009.
The striplines, with the series configuration used for the prototype testings, were transported to the TS
and used as striplines for series operation of horn-2 and horn-3.
After setup, in-situ current testing and also field measurements were performed on both horns.
The full installation of the T2K magnetic horn system was completed in October 2009.
Then, evacuation of the HV was performed and the device was filled with helium gas.
The T2K magnetic horns were operated in a helium atmosphere for the first time in November 2009.
However, a ground fault problem occurred during the current operation. After several investigations conducted over two weeks,
it was found that the voltage-to-ground at the secondary circuit of the transformers reached 3 kV and, therefore,
a breakdown occurred. By adding a neutral grounding scheme to the secondary circuit, the voltages were reduced
dramatically to several hundred volts, as described in Section \ref{sec:electrical}. 
Thereafter, no further ground fault problems occurred.

\subsection{Beam-operation experience}
\subsubsection{Beam-operation conditions}
\begin{figure}
\centering
\includegraphics[clip,width=0.45\textwidth,bb=30 20 700 322]{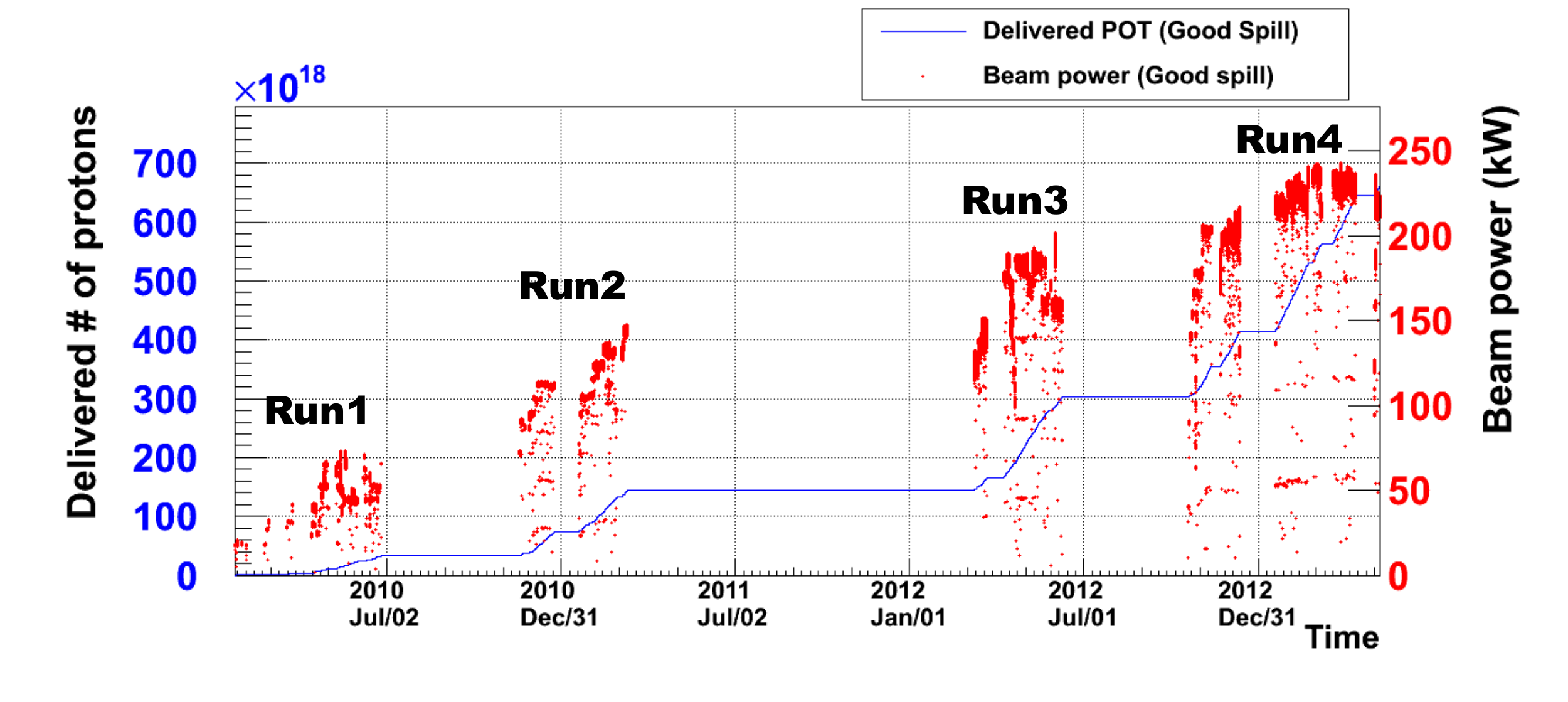}
\caption{History of integrated number of delivered POT (solid line) and beam power (dot).}
\label{fig:POT}
\end{figure}

The T2K experiment began its physics data-taking in January 2010.
For the physics data-taking operation, the magnetic horns were operated at 
250 kA\footnote{The expected neutrino flux at the Super-Kamiokande
detector at 250-kA decreases by 10\% compared to that at 320-kA.}.
The beam power at the beginning of the T2K experiment was 20 kW, which was gradually increased up to 240 kW as of May 2013.
The integrated number of delivered protons on target (POT) during the physics data-taking run is shown in Fig. \ref{fig:POT}.
The beam conditions for each run period are summarized in Table \ref{tab:run_condition}.
\begin{table}[htb]
\centering
\caption{Summary of beam conditions for each T2K run period.}
\small
\begin{tabular}{lrrrr}
\hline
Parameters & Run1 & Run2 & Run3 & Run4 \\ \hline
Run period (MM/YY) & 01/10- & 11/10- & 03/12- & 10/12- \\
& 06/10 & 03/11 & 06/12 & 05/13 \\
Beam power (kW) & 60 & 135 & 180 & 230 \\
\# of protons/pulse ($\times 10^{14}$) & 0.35 & 0.90 & 0.95 & 1.2 \\
Repetition cycle (s) & 3.52 & 3.20 & 2.56 & 2.48 \\
\# of bunches in pulse & 6 & 8 & 8 & 8 \\
P.O.T/run ($\times 10^{20}$) & 0.32 & 1.11 & 1.58 & 3.56 \\ 
\# of horn pulses/run ($\times 10^6$) & 2.2 & 2.5 & 3.0& 4.3 \\ \hline
\end{tabular}
\label{tab:run_condition}
\end{table}
The accelerator operation cycle was gradually shortened from 3.52 (Run1) to 2.48 s (Run4).

\subsubsection{Stability of horn operation current}
\begin{figure}
\centering
\includegraphics[clip,width=0.48\textwidth,bb=0 0 710 204]{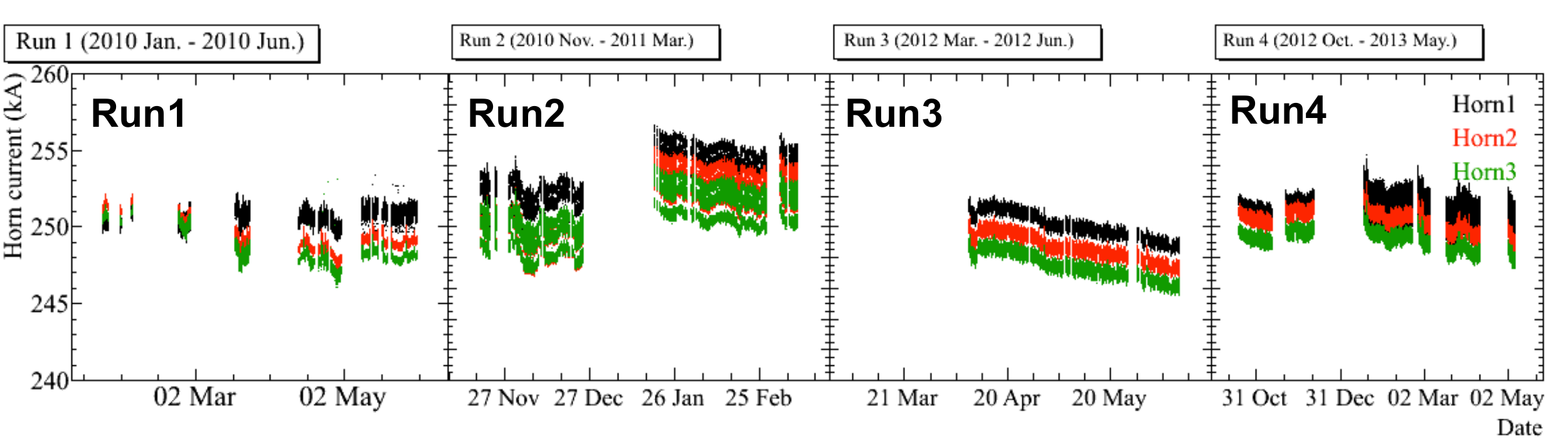}
\caption{Current stability for each horn during T2K run periods. In the early stage of Run3, the horns were
operated at 200 kA because of a power supply problem.}
\label{fig:cur_history}
\end{figure}

The stability of the operation current for each horn during each run period is shown in Fig. \ref{fig:cur_history}.
Although the measured current values for each horn varied by 2 \% during each run period, mainly because of
variation in the power cable resistance due to the external temperature, very stable operation of the T2K magnetic horns
was achieved and the observed operation current satisfied the requirement for physics analysis at 250 $\pm$ 5 kA.
By the end of Run4, the first set of T2K magnetic horns was operated over 1.2 $\times~10^7$ pulses. No significant damage
was observed at the magnetic horn conductors.

\subsubsection{Activation of cooling water}
The beam powers as of Run4 were still much lower than the design beam power of 750 kW, while the heat depositions at
the magnetic horns were significantly lower. However, the number of delivered protons per pulse was significantly higher than
that of the other proton beam facilities in the world\footnote{1.2$\times 10^{14}$ protons/pulse is the world record
for the number of extracted protons per pulse for proton synchrotrons.} and,
thus, many problems related to the high-intensity proton beams were experienced during the T2K run periods.
As the beam power gradually increased, many problems related to the water circulation system were observed.
The ion exchangers were not operated continuously during beam operation until the middle of Run4 and, thus,
radioactive $^7$Be ions were gradually accumulated.
As a result, a residual radiation level around the water tank of $\sim$300 $\mu$Sv/h (maximum) was observed
at 150-kW beam operation in Run3. This radiation level caused many difficulties during maintenance work
and many routine maintenance tasks such as a helium gas flushing, operation of the ion exchangers, and water drainage and
supply were automated and remotely performed by modifying the water circulation system.
As a result of the automation and remote operation, access to the water circulation units in the underground machine room 
(where access is prohibited during beam operation) is no longer necessary. 
This has resulted in less downtime for maintenance and a lower radiation dose
during maintenance work. Since the middle of Run4, ion exchangers have been continuously operated during beam operation, 
as will be described below.
As a result, the radiation level of the circulated water has been significantly reduced to the level of a few $\mu$Sv/h, 
even with 220-kW continuous operation. 
However, the ion exchangers were highly irradiated. Therefore, the ion exchangers were shielded with 10 cm-thick iron plates.
 
\subsubsection{Hydrogen production problem}
As described in Section \ref{sec:cooling}, hydrogen gas is produced inside the magnetic horns by water radiolysis,
with an expected production rate of 40 L/d for 750-kW operation. Hydrogen gas can combust explosively
under a hydrogen density of between 4 \% and 75 \% in an air environment. Since the hydrogen explosion limit in a helium
atmosphere contaminated with oxygen is unknown, the hydrogen density inside the magnetic horns should be reduced to 
less than 2 \% in the interests of safety.

\begin{figure}
\centering
\includegraphics[clip,width=0.5\textwidth,bb=80 30 770 580]{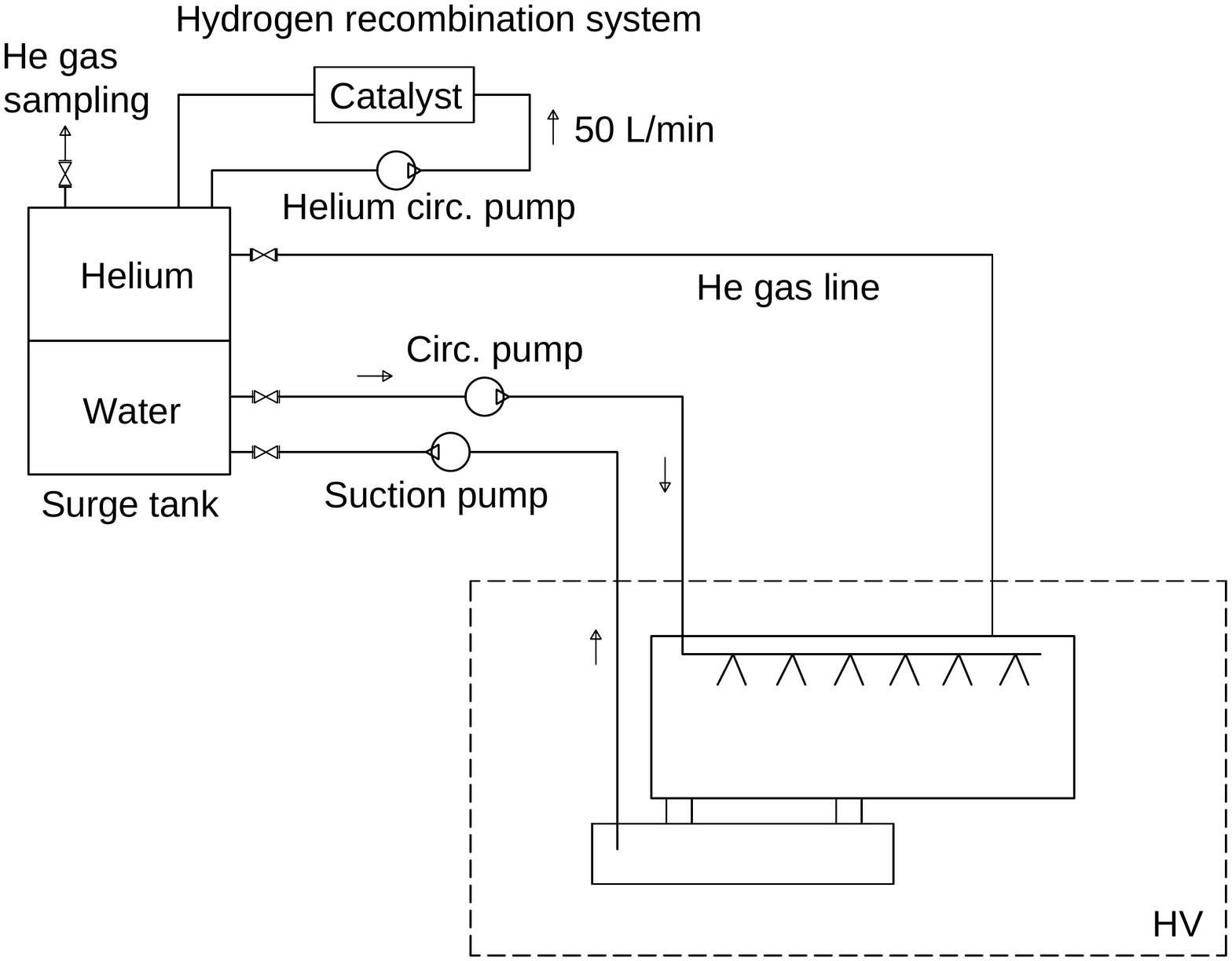}
\includegraphics[clip,width=0.5\textwidth,bb=70 30 770 600]{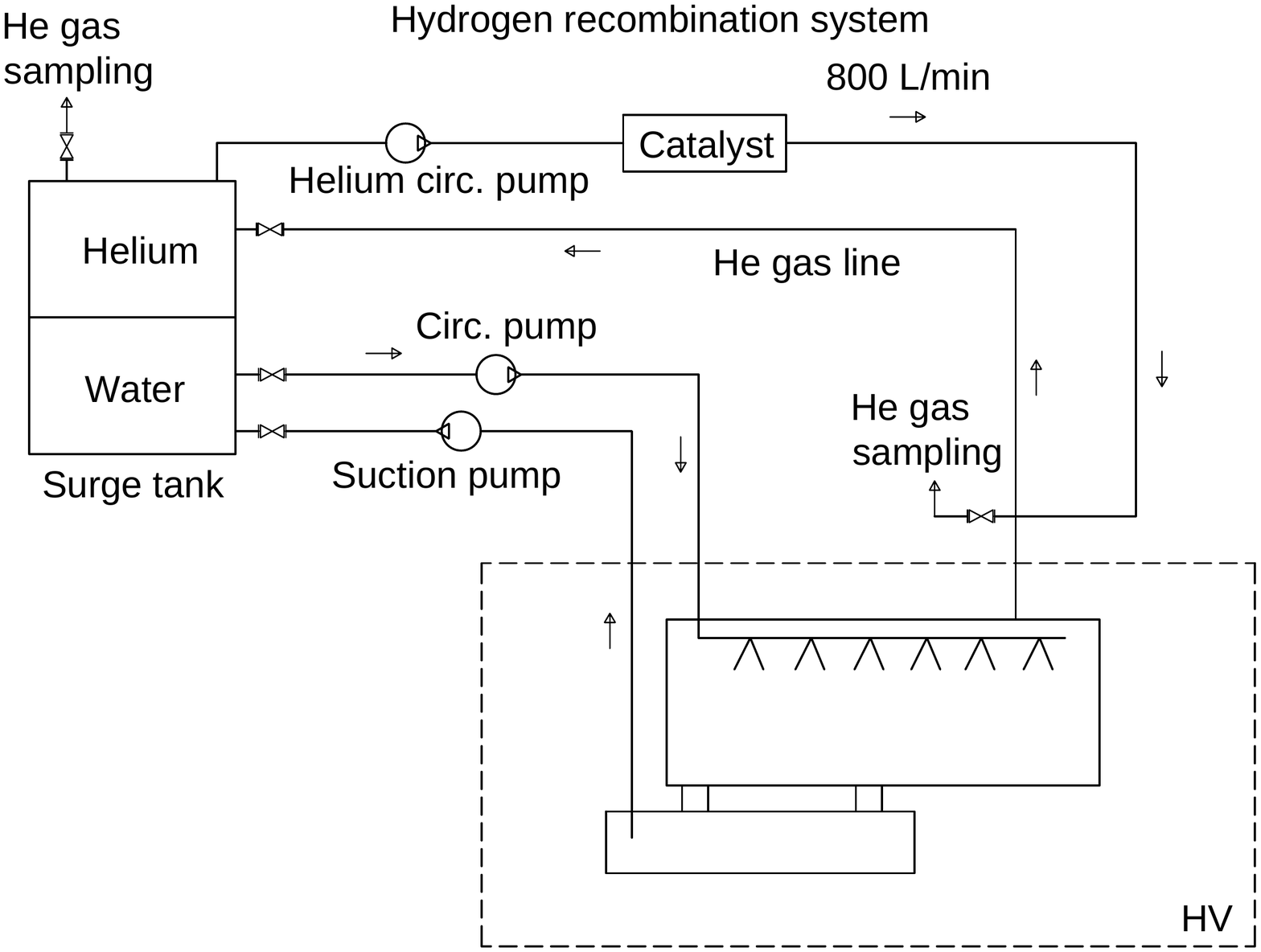}
\caption{Schematic diagrams of hydrogen recombination system. (a) First version (top) and
(b) modified version of the system (bottom).}
\label{fig:hydrogen_recomb}
\end{figure}

In the early days of T2K beam operation, with a beam
power of less than 50 kW, helium gas flushing in the water tank was performed once during every half-day beam stop for 
maintenance (roughly once a week), so as to reduce the hydrogen gas in the helium atmosphere.
For operation at a higher beam power, a hydrogen recombination system was introduced.
A helium circulation loop through a hydrogen recombination catalyst (alumina pellets with 0.5 \% Pd) was 
attached to the water tank, as shown in Fig. \ref{fig:hydrogen_recomb} (a), and the helium gas flowed at 50 L/min.
A helium gas sampling port was added to the water tank to measure the hydrogen density inside the tank.
Using this hydrogen recombination system, the measured hydrogen density in the water tank after 1 week of beam operation
at 150 kW was approximately 0.5 \%. For a beam power greater than 150 kW, 
the hydrogen density gradually increased because the helium flow rate to the catalysts was insufficient. 
In this system, the helium inside each magnetic horn was connected to the water tank by a single tube, thus,
the produced hydrogen gas was expected to reach the water tank by diffusion rather than forced circulation.
Since the actual hydrogen density inside the magnetic horns could not be measured, 
weekly helium gas flushing was still implemented so as to reduce the hydrogen density.
In order to improve the hydrogen recombination rate, the single path connection between the magnetic horns and 
water tank was partially modified to achieve forced circulation, as shown in Fig. \ref{fig:hydrogen_recomb} (b).
In this system, forced circulation was available outside the HV and the helium flow rate was greatly improved, to 800 L/min.
A helium gas sampling port was added near the HV to measure the hydrogen density near the magnetic horns.
During beam operation in Run4, the hydrogen density was measured near the HV after 1 week of continuous 
beam operation at 220 kW. The measured density was 1.6 \% at horn-1, however, the estimated hydrogen
production was 1.5 \%. This means that the majority of the hydrogen produced inside the magnetic horns did not diffuse
to the circulation line and, therefore, the hydrogen accumulated gradually.
From the middle of Run4, the helium gas flushing scheme was modified so that flushed helium gas could 
pass through the inner volume of the magnetic horns using the water supply line.
To achieve this, water circulation was stopped on weekly maintenance days and helium flushing through 
the magnetic horns was performed as a temporary solution.
Forced helium circulation inside the horns is necessary for higher-power beam operation.
This will be realized in the next version of the magnetic horns through the addition of a forced circulation line 
to the magnetic horns and support modules.
 
\subsubsection{Air contamination and acidification of cooling water}

The interiors of the magnetic horns and gas region of the water tank were filled with helium gas
and both were connected with a single tube.
Before helium gas was added to the interiors of the horns, the internal spaces were first evacuated
to a level of 1,000 Pa when the HV was evacuated. The interiors were then filled with helium gas\footnote{This 
procedure was conducted because the magnetic horn conductors are not designed
to withstand external pressure caused by evacuation of the interior.}. 
The air contamination was essentially 1 \% inside the horns. Filling of the helium
gas into the water tank was performed by helium gas flushing, because the walls of the water tank
were not sufficiently thick to withstand external pressure by evacuation. An air contamination level of a few percent was possible.
Such air contamination leads to acidification of the coolant water through the following process:
Beam exposure to nitrogens creates nitrogen oxides, which are dissolved in water,
and nitric acids can then be created, causing the water to become acidized.

During Run3, it was found that the coolant water was whitened after two months of beam operation. 
Through several investigations, the contents were found to be aluminum oxides with a density of 51 mg/L in the water. 
This corresponded to a total amount of 138 g in the entire water (2.7 m$^3$).
The hydrogen ion concentration (pH) of the whitened water was $\sim$3.8, which is strongly acidic.
Therefore, it is believed that the aluminum oxides in the water were created because of aluminum corrosion, 
since the corrosion rate of aluminum in strong acid (pH $<$ 4) is higher than in neutral conditions (5 $<$ pH $<$ 7).
The reduction in aluminum conductor thickness was estimated to be only 4.9 $\mu$m (0.16 \% of the original thickness), 
assuming corrosion occurred evenly along the entire conductor surface. 
This amount of thickness reduction did not affect the mechanical strength of the conductors.

\begin{figure}
\centering
\includegraphics[clip,width=0.47\textwidth,bb=20 10 724 471]{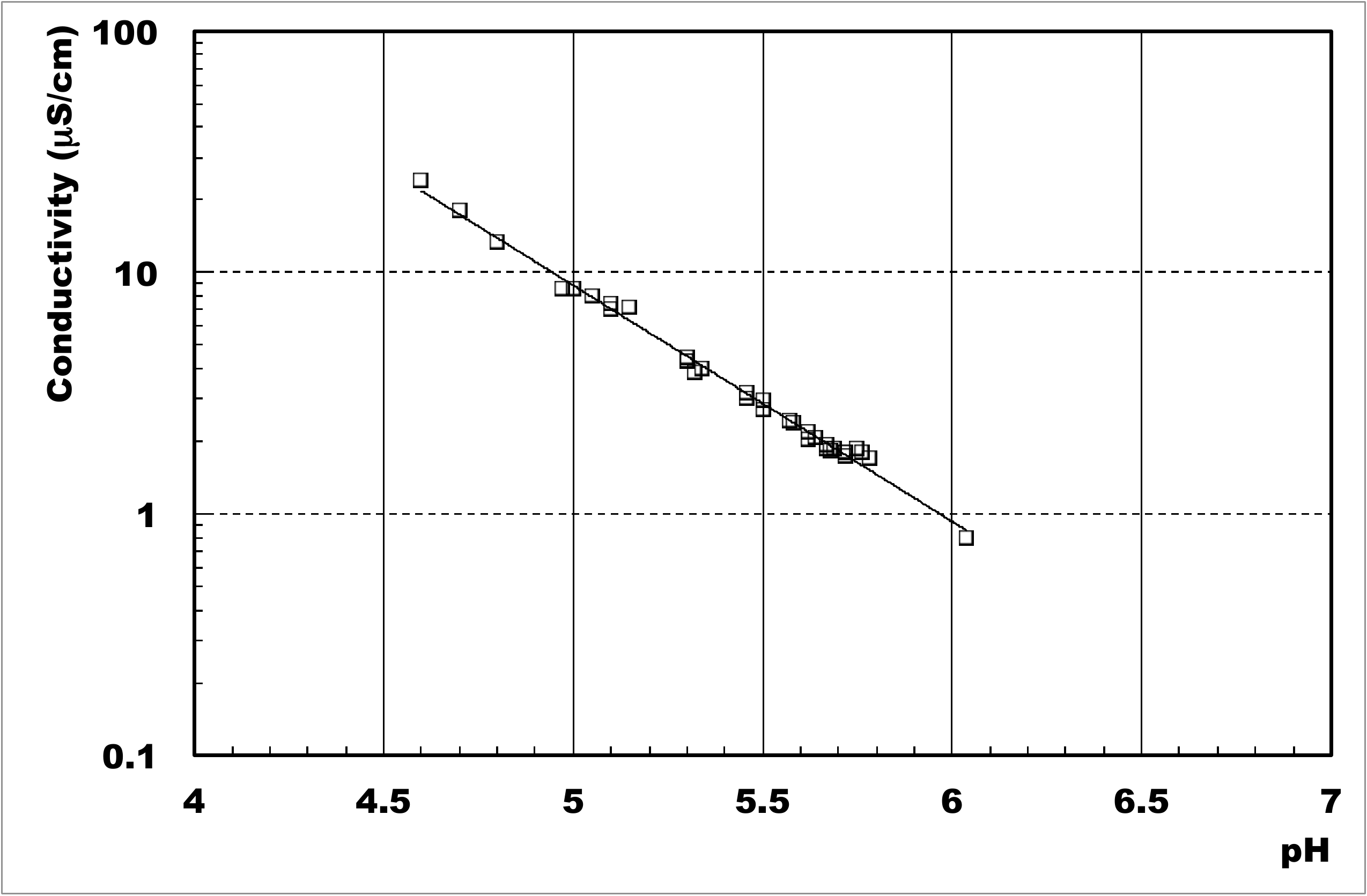}
\caption{Two-dimensional plot of pH (horizontal axis) and conductivity (vertical axis) of circulating water measured during Run4.}
\label{fig:pH_vs_conductivity}
\end{figure}

In the shutdown period between Run3 and Run4, a pH monitor system was adopted to the water circulation system.
This monitor system can measure the pH of the circulated water by sampling a few liters of the water rather than requiring
the circulating water to be continuously passed through the monitor\footnote{For precise measurement of pH of pure water, 
a solution including a few percent of KCl is usually used as a reference. Cl$^-$ cations dramatically increase the corrosion rate, 
thus, the measured water is not typically returned to the circulation line, but it is instead drained away.}. 
A series of pH measurement procedures can now be conducted automatically from a remote control panel at the ground floor of 
the TS.
Since Run4, a portion of the circulating water has been continuously passed through ion exchangers to remove nitric acid ions 
and to neutralize the water.
The pH of the circulating water has been frequently remotely measured, although approximately 4 L of the circulating water is 
drained each time. Even though the pH is not monitored continuously, it can be inferred from the water conductivity, 
which is monitored on a continuous basis.
Figure \ref{fig:pH_vs_conductivity} shows a two dimensional plot of the pH and conductivity of the circulating water 
measured during Run4.
There is a clear relationship between the pH and conductivity, therefore, the pH of the circulating water
can be estimated using the water conductivity. During continuous beam operation in Run4, the pH of the circulating water
remained at approximately 5.7, because of the ion exchanger path. Under these conditions, the aluminum corrosion rate is very small. 
Thereafter, the pH remained at $\sim$5.7 throughout Run4, which meant minor air contamination occurred continuously 
at some point in the helium gas plumbing.
However, this minor contamination did not lead to aluminum corrosion because of the continuous operation of the ion exchangers.

\subsection{Water leak}
\label{sec:leak}

During the Run4 period, it was found that the water level in the water tank gradually decreased at a rate of $\sim$10 L/d.
Since no leak was found in any of the exterior plumbing outside the HV, it was more likely that the magnetic horns were leaking. 
The leaked water was contained inside the HV and no external leaks occurred.
Since the helium gas inside the HV is circulated by a helium compressor, water vapor is condensed after passing
through the compressor and collected at a drain tank in the helium circulation line.
The rate of increase in the condensed water in the drain tank was almost the same as the water leakage rate, 
thus, the amount of accumulated leaked water at the bottom of the vessel was expected to have been very small 
compared to the total amount of leaked water. However, a small fraction of the leaked water still remained at the bottom
of the HV and actually corroded the bottom steel plate.

Water circulation tests were performed several times to investigate which horn was leaking.
The investigation was conducted by circulating water for one horn only and by checking the water level.
The results indicated that a decrease in the water level occurred at horn-3. 
In the shutdown period after Run4, an opportunity to open the top lid of the HV was provided.
Since no direct access to the horns is permitted, visual inspection was performed using a camera
controlled by a wireless network. The camera was attached to the bottom end of an 8-m-long rod, and the rod was 
lowered through 3-cm gaps between shielding blocks to obtain access to the level of the horns. 
This visual inspection was conducted for all three horns and it was found that horn-1 was leaking, while no leakage was observed
at horn-2 and horn-3. Water drops from upstream of the horn-1 conductors were visible, rather than from the water
nozzles or drain ports.  Upstream of the conductors, two aluminum knife-edge seals were used on both sides of the large 
ceramic ring. There are two possible causes of the leak: one is corrosion of the aluminum knife-edge seals,
which occurred when the water with pH$<$4 was circulated during Run3; 
the other may be that the seals were not sufficiently compressed when horn-1 was assembled.
In this case, when the axial forces of the aluminum bolts used to compress the aluminum knife-edge seals
were reduced by the thermal expansions due to beam heating, a leak would occur. During the shutdown period after Run4,
all three magnetic horns were replaced with second-version horns, primarily for upgrade purposes. Thus, the leak
problem was solved. For countermeasures against a water leak, the upstream aluminum knife-edge seals of the second
horn-1 were sufficiently compressed, which would decrease the probability of a leak developing.
Alternative seal designs are also studied for future horn-1 production.

\subsection{Power supply issues}

From Run1 to Run4, there were several changes in power supply and operation configuration.
At the beginning of the T2K experiment, two power supplies that were previously used for the K2K magnetic horns, 
which were manufactured in 1994 and 1998, were refurbished.
Horn-1 was operated by one power supply and horn-2 and horn-3, connected in series, were operated by another power supply.

A new power supply was fabricated in summer 2010 and began operation during Run2.
The new power supply was designed to operate all three magnetic horns simultaneously and had
a rated output current of 35 kA, a rated operation voltage of 10 kV, and a capacity of 7.5 mF.
It also had an energy recovery system. Details of this new power supply are given in \cite{NKCPS}.
Because the total resistance and inductance are large due to the three-horn operation powered by one power supply, 
the pulse width expanded to 4.6 ms.
This power supply was operated at 250-kA for physics data-taking throughout Run2, without any complications.

The power supply experienced a serious problem due to a malfunction in December 2011. Several IGBT switches
located between the charging unit and the capacitor bank were broken as a result of this problem. 
One of the K2K power supplies was reinstalled for operation of the three horns in series. 
Since the three-horn operation is a heavy requirement for the K2K power supply, several measures 
were taken to ensure safe operation. The power supply was operated with a charging voltage of 6.7 kV and charging current of 20 A,
both of which are close to the rated values. Three-horn operation powered by the K2K power supply was initiated
at the beginning of Run3 in early March 2012. 
Several days later, the power supply suddenly stopped as a result of a thyristor problem. One of three thyristors with
a rated voltage of 4 kV was shorted by some fault. After replacement with a spare thyristor stack with the same specifications 
as the broken device, the power supply began operation with a reduced current of 200 kA for safety.
It was thought that, because of the high-voltage operation
at 6.7 kV, a higher voltage than specified may have been applied to one of the thyristors as a result of an unexpected voltage 
unbalance among the thyristors. 
Two weeks later, the thyristor stack was replaced with a stack with a higher rated voltage (7 kV), which can
withstand such a voltage unbalance, even with one thyristor.
After the replacement, the power supply was operated without any difficulty for a year (the remainder of Run3 and half of Run4).

The new power supply was reintroduced in January 2013, a year after the initial problem. 
Then, the horn operation was returned to a two-power-supply configuration, with
the new supply being used for horn-1 and the old supply powering horn-2 and horn-3.
This was to reduce the operation voltages, which could greatly help to lower the risk of failure.
The two-power-supply configuration continued for the latter half of Run4. Table \ref{tab:PS_condition} summarizes 
the operation configurations for the Run1-4 periods.
\begin{table}[htb]
\centering
\caption{Summary of operation configuration for Run1-4 periods.}
\small
\begin{tabular}{lllll}
\hline
Period & Config. & Current & Voltage & Width \\ \hline
Run1 & horn-1 (K2K) & 250 kA & 4.5 kV & 2.4 ms\\
& horn-2+3 (K2K) & 250 kA & 5.4 kV & 3.6 ms\\
Run2 & horn-1+2+3 (New) & 250 kA & 6.2 kV & 4.6 ms\\
Run3 (early) & horn-1+2+3 (K2K) & 200 kA & 5.4 kV & 4.3 ms \\
Run3 (late) & horn-1+2+3 (K2K) & 250 kA & 6.7 kV & 4.3 ms \\
Run4 (early) & horn-1+2+3 (K2K) & 250 kA & 6.7 kV & 4.3 ms\\
Run4 (late)   & horn-1 (New) & 250 kA & 4.1 kV& 3.1 ms \\
& horn-2+3 (K2K) & 250 kA & 5.4 kV & 3.6 ms\\ \hline
\end{tabular} 
\label{tab:PS_condition}
\end{table} 


%
%
\section{Conclusion}
\label{sec:summary}

A magnetic horn system to realize a high-intensity neutrino beam was developed for the T2K experiment,
and an electrical system for 320-kA operation was constructed. 
The heat transfer coefficient at the inner conductor surface following the use of sprayed water as a coolant 
is as high as 7.9 kW/m$^2\cdot$K.
This is sufficiently high to maintain maximum temperatures of below 50$^\circ$C.
However, for the stripline cooling by helium, the required flow rate was not achieved because of a helium leak 
around the remote clamp, which will be modified in the next version of this device.
The conductor shapes are optimized using FEM to ensure that the stress is maintained below an allowed value,
and the measured vibrations are consistent with the predictions of the FEM analyses.
Operation experience acquired during the T2K beam periods is also described.
The T2K magnetic horns have been operated stably at 250 $\pm$ 5 kA overall, however,
several problems such as hydrogen production and air contamination were observed.
The majority of these issues were solved, but countermeasures against the hydrogen problem are still under development.
The first set of T2K magnetic horns has been operated for over 12 million pulses during four
years of operation, from 2010 to 2013, under a maximum beam power of 230 kW.
No significant damage has been observed during this period. 
This successful operation has led to the discovery of the
$\nu_{\mu}\rightarrow\nu_e$ oscillation by the T2K experiment in 2013 \cite{nue_app}.


\section*{Acknowledgments}

The authors thank the Japanese Ministry of Education, Culture, Sports, Science, and Technology (MEXT) for their support
of the T2K magnetic horn development. The T2K magnetic horns have been built and operated using funds provided by
MEXT, Japan, and the Department of Energy, U.S.A.
In addition, the development of the T2K magnetic horns was partly supported by Grands-in-Aid for Scentific Research in Japan.

The authors would like to acknowledge the tremendous support and effort from the J-PARC accelerator division,
and would also like to thank Prof. Yoshioka, Prof. Kobayashi, and Prof. Matsumoto from KEK for their kind support
and useful advice. Finally, the authors would like to appreciate many discussions with researchers, who work for neutrino
beamlines in the world, and useful information from them.




\begin{thebibliography}{50}

\bibitem{T2K-LOI}
Y. Itow \textit{et al}., arXiv:hep-ex/0106019 (2001).

\bibitem{J-PARC}
Y. Yamazaki \textit{et al}., KEK Report 2002-13 (2003); Y. Yamazaki \textit{et al}., JAERI-Tech 2003-44 (2003);
Y. Yamazaki \textit{et al}., J-PARC-03-01 (2003).

\bibitem{SK-NIM}
S. Fukuda \textit{et al}., 
Nucl. Instr. Meth. \textbf{A501}, 418 (2003).

\bibitem{off-axis}
D. Beavis, A. Carroll, I. Chiang \textit{et al}., Proposal of BNL AGS E-889, (1995).

\bibitem{T2K-NIM}
K. Abe \textit{et al}., 
Nucl. Instr. Meth. \textbf{A659}, 106 (2011).

\bibitem{T2KhornDesign}
A.K. Ichikawa, 
Nucl. Instr. Meth. \textbf{A690}, 27 (2012).

\bibitem{Alhandbook}
Japan Aluminum Association \textit{et al}., Aluminum Handbook, 7th edition (2007), in Japanese.

\bibitem{K2K}
S.H. Ahn \textit{et al}.,
Phys. Lett. \textbf{B511}, 178 (2001).

\bibitem{nuflux}
K. Abe \textit{et al}.,
Phys. Rev. D \textbf{87}, 012001 (2013).

\bibitem{rad_water}
M. Hagiwara \textit{et al}., Prog. Nucl. Sci. Tech., \textbf{3}, 56 (2012).

\bibitem{water_radioactivity}
K. Bessho \textit{et al}.,
Proceedings of the Eleventh Meeting of the Task-Force on Shielding Aspects of Accelerators,
Targets and Irradiation Facilities (SATIF11), page 61 (2013).

\bibitem{Oyama_NBI2012}
Y. Oyama, Presentation at 8th International Workshop on Neutrino Beams and Instrumentation,
CERN, Geneva, Switzerland, November 2012.

\bibitem{NitricAcid}
Y. Kanda, T. Momose, M. Taira, Radiat. Phys. Chem., \textbf{48}, 49 (1996).

\bibitem{MARS}
MARS Code System, http://www-ap.fnal.gov/MARS/ .

\bibitem{ANSYS}
ANSYS, http://www.ansys.com .

\bibitem{NuMI}
NuMI Technical Design Handbook, 
http://www-numi.fnal.gov/numwork/tdh/tdh\_index.html

\bibitem{Ceramics}
Characteristic of Kyocera Technical Ceramics,
http://global.kyocera.com/prdct/fc/product/pdf/material.pdf 

\bibitem{NKCPS}
K. Koseki, Nucl. Instr. Meth. \textbf{A735}, 633 (2014).

\bibitem{nue_app}
K. Abe \textit{et al}., 
Phys. Rev. Lett. \textbf{112}, 061802 (2014)

\end{thebibliography}
\end{document}